\documentclass[aps, prl, twocolumn, superscriptaddress]{revtex4-1} 
\usepackage{amssymb, amsmath, amsthm, bbm}
\usepackage{xcolor}
\usepackage{graphicx}
\usepackage{hyperref}
\usepackage{comment}
\usepackage{blkarray}
\usepackage{mathtools}
\usepackage{tikz}
\usetikzlibrary{decorations.pathreplacing}
\usepackage{dsfont}
\usepackage{braket}

\usepackage{graphicx}
\usepackage{dcolumn}
\usepackage{bm}

\newcommand{\prlsection}[1]{{\em {#1}.---~}}
\newcommand{\Tr}[1]{\mathrm{Tr}}
\newcommand{\be}{\begin{equation}}
\newcommand{\ee}{\end{equation}}
\newcommand{\1}{\mathbbm{1}}

\newcommand{\titleinfo}{
	Integrable Digital Quantum Simulation:\\
	Generalized Gibbs Ensembles and Trotter Transitions}
\begin{document}
	
	\preprint{APS/123-QED}
	
	\title{\titleinfo
	}
	
    \author{Eric Vernier}
	\affiliation{CNRS \& Universit\'e Paris Cit\'e, Laboratoire de Probabilit\'es, Statistique et Mod\'elisation, F-75013 Paris, France}
	
	\author{Bruno Bertini}
	\affiliation{School of Physics and Astronomy, University of Nottingham, Nottingham, NG7 2RD, UK}
	\affiliation{Centre for the Mathematics and Theoretical Physics of Quantum Non-Equilibrium Systems, University of Nottingham, Nottingham, NG7 2RD, UK}

	\author{Giuliano Giudici}
	\affiliation{Arnold Sommerfeld Center for Theoretical Physics, University of Munich, Theresienstr. 37, 80333 M\"unchen, Germany}
    \affiliation{Munich Center for Quantum Science and Technology (MCQST), Schellingstra\ss{}e~4, 80799 M\" {u}nchen, Germany}

	\author{Lorenzo Piroli}
	\affiliation{Philippe Meyer Institute, Physics Department, \'Ecole Normale Sup\'erieure (ENS), Universit\'e PSL, 24 rue Lhomond, F-75231 Paris, France}

	
	\begin{abstract}
		The Trotter-Suzuki decomposition is a promising avenue for digital quantum simulation (DQS), approximating continuous-time dynamics by discrete Trotter steps of duration $\tau$. Recent work suggested that DQS is typically characterized by a sharp Trotter transition: when $\tau$ is increased beyond a threshold value, approximation errors become uncontrolled at large times due to the onset of quantum chaos. Here we contrast this picture with the case of \emph{integrable} DQS. We focus on a simple quench from a spin-wave state in the prototypical XXZ Heisenberg spin chain, and study its integrable Trotterized evolution as a function of $\tau$. Due to its exact local conservation laws, the system does not heat up to infinite temperature and the late-time properties of the dynamics are captured by a discrete Generalized Gibbs Ensemble (dGGE). By means of exact calculations we find that, for small $\tau$, the dGGE depends analytically on the Trotter step, implying that discretization errors remain bounded even at infinite times. Conversely, the dGGE changes abruptly at a threshold value $\tau_{\rm th}$, signaling a novel type of Trotter transition. We show that the latter can be detected locally, as it is associated with the appearance of a non-zero staggered magnetization with a subtle dependence on $\tau$. We highlight the differences between continuous and discrete GGEs, suggesting the latter as novel interesting nonequilibrium states exclusive to digital platforms. 
	\end{abstract}
	\maketitle

	 \prlsection{Introduction} The intrinsic limitations in the classical simulation of quantum many-body dynamics could be overcome using a quantum computer, adopting the logic of digital quantum simulation (DQS)~\cite{feynman1982simulating,georgescu2014quantum}. As realized early on~\cite{lloyd1996universal}, the Trotter-Suzuki decomposition~\cite{trotter1959product,suzuki1991general} allows one to approximate the continuous-time evolution of a target system by a sequence of elementary steps, which could be implemented as quantum \emph{gates} acting on neighboring qubits. From the experimental point of view, DQS is at an early stage if compared to \emph{analog} quantum simulation~\cite{preskill2018quantum,daley2022practical}. Yet, the past few years have witnessed remarkable progress in the experimental control of platforms for DQS such as trapped-ions~\cite{lanyon2011universal,barreiro2011open,blatt2012quantum, monroe2021programmable} and superconducting circuits~\cite{salathe2015digital,barends2015digital,langford2017experimentally,wendin2017quantum, kjaergaard2020superconducting, bravyi2022future}, motivating much ongoing theoretical research on the subject.
	
	Neglecting noise, the accuracy of DQS depends on the \emph{Trotter step} $\tau$, which controls the number of gates applied per time unit. While many bounds on approximation errors for the system wavefunction exist~\cite{lloyd1996universal,childs2018toward,childs2019nearly,tran2020destructive,childs2021theory,layden2022first}, recent work~\cite{heyl2019quantum} focused on the dynamics of local observables, putting forward the existence of a sharp \emph{Trotter transition}, see also~\cite{ishii2018heating,sieberer2019digital,kargi2021quantum,chinni2022trotter}. In agreement with general results on periodically driven systems~\cite{abanin2015exponentially,abanin2017effective,abanin2017rigorous,mori2016rigorous, kuwahara2016floquet}, it was found that if $\tau$ is small, the discrete and continuous dynamics remain close to one another for a time which is at least exponentially long in $\tau_0/\tau$, $\tau_0$ being some dimensionful constant. Conversely, if $\tau$ increased beyond a threshold value, approximation errors become uncontrolled, corresponding to the onset of quantum chaos~\cite{ishii2018heating,sieberer2019digital,kargi2021quantum}. Such transitions were found also for integrable Hamiltonians~\cite{sieberer2019digital}, as typical Trotterizations break integrability. 
	
	\begin{figure}
		\includegraphics[scale=0.19]{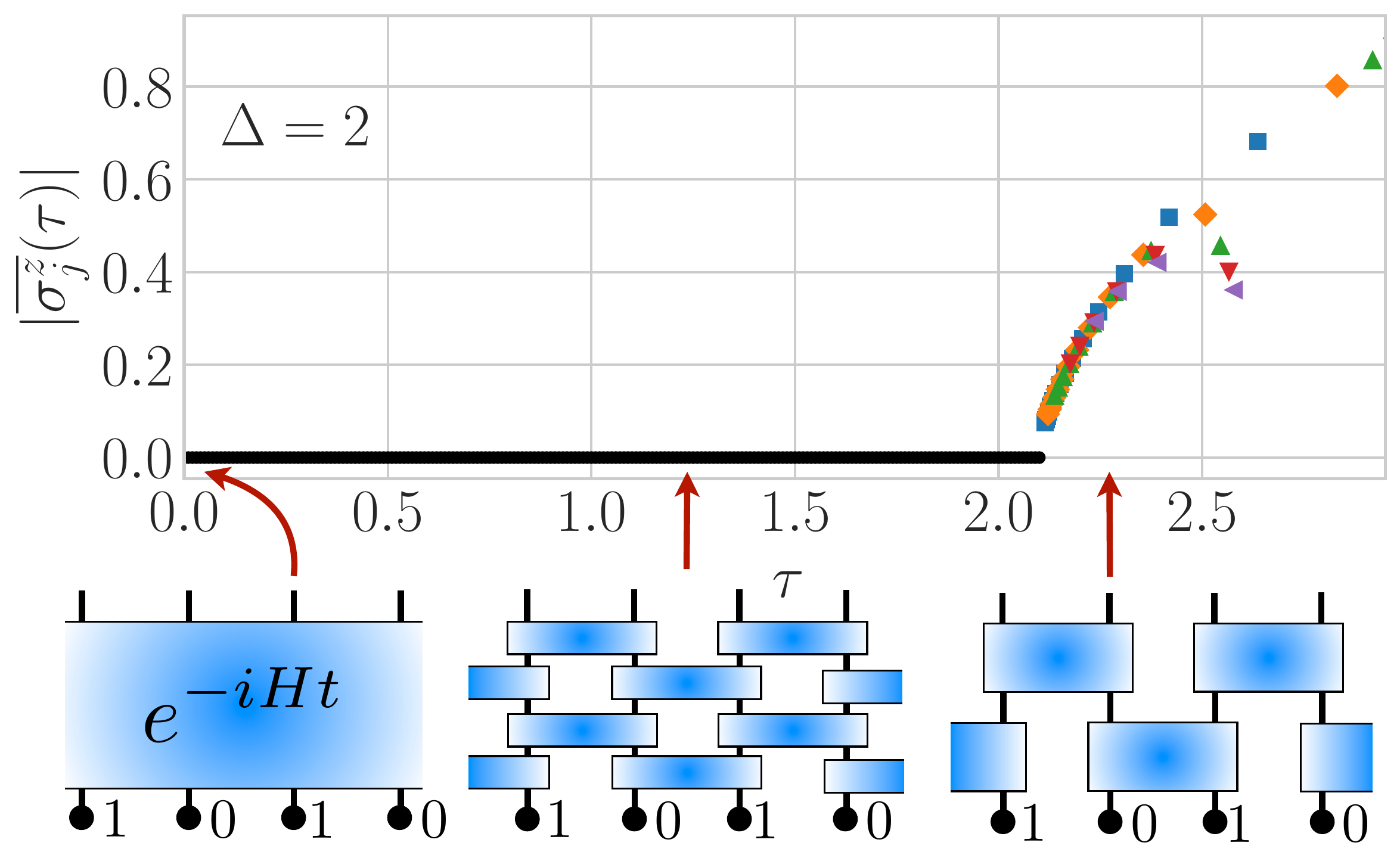}
		\caption{Asymptotic local magnetization $|\overline{\sigma^z_j}(\tau)|$ as a function of the Trotter step $\tau$, after a quench from the N\'eel state. $\overline{\sigma^z_j}(\tau)$, defined in~\eqref{eq:expectation_values}, is zero for the continuous dynamics and up to a threshold value $\tau_{\rm th}$. For $\tau>\tau_{\rm th}$, the dGGE changes abruptly, and the asymptotic local magnetization depends nontrivially on $\tau$. Symbols correspond to the results of analytic computations at special values of $\tau$ cf. the main text.}
		\label{fig:sketch}
	\end{figure}
	
	Here, we contrast this generic picture with the case of \emph{integrable} DQS, focusing on the quench dynamics~\cite{calabrese2006time,calabrese2007quantum} of a prototypical integrable model, the XXZ Heisenberg chain~\cite{korepin1997quantum}, and its integrable Trotterized evolution~\cite{vanicat2018integrable,ljubotina2019ballistic,medenjak2020rigorous}. Due to the local conservation laws, late-time physics is captured by a Generalized Gibbs Ensemble (GGE)~\cite{rigol2007relaxation,vidmar2016generalized,essler2016quench}. We call it \emph{discrete} to distinguish it from the one arising in continuous dynamics --- we will show that it displays qualitatively different features. One may ask how the properties of the discrete GGE (dGGE) depend on $\tau$. We show that, while near the continuous-time limit such dependence is analytical, the dGGE changes abruptly at a threshold value $\tau_{\rm th}$, signaling a novel type of Trotter transition. We anticipate that this transition is due to a sudden change in the structure of local conservation laws, and is therefore different from traditional quantum phase transitions. Our main result is that this novel transition can be detected locally, as it is associated with the emergence of a non-zero staggered magnetization for quenches from a class of initial states, cf. Fig.~\ref{fig:sketch}. Our results highlight dGGEs as novel nonequilibrium states exclusive to DQS platforms, and could be relevant for recent experiments implementing integrable Trotterized dynamics in superconducting quantum processors~\cite{morvan2022formation,maruyoshi2022conserved,keenan2022evidence}.

	 \prlsection{The model} We consider the XXZ Heisenberg model
	\begin{equation}\label{eq:xxz_hamiltonian}
		H=\frac{1}{4}\sum_{j=1}^{L}\left[\sigma^{x}_j\sigma^{x}_{j+1}+\sigma^{y}_j\sigma^{y}_{j+1}+\Delta(\sigma^{z}_j\sigma^{z}_{j+1}-1)\right]\,,
	\end{equation}
	where $L$ is the system size, which we take to be even from now on, $\Delta$ is the anisotropy parameter, while $\sigma^\alpha_j$ are the Pauli matrices acting at position $j$, with $\sigma^{\alpha}_{L+1}=\sigma^{\alpha}_{1}$. In this work we focus on the gapped regime $\Delta> 1$~\cite{korepin1997quantum}, and discuss at the end how our results depend on this choice. The Hamiltonian~\eqref{eq:xxz_hamiltonian} is integrable, with an extensive number of local and quasi-local conservation laws~\cite{ilievski2016quasilocal}. As discussed earlier, the logic behind DQS is to approximate the continuous time evolution $e^{-iHt}$ as a sequence of $M$ unitary operators $U(t/M)$ which can be implemented via a local quantum circuit. This procedure is not unique and, in general, the discretized dynamics does not feature exact local conservation laws~\cite{sieberer2019digital}. Here we consider a special, yet very natural, decomposition introduced in Ref.~\cite{vanicat2018integrable,ljubotina2019ballistic}, which preserves integrability. It is defined by the repeated application of the unitary operator $U(\tau)=U_{e}(\tau)U_{o}(\tau)$
	with
	\begin{equation}\label{eq:floquet_even_odd}
		U_{o}(\tau)=\prod_{n=1}^{L / 2} V_{2 n, 2 n+1}(\tau),\,\, U_{e}(\tau)=\prod_{n=1}^{L/ 2} V_{2 n-1,2 n}(\tau)\,,
	\end{equation}
	and
	\begin{equation}
		V_{n, n+1}(\tau)=e^{-i \frac{\tau}{4}\left[\sigma_n^x \sigma_{n+1}^x+\sigma_n^y \sigma_{n+1}^y+ \Delta (\sigma_n^z \sigma_{n+1}^z-\openone)\right]}\,,
		\label{eq:twositegate}
	\end{equation}
	where $\tau\in\mathbb{R}$ is the Trotter step. The continuous evolution is recovered in the limit $e^{-iHt}=\lim_{M\to\infty}U(t/M)^{M}$.
	
	For finite $t/M=\tau$, we a have brickwork quantum circuit, cf. Fig.~\ref{fig:sketch}, which can be thought of as a discrete dynamics generated by the Floquet operator $U(\tau)$. The latter is integrable: although $U(\tau)$ is not generated by a local Hamiltonian, it features an extensive number of local and quasi-local conserved operators, or \emph{charges}, which can be constructed using a standard transfer-matrix approach~\cite{vanicat2018integrable,ljubotina2019ballistic,medenjak2020rigorous}. For small $\tau$, the charges may be thought of as a deformation of those of the Hamiltonian~\eqref{eq:xxz_hamiltonian}. More precisely, for each charge $Q_k$, with $[Q_k,H]=0$, we have two new operators $\tilde{Q}^{\pm}_k(\tau)$ with $[\tilde{Q}^{\pm}_k(\tau),U(\tau)]=0$. The charges $\tilde{Q}^{\pm}_k(\tau)$ break the single-site translation symmetry $\mathcal{T}$, which map one onto the other, $\mathcal{T}\tilde{Q}^{\pm}_k(\tau) \mathcal{T}^{\dagger}=\tilde{Q}^{\mp}_k(\tau)$, whereas both $U(\mathcal{\tau})$ and $\tilde{Q}^{\pm}_k(\tau)$ are invariant under a shift of two sites, $\mathcal{T}^2$. The first pair of such charges
	\begin{equation}\label{eq:deformed_ham}
		Q^{\pm}_{1}=:\tilde{H}^{\pm}(\tau)=\sum_{j} h^{\pm}_{2j,2j+1,2j+2}(\tau)\,,
	\end{equation}
	map to the Hamiltonian~\eqref{eq:xxz_hamiltonian}, \emph{i.e.} $\tilde{H}^{\pm}(\tau)\to H$ in the limit $\tau\to0$. Here $h^{\pm}_{2j,2j+1,2j+2}$ is an operator supported over three neighboring spins, cf.~\cite{SM} for the exact expression. 
	
 \prlsection{The quench protocol} Most of existing works studying integrable Trotterizations focused on transport~\cite{vanicat2018integrable,ljubotina2019ballistic}. Here we are interested in the quench dynamics from simple initial states, see also~\cite{medenjak2020rigorous, giudice2022temporal}, in which linear response theory does not apply. It is natural to consider states respecting the two-site translation symmetry of the brickwork circuit $U(\tau)$. We will focus on the N\'eel state
	\begin{equation}\label{eq:neel}
		\ket{\Psi_0}=\ket{0}_1\otimes \ket{1}_2\otimes \cdots \otimes \ket{0}_{L-1}\otimes \ket{1}_{L}\,,
	\end{equation}
	where $\ket{0}_x$, $\ket{1}_x$ are the basis elements of the space at position $x$. This state breaks both the translation ($\mathcal{T}$) and the spin-flip ($\mathcal{S}$) symmetry $\sigma_j^z\to-\sigma_j^z$, $\sigma_j^{\pm}\to\sigma_j^{\mp}$. However, it is invariant under the joint action $\mathcal{T}\mathcal{S}\ket{\Psi_0}=\ket{\Psi_0}$. Later, we will discuss more general initial states. We will be interested in the thermodynamic limit, and focus on local observables at late times after the quench, namely
	\begin{equation}\label{eq:expectation_values}
		\bar{\mathcal{O}}_x(\tau):=\lim_{t\to\infty}\lim_{L\to\infty}\braket{\Psi_0| [U^{\dagger}(\tau)]^{t/\tau} \mathcal{O}_x U(\tau)^{t/\tau} |\Psi_0}\,,
	\end{equation}
	where $\mathcal{O}_x$ is an operator with support at position $x$. Quantum quenches in the model~\eqref{eq:xxz_hamiltonian} have been studied extensively in the continuous-time limit. It is known that the expectation values~\eqref{eq:expectation_values} is captured by a GGE~\cite{rigol2007relaxation,vidmar2016generalized}, generalizing the thermal Gibbs density matrix: It is the ensemble maximizing the entropy such that expectation values of the charges match those in the initial state. This construction straightforwardly extends to the discrete dynamics~\eqref{eq:floquet_even_odd} and defines the dGGE. Obtaining a quantitative description of the GGE is a notoriously difficult problem~\cite{caux2016quench}, which has been solved only in some cases for continuous-time evolution. As our first main result, we will generalize the tools developed in theory of quantum quenches in integrable models~\cite{essler2016quench,caux2016quench,calabrese2016introduction,piroli2017integrable} and provide an analytic description of the dGGE for the N\'eel state~\eqref{eq:neel} (our techniques apply to a broader class of initial states, as explained later). Before proceeding, we recall some basic facts about the Floquet operator~\eqref{eq:floquet_even_odd}.
	
	 \prlsection{The quasiparticle picture} The spectrum of the Floquet operator $U(\tau)$ can be found analytically via the Bethe ansatz~\cite{aleiner2021bethe,pieter2022correlations,miao2022floquet}. Here we present the aspects which are directly relevant for us, and refer to Refs.~\cite{aleiner2021bethe,SM} for more detail. Introducing the parameters~\cite{ljubotina2019ballistic}
	 \begin{subequations}
	\begin{align}\label{eq:def_x}
	\gamma&=\arccos \left[\sin (\Delta \tau/2)/\sin (\tau/2)\right]\,,\\
	x&=i \operatorname{arcsinh}\left[\sin (\gamma) \tan (\tau/2)\right]\,,
	\end{align}
	 \end{subequations}
	there are two cases. First, if $\gamma=i\eta$, with $\eta\in\mathbb{R}$, then $x\in\mathbb{R}$ and $\tilde{H}^{\pm}(\tau)$ in~\eqref{eq:deformed_ham} are gapped. Conversely, if $\gamma\in\mathbb{R}$ then $x$ is purely imaginary, and $\tilde{H}^{\pm}(\tau)$ are gapless. We will refer to these cases as gapped and gapless, respectively, although we stress that $\tilde{H}^{\pm}(\tau)$ does not generate the dynamics, \emph{i.e.} $U(\tau)\neq e^{-i\tau \tilde{H}^{\pm}(\tau)}$. The spectrum of $U(\tau)$ is organized into sectors labeled by the number of ``magnons'' $M$, namely $M$ is the quantum number associated with the conserved operator $\hat{M}=\sum_j (\openone-\sigma_j^z)/2$. The eigenstates are parametrized by sets of complex numbers $\{p_j\}_{j=1}^M$, satisfying the quantization conditions~\cite{aleiner2021bethe} 
	\begin{equation}\label{eq:bethe_eq}
		\left[\frac{f_x^{+}(p_i)}{f_x^{-}(p_i)}\right]^{\frac{L}{2}}=\prod_{k \neq j}^m \frac{\sinh \left(p_j-p_k+i \gamma\right)}{\sinh \left(p_j-p_k-i \gamma\right)}\,,
	\end{equation}
	where $f_x^{\pm}(p)=\sinh (p+ i \frac{x}{2}\pm i \frac{\gamma}{2})\sinh (p- i\frac{x}{2}\pm i \frac{\gamma}{2})$. Physically, $p_j$ are related to the quasimomenta $\lambda_j$, or \emph{rapidities}, of the quasiparticles: We have $\lambda_j=p_j\in \mathbb{R}$ and $\lambda_j=ip_j\in[-\pi/2,\pi/2]$ in the gapless and gapped regimes, respectively. When $x=0$, Eqs.~\eqref{eq:bethe_eq} coincide with the standard Bethe equations for the model~\eqref{eq:xxz_hamiltonian}. In this case, we have a simplification in the thermodynamic limit $L,M\to\infty$, with the density $D=M/L$ kept fixed: The string hypothesis~\cite{takahashi2005thermodynamics} states that the rapidities organize themselves into sets of $n$ elements forming a ``string'', which is interpreted as a bound state of $n$ quasiparticles.	Each one is associated with a string center $\lambda$, corresponding to the bound-state quasimomentum. Accordingly, macrostates are described by the functions $\rho_n(\lambda)$: in a large volume $L$, $L\rho_n(\lambda) d\lambda$ yields the number of $n$-quasiparticle bound states with rapidities in the interval $[\lambda,\lambda+d\lambda]$~\cite{takahashi2005thermodynamics}. Analogously to the case of free quantum gases, one also introduces the distribution function $\rho_{n}^{h}(\lambda)$ for the quasiparticles \emph{holes}, \emph{i.e.} the allowed values of the rapidities which are not occupied~\cite{takahashi2005thermodynamics}. In the following, we will assume that the string hypothesis also holds for $x\neq 0$, extending this thermodynamic description to the discrete dynamics~\footnote{The validity of this assumption is verified a posteriori, based on the agreement of our predictions with numerical computations}.

	\begin{figure}
		\includegraphics[scale=0.41]{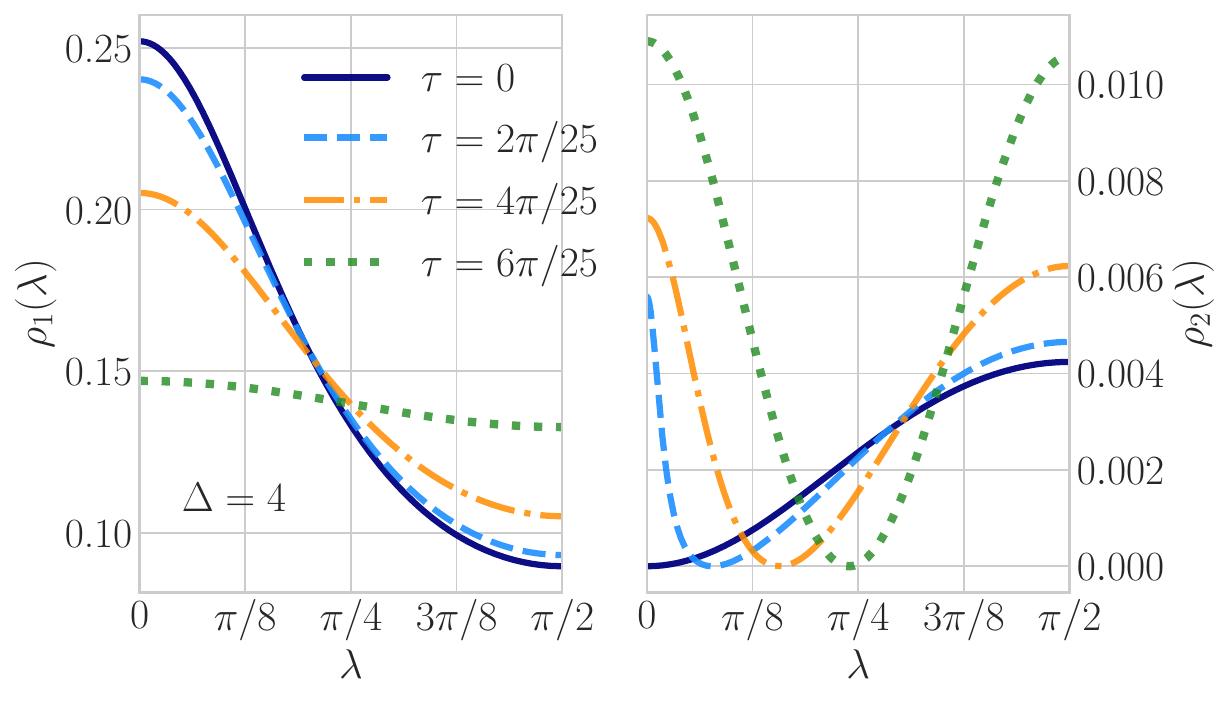}
		\caption{dGGE quasiparticle distribution functions $\rho_n(\lambda)$ for $n=1,2$ and different values of $\tau$ in the gapped phase. The functions are symmetric with respect to $\lambda=0$, and the plot only shows the region $[0,\pi/2]$.}
		\label{fig:root_densities}
	\end{figure}
		
	\prlsection{The dGGE} In order to provide a quantitative description of the dGGE, we compute exactly the corresponding set of functions $\rho_n(\lambda)$ and $\rho_n^h(\lambda)$. This is a hard problem that, in the continuous-time limit, was first solved in Refs.~\cite{wouters2014quenching,pozsgay2014correlations,brockmann2014quench,mestyan2015quenching} via the so-called quench-action approach~\cite{caux2013time,caux2016quench}. Here we follow a different strategy, developed in Refs.~\cite{Pozsgay_2013, piroli2017integrable,Piroli_Pozsgay_Vernier_2017,Piroli_Pozsgay_Vernier_2018}, that can be applied analytically for certain classes of ``integrable'' initial states~\cite{piroli2017integrable,pozsgay2019integrable}. It is based on the study of the so-called quantum transfer matrix, generating a suitably-defined space-time rotated dynamics~\cite{klumper_integrability_2004,piroli2017integrable}. This approach can be naturally extended to the discrete setting consisted here. This step, however, is technical and we report it in the Supplemental Material~\cite{SM}, see also~\cite{InPrep}. Here, we simply present the final result of our analysis. 
	
	The structure of the solution depends on the value of $\gamma$. For definiteness, let us consider the gapped regime $\gamma=i\eta$, with $\eta\in\mathbb{R}$, which holds for small $\tau$. Introducing the standard notation~\cite{takahashi2005thermodynamics} $\eta_n(\lambda)=\rho_n^h(\lambda)/\rho_n(\lambda)$, we derive the following analytic expression for the dGGE
    \begin{equation}  
    \!\!\!\eta_1(\lambda)  = -1+\left(1 + \mathfrak{a}(\lambda-i\eta/2) \right)
    \left(1 + 1/\mathfrak{a}(\lambda+i\eta/2) \right), 
    \label{eq:eta1gapped}
    \end{equation} 
    where 
    \begin{equation} 
    \!\!\!\!\!\mathfrak{a}(\lambda) =  \frac{\sin(2\lambda+i\eta)}{\sin(2\lambda-i\eta)}
    \frac{\sin(\lambda- x/2-i\eta)}{\sin(\lambda+ x/2+i\eta)}
    \frac{\sin(\lambda- x/2)}{\sin(\lambda+ x/2)},
    \label{eq:afrakgapped}
    \end{equation} 
	while $\eta_n(\lambda)$ for $n>1$ are defined by 
	\begin{equation}\label{eq:higher_n}
	\eta_{n+1}(\lambda)=-1+\eta_n(\lambda+i\eta/2)\eta_n(\lambda-i\eta/2)/[1+\eta_{n-1}(\lambda)]
	\end{equation}
 with $\eta_0(\lambda)\equiv 0$. Eqs.~\eqref{eq:eta1gapped}--\eqref{eq:higher_n} are our first main results. 
 
For a given solution $\eta_n(\lambda)$ of the above equations, one can obtain the functions $\rho_n(\lambda)$ via the following integral equations 
	\begin{equation}
		\rho_n(\lambda)[1+\eta_{n}(\lambda)]=a^{(x/2)}_n(\lambda)-\sum_{m=1}^{\infty}\left(a_{n m} \ast \rho_m\right)(\lambda),
		\label{eq:thermo_bethe}
	\end{equation}
which are obtained from the thermodynamic limit of Eqs.~\eqref{eq:bethe_eq}~\cite{SM}. Here, $(f\ast g)(\lambda):=\int_{-\pi/2}^{\pi/2}d\mu f(\mu-\lambda)g(\mu)$, while we introduced the notation $f^{(x)}(\lambda)=(f(\lambda+x)+f(\lambda-x))/2$, and defined
	\begin{align}
		a_{nm}(\lambda)=&(1-\delta_{nm})a_{|n-m|}(\lambda)+2a_{|n-m|+2}(\lambda)\nonumber\\
		+&\ldots +2a_{n+m-2}(\lambda)+a_{n+m}(\lambda)\,,
	\end{align}
	with  $a_n(\lambda)=\pi^{-1}\sinh\left( n\eta\right)/[\cosh (n \eta) - \cos( 2 \lambda)]$. Eqs.~\eqref{eq:thermo_bethe} can be solved numerically by standard iterative approaches~\cite{takahashi2005thermodynamics}. An example of the solution is reported in Fig.~\ref{fig:root_densities} for different values of $\tau$. 
	
	Similar analytic solutions can be obtained from any of the integrable states of the XXZ Hamiltonian~\cite{piroli2017integrable}. These include, among others, all product states $\ket{\Psi_0}=\ket{\psi}^{\otimes L/2}$, where $\ket{\psi}$ is an arbitrary two-qubit state. The derivation is non-trivial, and will be reported elsewhere~\cite{InPrep}.	
	
	\begin{figure}
		\hspace*{-3mm} \includegraphics[scale=0.48]{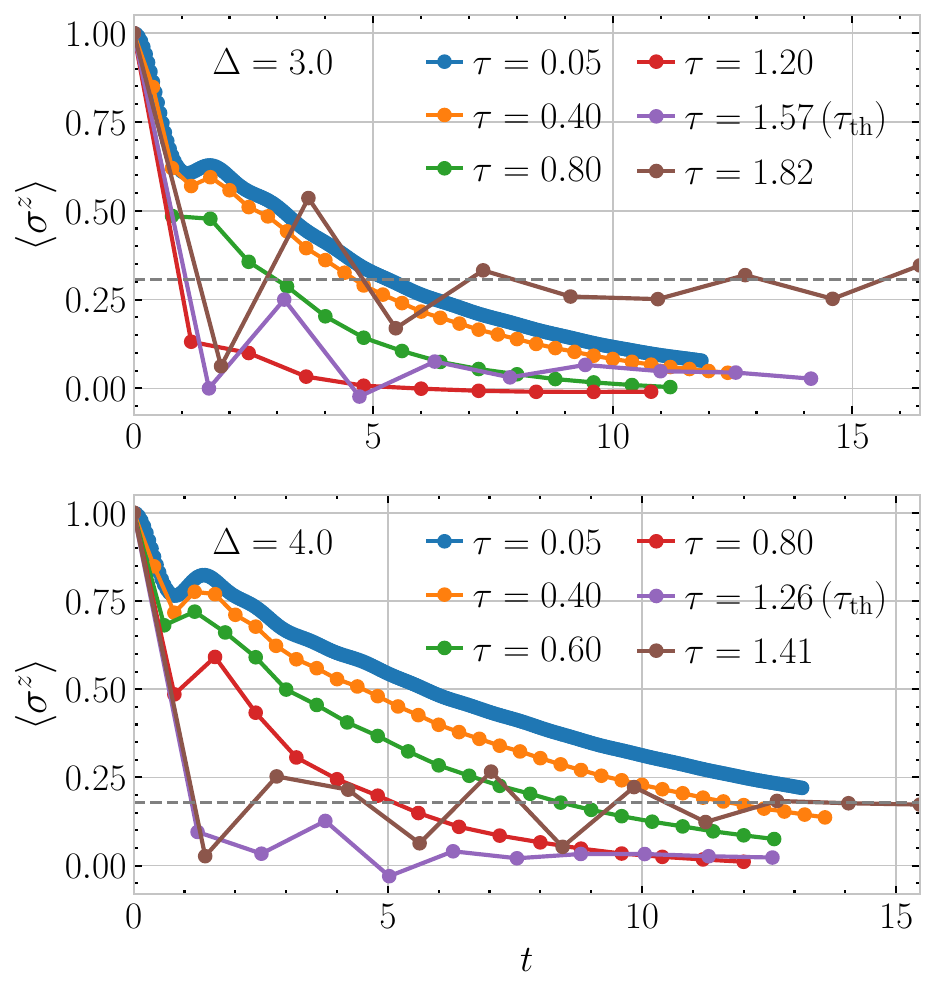}
		\caption{Time evolution of the staggered magnetization. The plots are consistent with a vanishing value of $\overline{\sigma_j^{z}}(\tau)$ for $\tau<\tau_{\rm th}(\Delta)$. We also plot the dynamics for one value of $\tau>\tau_{\rm th}(\Delta)$, corresponding to $\gamma=\pi/3$ (brown lines). We see very good agreement with our analytic prediction~\eqref{eq:finalresultmain} (dashed lines).}
		\label{fig:realtime}
	\end{figure}

\prlsection{The Trotter transition} Eqs.~\eqref{eq:eta1gapped}--\eqref{eq:higher_n} describe the dGGE in the gapped phase, corresponding to $\Delta>1$ and small $\tau$. When the Trotter step is increased beyond the threshold value
\begin{equation}\label{eq:transition}
	\tau_{\rm th}(\Delta)=\frac{2\pi}{\Delta+1}\,,
\end{equation} 
the system enters the gapless regime, and the solution for $\gamma$ in Eq.~\eqref{eq:def_x} becomes real.~\footnote{Interestingly, for $\tau$ approaching $\tau_{\rm th}(\Delta)$ from below, the stationary state described by Eqs.~\eqref{eq:eta1gapped}--\eqref{eq:higher_n} approaches the one of the isotropic Heisenberg chain in continuous time}. This phase persists up to $\tau=2\pi/\Delta$, after which further phase transitions appear~\cite{SM}. We will restrict ourselves to this first gapless phase, but similar analyses can be carried out in the other cases.

The structure of the quasiparticle spectrum in the gapless regime is complicated~\cite{aleiner2021bethe}. Similarly to the Hamiltonian case~\cite{takahashi2005thermodynamics}, however, simplifications occur at the special points ${\gamma}/{\pi}\in \mathbb Q$ known as roots of unity. In this case, the string hypothesis still holds, but there are a finite number $N_b<\infty$ of bound-state types. They are described by the distribution functions $\{\rho_n(\lambda)\}_{n=1}^{N_b}$, with $\lambda\in(-\infty,\infty)$, satisfying a suitable modification of Eqs.~\eqref{eq:bethe_eq}~\cite{SM}.  For these values of $\gamma$, we are able to extend our results~\eqref{eq:eta1gapped}--\eqref{eq:higher_n}, and obtain the distribution functions $\rho_n(\lambda)$ corresponding to the dGGE~\cite{SM}. 

The quasiparticle description of the dGGE thus changes abruptly for $\tau>\tau_{\rm th}(\Delta)$. An important question, however, is whether this transition is visible in the correlation functions. One can expect this to be the case because the distribution functions~$\rho_n(\lambda)$ completely specify the expectation values of local operators~\cite{mestyan2014short,exact2016piroli,correlation2016piroli,bastianello2018exact}. In order to identify which observables could detect the transition, we leverage the results of Refs.~\cite{vanicat2018integrable,ljubotina2019ballistic}, studying the structure of conservation laws for the discrete dynamics~\eqref{eq:floquet_even_odd}. There, it was found that at the root-of-unity points the system displays additional conservation laws breaking the spin-flip symmetry. This suggests that the transition should be visible in the late-time limit $\overline{\sigma_j^{z}}(\tau)$, as defined in Eq.~\eqref{eq:expectation_values}. 

Due to the symmetries of the initial state, $\overline{\sigma_{2j}^{z}}(\tau)=-\overline{\sigma_{2j+1}^{z}}(\tau)$, so that $\overline{\sigma_j^{z}}(\tau)$ coincides with the intensive value of the staggered magnetization. In order to compute it, we exploit the microcanonical interpretation  of the dGGE~\cite{caux2013time}, in which $\rho_n(\lambda)$ are seen as the rapidity distribution functions of a typical eigenstate in the ensemble. Considering the known finite-size formula for the expectation value of $\sigma^z_{2k}$ in an eigenstate of a suitable XXZ transfer matrix with arbitrary inhomogeneities~\cite{kitanine1999form,kitanine2000correlation}, and specialising it to our case, we find~\cite{SM} 
\be
\braket{\{\lambda_j\}|\sigma_{2k}^z|\{\lambda_j\}}= 1 + 2  w^T G^{-1}  v.
\label{eq:finalresultfinitevolume}
\ee
Here $\ket{\{\lambda_j\}}$ is a normalized eigenstate  of the Floquet operator $U(\tau)$ and we introduced the Gaudin matrix $G_{ij}=L\delta_{ij}[a^{(x/2)}_2(\lambda_i)-\sum_{k}a_2(\lambda_i-\lambda_j)/L]+a_2(\lambda_i-\lambda_j)$ together with the two vectors $w_{i}=1$ and $v_{i}=-a_2(\lambda_i- x/2)$. The expectation value of $\sigma_{2k}^z$ in the dGGE is finally obtained by taking the thermodynamic limit of Eq.~\eqref{eq:finalresultfinitevolume}, assuming that the rapidities distribute according to $\rho_n(\lambda)$. In the gapped regime, this yields~\cite{SM}
\begin{equation}
\overline{\sigma_j^{z}}(\tau)=1 -2  \sum_{n=1}^{\infty} n \int_{-\pi/2}^{\pi/2} \!\!{{\rm d}\lambda}\,  [1+\eta_n(\lambda)]^{-1} b^{{\rm eff}}_{n}(\lambda)\,,
\label{eq:finalresultmain}
\end{equation}
where we defined $b^{\rm eff}_{n}(\lambda)$ as the solution to the equations $b_n^{\rm eff}(\lambda) = b_n(\lambda) - \sum_{m} [a_{nm}\ast(1+\eta_m)^{-1}b_m^{\rm eff}](\lambda) $, while $b_n(\lambda)=a_n(\lambda -  x/2)$. An analogous expression can be derived in the gapless regime for root-of-unity points~\cite{SM}.

Plugging the exact rapidity distribution functions of the dGGE into~\eqref{eq:finalresultmain}, we obtain an analytic prediction for the asymptotic staggered magnetization. For $\tau<\tau_{\rm th}(\Delta)$, we find $\overline{\sigma_j^{z}}(\tau)\equiv 0$. We note that, in the continuous-time limit, this simply follows from the fact that the initial state has zero magnetization and from the late-time restoration of translation-symmetry~\cite{fagotti2014relaxation}. For $\tau>\tau_{\rm th}(\Delta)$, we find $\overline{\sigma_j^{z}}(\tau)\neq 0$. An example of our data is reported in Fig.~\ref{fig:sketch}, showing a clear transition at $\tau_{\rm th}(\Delta)$. We have tested our predictions against iTEBD numerical calculations~\cite{schollwock2011density,SM}, as reported in Fig.~\ref{fig:realtime}. The plots are consistent with a vanishing value of $\overline{\sigma_j^{z}}(\tau)$ for $\tau<\tau_{\rm th}(\Delta)$, while they show very good agreement with our analytic result at the values of $\tau$ corresponding to root-of-unity points. These quantitative predictions for $\overline{\sigma_j^{z}}(\tau)$ are the second main result of our work.

Since we only study rational $\gamma/\pi$, we can not provide predictions for all $\tau>\tau_{\rm th}(\Delta)$. In practice, the numerical evaluation of $\overline{\sigma_j^{z}}(\tau)$ in the gapless phase becomes harder as the number of strings $N_b$ increases. This is why we report only a finite number of points in Fig.~\ref{fig:sketch}. In general, it appears that $\overline{\sigma_j^{z}}(\tau)\to0$ in a non-analytic way as $\tau\to\tau_{\rm th}(\Delta)$. In fact, our data may be consistent with a nowhere continuous dependence of $\overline{\sigma_j^{z}}(\tau)$ on $\tau$. This behavior would be analogous to that of the so-called Drude weight~\cite{bulchandani2021superdiffusion,bertini2021finite} characterizing transport in the gapless XXZ chain, both in the continuous-\cite{prosen2011open,prosen2013families,collura2018analytic} and discrete-time setting~\cite{ljubotina2019ballistic}. We leave the study of the full dependence of $\overline{\sigma_j^{z}}(\tau)$ on $\tau$ as an interesting open question.

Our predictions were derived for the N\'eel state, but they hold more generally. The underlying mechanism for the transition is a sudden change in the structure of the conserved charges: for $\tau>\tau_{\rm th}(\tau)$ additional charges breaking spin-flip symmetry appear. In addition, the symmetries of the initial states are important to make the transition visible. In~\cite{SM} we show that the same transition is expected for initial states with the same symmetries of the N\'eel state. Namely, we claim that a discontinuous behavior in the staggered magnetization appears for any initial state $\ket{\Psi_0}$ breaking $\mathcal{T}$ and $\mathcal{S}$ individually, but preserving the combined symmetry $\mathcal{T}\mathcal{S}$~\cite{SM}.

Finally, we comment on the dependence of our results on the choice $\Delta>1$. It is easy to see that for $0<\Delta<1$, a transition still takes place at the Trotter step~\eqref{eq:transition}, but $\overline{\sigma_j^{z}}(\tau)$ is identically zero above $\tau_{\rm th}(\Delta)$, rather then below it. The points $\Delta=0$ and $\Delta=1$ are special, since in this case the system remains in the gapless regime for all $\tau$. For $\Delta=0$ the dynamics maps to to free fermions, and can be studied exactly, as we show in the Supplemental Material~\cite{SM}. Conversely, the case $\Delta=1$ requires a dedicated analysis, which we leave for future research~\cite{InPrep}.

\prlsection{Outlook} 
Our work opens several directions. First, our quasiparticle description of the dGGE lays the basis to study entanglement dynamics at large space-time scales, extending the results of  Ref.~\cite{alba2017entanglement, bertini2022growth} to the discrete setting. Similarly, it also paves the way to the application of the so-called Generalized Hydrodynamic theory~\cite{bertini2016transport,castro2016emergent} to integrable quantum circuits. Using these tools, it would be particularly interesting to understand how the Trotter transitions studied here affect the coarse-grained dynamics of entanglement and local observables. 
	
\prlsection{Acknowledgements} We thank Pasquale Calabrese for useful discussions. B.B. was supported by the Royal Society through the University Research Fellowship No. 201101. G. G. acknowledges support from the Deutsche Forschungsgemeinschaft (DFG, German Research Foundation) under Germany's Excellence Strategy -- EXC-2111 -- 390814868 and from the ERC grant QSIMCORR, ERC-2018-COG, No. 771891.
	
	\let\oldaddcontentsline\addcontentsline
	\renewcommand{\addcontentsline}[3]{}
	\bibliography{bibliography}

\begin{thebibliography}{92}%
\makeatletter
\providecommand \@ifxundefined [1]{%
 \@ifx{#1\undefined}
}%
\providecommand \@ifnum [1]{%
 \ifnum #1\expandafter \@firstoftwo
 \else \expandafter \@secondoftwo
 \fi
}%
\providecommand \@ifx [1]{%
 \ifx #1\expandafter \@firstoftwo
 \else \expandafter \@secondoftwo
 \fi
}%
\providecommand \natexlab [1]{#1}%
\providecommand \enquote  [1]{``#1''}%
\providecommand \bibnamefont  [1]{#1}%
\providecommand \bibfnamefont [1]{#1}%
\providecommand \citenamefont [1]{#1}%
\providecommand \href@noop [0]{\@secondoftwo}%
\providecommand \href [0]{\begingroup \@sanitize@url \@href}%
\providecommand \@href[1]{\@@startlink{#1}\@@href}%
\providecommand \@@href[1]{\endgroup#1\@@endlink}%
\providecommand \@sanitize@url [0]{\catcode `\\12\catcode `\$12\catcode
  `\&12\catcode `\#12\catcode `\^12\catcode `\_12\catcode `\%12\relax}%
\providecommand \@@startlink[1]{}%
\providecommand \@@endlink[0]{}%
\providecommand \url  [0]{\begingroup\@sanitize@url \@url }%
\providecommand \@url [1]{\endgroup\@href {#1}{\urlprefix }}%
\providecommand \urlprefix  [0]{URL }%
\providecommand \Eprint [0]{\href }%
\providecommand \doibase [0]{http://dx.doi.org/}%
\providecommand \selectlanguage [0]{\@gobble}%
\providecommand \bibinfo  [0]{\@secondoftwo}%
\providecommand \bibfield  [0]{\@secondoftwo}%
\providecommand \translation [1]{[#1]}%
\providecommand \BibitemOpen [0]{}%
\providecommand \bibitemStop [0]{}%
\providecommand \bibitemNoStop [0]{.\EOS\space}%
\providecommand \EOS [0]{\spacefactor3000\relax}%
\providecommand \BibitemShut  [1]{\csname bibitem#1\endcsname}%
\let\auto@bib@innerbib\@empty
\bibitem [{\citenamefont {Feynman}(1982)}]{feynman1982simulating}%
  \BibitemOpen
  \bibfield  {author} {\bibinfo {author} {\bibfnamefont {R.~P.}\ \bibnamefont
  {Feynman}},\ }\href {\doibase 10.1007/BF02650179} {\bibfield  {journal}
  {\bibinfo  {journal} {Int. J. Theor. Phys.}\ }\textbf {\bibinfo {volume}
  {21}},\ \bibinfo {pages} {467} (\bibinfo {year} {1982})}\BibitemShut
  {NoStop}%
\bibitem [{\citenamefont {Georgescu}\ \emph {et~al.}(2014)\citenamefont
  {Georgescu}, \citenamefont {Ashhab},\ and\ \citenamefont
  {Nori}}]{georgescu2014quantum}%
  \BibitemOpen
  \bibfield  {author} {\bibinfo {author} {\bibfnamefont {I.~M.}\ \bibnamefont
  {Georgescu}}, \bibinfo {author} {\bibfnamefont {S.}~\bibnamefont {Ashhab}}, \
  and\ \bibinfo {author} {\bibfnamefont {F.}~\bibnamefont {Nori}},\ }\href
  {\doibase 10.1103/RevModPhys.86.153} {\bibfield  {journal} {\bibinfo
  {journal} {Rev. Mod. Phys.}\ }\textbf {\bibinfo {volume} {86}},\ \bibinfo
  {pages} {153} (\bibinfo {year} {2014})}\BibitemShut {NoStop}%
\bibitem [{\citenamefont {Lloyd}(1996)}]{lloyd1996universal}%
  \BibitemOpen
  \bibfield  {author} {\bibinfo {author} {\bibfnamefont {S.}~\bibnamefont
  {Lloyd}},\ }\href {\doibase 10.1126/science.273.5278.1073} {\bibfield
  {journal} {\bibinfo  {journal} {Science}\ }\textbf {\bibinfo {volume}
  {273}},\ \bibinfo {pages} {1073} (\bibinfo {year} {1996})}\BibitemShut
  {NoStop}%
\bibitem [{\citenamefont {Trotter}(1959)}]{trotter1959product}%
  \BibitemOpen
  \bibfield  {author} {\bibinfo {author} {\bibfnamefont {H.~F.}\ \bibnamefont
  {Trotter}},\ }\href
  {https://www.ams.org/journals/proc/1959-010-04/S0002-9939-1959-0108732-6/}
  {\bibfield  {journal} {\bibinfo  {journal} {Proc. American Math. Soc.}\
  }\textbf {\bibinfo {volume} {10}},\ \bibinfo {pages} {545} (\bibinfo {year}
  {1959})}\BibitemShut {NoStop}%
\bibitem [{\citenamefont {Suzuki}(1991)}]{suzuki1991general}%
  \BibitemOpen
  \bibfield  {author} {\bibinfo {author} {\bibfnamefont {M.}~\bibnamefont
  {Suzuki}},\ }\href {\doibase 10.1063/1.529425} {\bibfield  {journal}
  {\bibinfo  {journal} {J. Math. Phys.}\ }\textbf {\bibinfo {volume} {32}},\
  \bibinfo {pages} {400} (\bibinfo {year} {1991})}\BibitemShut {NoStop}%
\bibitem [{\citenamefont {Preskill}(2018)}]{preskill2018quantum}%
  \BibitemOpen
  \bibfield  {author} {\bibinfo {author} {\bibfnamefont {J.}~\bibnamefont
  {Preskill}},\ }\href {\doibase 10.22331/q-2018-08-06-79} {\bibfield
  {journal} {\bibinfo  {journal} {Quantum}\ }\textbf {\bibinfo {volume} {2}},\
  \bibinfo {pages} {79} (\bibinfo {year} {2018})}\BibitemShut {NoStop}%
\bibitem [{\citenamefont {Daley}\ \emph {et~al.}(2022)\citenamefont {Daley},
  \citenamefont {Bloch}, \citenamefont {Kokail}, \citenamefont {Flannigan},
  \citenamefont {Pearson}, \citenamefont {Troyer},\ and\ \citenamefont
  {Zoller}}]{daley2022practical}%
  \BibitemOpen
  \bibfield  {author} {\bibinfo {author} {\bibfnamefont {A.~J.}\ \bibnamefont
  {Daley}}, \bibinfo {author} {\bibfnamefont {I.}~\bibnamefont {Bloch}},
  \bibinfo {author} {\bibfnamefont {C.}~\bibnamefont {Kokail}}, \bibinfo
  {author} {\bibfnamefont {S.}~\bibnamefont {Flannigan}}, \bibinfo {author}
  {\bibfnamefont {N.}~\bibnamefont {Pearson}}, \bibinfo {author} {\bibfnamefont
  {M.}~\bibnamefont {Troyer}}, \ and\ \bibinfo {author} {\bibfnamefont
  {P.}~\bibnamefont {Zoller}},\ }\href {\doibase 10.1038/s41586-022-04940-6}
  {\bibfield  {journal} {\bibinfo  {journal} {Nature}\ }\textbf {\bibinfo
  {volume} {607}},\ \bibinfo {pages} {667} (\bibinfo {year}
  {2022})}\BibitemShut {NoStop}%
\bibitem [{\citenamefont {Lanyon}\ \emph {et~al.}(2011)\citenamefont {Lanyon},
  \citenamefont {Hempel}, \citenamefont {Nigg}, \citenamefont {M{\"u}ller},
  \citenamefont {Gerritsma}, \citenamefont {Z{\"a}hringer}, \citenamefont
  {Schindler}, \citenamefont {Barreiro}, \citenamefont {Rambach}, \citenamefont
  {Kirchmair} \emph {et~al.}}]{lanyon2011universal}%
  \BibitemOpen
  \bibfield  {author} {\bibinfo {author} {\bibfnamefont {B.~P.}\ \bibnamefont
  {Lanyon}}, \bibinfo {author} {\bibfnamefont {C.}~\bibnamefont {Hempel}},
  \bibinfo {author} {\bibfnamefont {D.}~\bibnamefont {Nigg}}, \bibinfo {author}
  {\bibfnamefont {M.}~\bibnamefont {M{\"u}ller}}, \bibinfo {author}
  {\bibfnamefont {R.}~\bibnamefont {Gerritsma}}, \bibinfo {author}
  {\bibfnamefont {F.}~\bibnamefont {Z{\"a}hringer}}, \bibinfo {author}
  {\bibfnamefont {P.}~\bibnamefont {Schindler}}, \bibinfo {author}
  {\bibfnamefont {J.~T.}\ \bibnamefont {Barreiro}}, \bibinfo {author}
  {\bibfnamefont {M.}~\bibnamefont {Rambach}}, \bibinfo {author} {\bibfnamefont
  {G.}~\bibnamefont {Kirchmair}},  \emph {et~al.},\ }\href {\doibase
  10.1126/science.1208001} {\bibfield  {journal} {\bibinfo  {journal}
  {Science}\ }\textbf {\bibinfo {volume} {334}},\ \bibinfo {pages} {57}
  (\bibinfo {year} {2011})}\BibitemShut {NoStop}%
\bibitem [{\citenamefont {Barreiro}\ \emph {et~al.}(2011)\citenamefont
  {Barreiro}, \citenamefont {M{\"u}ller}, \citenamefont {Schindler},
  \citenamefont {Nigg}, \citenamefont {Monz}, \citenamefont {Chwalla},
  \citenamefont {Hennrich}, \citenamefont {Roos}, \citenamefont {Zoller},\ and\
  \citenamefont {Blatt}}]{barreiro2011open}%
  \BibitemOpen
  \bibfield  {author} {\bibinfo {author} {\bibfnamefont {J.~T.}\ \bibnamefont
  {Barreiro}}, \bibinfo {author} {\bibfnamefont {M.}~\bibnamefont
  {M{\"u}ller}}, \bibinfo {author} {\bibfnamefont {P.}~\bibnamefont
  {Schindler}}, \bibinfo {author} {\bibfnamefont {D.}~\bibnamefont {Nigg}},
  \bibinfo {author} {\bibfnamefont {T.}~\bibnamefont {Monz}}, \bibinfo {author}
  {\bibfnamefont {M.}~\bibnamefont {Chwalla}}, \bibinfo {author} {\bibfnamefont
  {M.}~\bibnamefont {Hennrich}}, \bibinfo {author} {\bibfnamefont {C.~F.}\
  \bibnamefont {Roos}}, \bibinfo {author} {\bibfnamefont {P.}~\bibnamefont
  {Zoller}}, \ and\ \bibinfo {author} {\bibfnamefont {R.}~\bibnamefont
  {Blatt}},\ }\href {\doibase 10.1038/nature09801} {\bibfield  {journal}
  {\bibinfo  {journal} {Nature}\ }\textbf {\bibinfo {volume} {470}},\ \bibinfo
  {pages} {486} (\bibinfo {year} {2011})}\BibitemShut {NoStop}%
\bibitem [{\citenamefont {Blatt}\ and\ \citenamefont
  {Roos}(2012)}]{blatt2012quantum}%
  \BibitemOpen
  \bibfield  {author} {\bibinfo {author} {\bibfnamefont {R.}~\bibnamefont
  {Blatt}}\ and\ \bibinfo {author} {\bibfnamefont {C.~F.}\ \bibnamefont
  {Roos}},\ }\href {\doibase 10.1038/nphys2252} {\bibfield  {journal} {\bibinfo
   {journal} {Nature Physics}\ }\textbf {\bibinfo {volume} {8}},\ \bibinfo
  {pages} {277} (\bibinfo {year} {2012})}\BibitemShut {NoStop}%
\bibitem [{\citenamefont {Monroe}\ \emph {et~al.}(2021)\citenamefont {Monroe},
  \citenamefont {Campbell}, \citenamefont {Duan}, \citenamefont {Gong},
  \citenamefont {Gorshkov}, \citenamefont {Hess}, \citenamefont {Islam},
  \citenamefont {Kim}, \citenamefont {Linke}, \citenamefont {Pagano},
  \citenamefont {Richerme}, \citenamefont {Senko},\ and\ \citenamefont
  {Yao}}]{monroe2021programmable}%
  \BibitemOpen
  \bibfield  {author} {\bibinfo {author} {\bibfnamefont {C.}~\bibnamefont
  {Monroe}}, \bibinfo {author} {\bibfnamefont {W.~C.}\ \bibnamefont
  {Campbell}}, \bibinfo {author} {\bibfnamefont {L.-M.}\ \bibnamefont {Duan}},
  \bibinfo {author} {\bibfnamefont {Z.-X.}\ \bibnamefont {Gong}}, \bibinfo
  {author} {\bibfnamefont {A.~V.}\ \bibnamefont {Gorshkov}}, \bibinfo {author}
  {\bibfnamefont {P.~W.}\ \bibnamefont {Hess}}, \bibinfo {author}
  {\bibfnamefont {R.}~\bibnamefont {Islam}}, \bibinfo {author} {\bibfnamefont
  {K.}~\bibnamefont {Kim}}, \bibinfo {author} {\bibfnamefont {N.~M.}\
  \bibnamefont {Linke}}, \bibinfo {author} {\bibfnamefont {G.}~\bibnamefont
  {Pagano}}, \bibinfo {author} {\bibfnamefont {P.}~\bibnamefont {Richerme}},
  \bibinfo {author} {\bibfnamefont {C.}~\bibnamefont {Senko}}, \ and\ \bibinfo
  {author} {\bibfnamefont {N.~Y.}\ \bibnamefont {Yao}},\ }\href {\doibase
  10.1103/RevModPhys.93.025001} {\bibfield  {journal} {\bibinfo  {journal}
  {Rev. Mod. Phys.}\ }\textbf {\bibinfo {volume} {93}},\ \bibinfo {pages}
  {025001} (\bibinfo {year} {2021})}\BibitemShut {NoStop}%
\bibitem [{\citenamefont {Salath\'e}\ \emph {et~al.}(2015)\citenamefont
  {Salath\'e}, \citenamefont {Mondal}, \citenamefont {Oppliger}, \citenamefont
  {Heinsoo}, \citenamefont {Kurpiers}, \citenamefont
  {Poto\ifmmode~\check{c}\else \v{c}\fi{}nik}, \citenamefont {Mezzacapo},
  \citenamefont {Las~Heras}, \citenamefont {Lamata}, \citenamefont {Solano},
  \citenamefont {Filipp},\ and\ \citenamefont {Wallraff}}]{salathe2015digital}%
  \BibitemOpen
  \bibfield  {author} {\bibinfo {author} {\bibfnamefont {Y.}~\bibnamefont
  {Salath\'e}}, \bibinfo {author} {\bibfnamefont {M.}~\bibnamefont {Mondal}},
  \bibinfo {author} {\bibfnamefont {M.}~\bibnamefont {Oppliger}}, \bibinfo
  {author} {\bibfnamefont {J.}~\bibnamefont {Heinsoo}}, \bibinfo {author}
  {\bibfnamefont {P.}~\bibnamefont {Kurpiers}}, \bibinfo {author}
  {\bibfnamefont {A.}~\bibnamefont {Poto\ifmmode~\check{c}\else
  \v{c}\fi{}nik}}, \bibinfo {author} {\bibfnamefont {A.}~\bibnamefont
  {Mezzacapo}}, \bibinfo {author} {\bibfnamefont {U.}~\bibnamefont
  {Las~Heras}}, \bibinfo {author} {\bibfnamefont {L.}~\bibnamefont {Lamata}},
  \bibinfo {author} {\bibfnamefont {E.}~\bibnamefont {Solano}}, \bibinfo
  {author} {\bibfnamefont {S.}~\bibnamefont {Filipp}}, \ and\ \bibinfo {author}
  {\bibfnamefont {A.}~\bibnamefont {Wallraff}},\ }\href {\doibase
  10.1103/PhysRevX.5.021027} {\bibfield  {journal} {\bibinfo  {journal} {Phys.
  Rev. X}\ }\textbf {\bibinfo {volume} {5}},\ \bibinfo {pages} {021027}
  (\bibinfo {year} {2015})}\BibitemShut {NoStop}%
\bibitem [{\citenamefont {Barends}\ \emph {et~al.}(2015)\citenamefont
  {Barends}, \citenamefont {Lamata}, \citenamefont {Kelly}, \citenamefont
  {Garc{\'\i}a-{\'A}lvarez}, \citenamefont {Fowler}, \citenamefont {Megrant},
  \citenamefont {Jeffrey}, \citenamefont {White}, \citenamefont {Sank},
  \citenamefont {Mutus} \emph {et~al.}}]{barends2015digital}%
  \BibitemOpen
  \bibfield  {author} {\bibinfo {author} {\bibfnamefont {R.}~\bibnamefont
  {Barends}}, \bibinfo {author} {\bibfnamefont {L.}~\bibnamefont {Lamata}},
  \bibinfo {author} {\bibfnamefont {J.}~\bibnamefont {Kelly}}, \bibinfo
  {author} {\bibfnamefont {L.}~\bibnamefont {Garc{\'\i}a-{\'A}lvarez}},
  \bibinfo {author} {\bibfnamefont {A.~G.}\ \bibnamefont {Fowler}}, \bibinfo
  {author} {\bibfnamefont {A.}~\bibnamefont {Megrant}}, \bibinfo {author}
  {\bibfnamefont {E.}~\bibnamefont {Jeffrey}}, \bibinfo {author} {\bibfnamefont
  {T.~C.}\ \bibnamefont {White}}, \bibinfo {author} {\bibfnamefont
  {D.}~\bibnamefont {Sank}}, \bibinfo {author} {\bibfnamefont {J.~Y.}\
  \bibnamefont {Mutus}},  \emph {et~al.},\ }\href {\doibase 10.1038/ncomms8654}
  {\bibfield  {journal} {\bibinfo  {journal} {Nature Comm.}\ }\textbf {\bibinfo
  {volume} {6}},\ \bibinfo {pages} {1} (\bibinfo {year} {2015})}\BibitemShut
  {NoStop}%
\bibitem [{\citenamefont {Langford}\ \emph {et~al.}(2017)\citenamefont
  {Langford}, \citenamefont {Sagastizabal}, \citenamefont {Kounalakis},
  \citenamefont {Dickel}, \citenamefont {Bruno}, \citenamefont {Luthi},
  \citenamefont {Thoen}, \citenamefont {Endo},\ and\ \citenamefont
  {DiCarlo}}]{langford2017experimentally}%
  \BibitemOpen
  \bibfield  {author} {\bibinfo {author} {\bibfnamefont {N.~K.}\ \bibnamefont
  {Langford}}, \bibinfo {author} {\bibfnamefont {R.}~\bibnamefont
  {Sagastizabal}}, \bibinfo {author} {\bibfnamefont {M.}~\bibnamefont
  {Kounalakis}}, \bibinfo {author} {\bibfnamefont {C.}~\bibnamefont {Dickel}},
  \bibinfo {author} {\bibfnamefont {A.}~\bibnamefont {Bruno}}, \bibinfo
  {author} {\bibfnamefont {F.}~\bibnamefont {Luthi}}, \bibinfo {author}
  {\bibfnamefont {D.~J.}\ \bibnamefont {Thoen}}, \bibinfo {author}
  {\bibfnamefont {A.}~\bibnamefont {Endo}}, \ and\ \bibinfo {author}
  {\bibfnamefont {L.}~\bibnamefont {DiCarlo}},\ }\href {\doibase
  10.1038/s41467-017-01061-x} {\bibfield  {journal} {\bibinfo  {journal}
  {Nature Comm.}\ }\textbf {\bibinfo {volume} {8}},\ \bibinfo {pages} {1}
  (\bibinfo {year} {2017})}\BibitemShut {NoStop}%
\bibitem [{\citenamefont {Wendin}(2017)}]{wendin2017quantum}%
  \BibitemOpen
  \bibfield  {author} {\bibinfo {author} {\bibfnamefont {G.}~\bibnamefont
  {Wendin}},\ }\href {\doibase 10.1088/1361-6633/aa7e1a} {\bibfield  {journal}
  {\bibinfo  {journal} {Rep. Progr. Phys.}\ }\textbf {\bibinfo {volume} {80}},\
  \bibinfo {pages} {106001} (\bibinfo {year} {2017})}\BibitemShut {NoStop}%
\bibitem [{\citenamefont {Kjaergaard}\ \emph {et~al.}(2020)\citenamefont
  {Kjaergaard}, \citenamefont {Schwartz}, \citenamefont {Braum\"{u}ller},
  \citenamefont {Krantz}, \citenamefont {Wang}, \citenamefont {Gustavsson},\
  and\ \citenamefont {Oliver}}]{kjaergaard2020superconducting}%
  \BibitemOpen
  \bibfield  {author} {\bibinfo {author} {\bibfnamefont {M.}~\bibnamefont
  {Kjaergaard}}, \bibinfo {author} {\bibfnamefont {M.~E.}\ \bibnamefont
  {Schwartz}}, \bibinfo {author} {\bibfnamefont {J.}~\bibnamefont
  {Braum\"{u}ller}}, \bibinfo {author} {\bibfnamefont {P.}~\bibnamefont
  {Krantz}}, \bibinfo {author} {\bibfnamefont {J.~I.-J.}\ \bibnamefont {Wang}},
  \bibinfo {author} {\bibfnamefont {S.}~\bibnamefont {Gustavsson}}, \ and\
  \bibinfo {author} {\bibfnamefont {W.~D.}\ \bibnamefont {Oliver}},\ }\href
  {\doibase 10.1146/annurev-conmatphys-031119-050605} {\bibfield  {journal}
  {\bibinfo  {journal} {Annual Rev. Cond. Matt. Phys.}\ }\textbf {\bibinfo
  {volume} {11}},\ \bibinfo {pages} {369} (\bibinfo {year} {2020})}\BibitemShut
  {NoStop}%
\bibitem [{\citenamefont {Bravyi}\ \emph {et~al.}(2022)\citenamefont {Bravyi},
  \citenamefont {Dial}, \citenamefont {Gambetta}, \citenamefont {Gil},\ and\
  \citenamefont {Nazario}}]{bravyi2022future}%
  \BibitemOpen
  \bibfield  {author} {\bibinfo {author} {\bibfnamefont {S.}~\bibnamefont
  {Bravyi}}, \bibinfo {author} {\bibfnamefont {O.}~\bibnamefont {Dial}},
  \bibinfo {author} {\bibfnamefont {J.~M.}\ \bibnamefont {Gambetta}}, \bibinfo
  {author} {\bibfnamefont {D.}~\bibnamefont {Gil}}, \ and\ \bibinfo {author}
  {\bibfnamefont {Z.}~\bibnamefont {Nazario}},\ }\href {\doibase
  10.1063/5.0082975} {\bibfield  {journal} {\bibinfo  {journal} {J. Appl.
  Phys.}\ }\textbf {\bibinfo {volume} {132}},\ \bibinfo {pages} {160902}
  (\bibinfo {year} {2022})}\BibitemShut {NoStop}%
\bibitem [{\citenamefont {Childs}\ \emph {et~al.}(2018)\citenamefont {Childs},
  \citenamefont {Maslov}, \citenamefont {Nam}, \citenamefont {Ross},\ and\
  \citenamefont {Su}}]{childs2018toward}%
  \BibitemOpen
  \bibfield  {author} {\bibinfo {author} {\bibfnamefont {A.~M.}\ \bibnamefont
  {Childs}}, \bibinfo {author} {\bibfnamefont {D.}~\bibnamefont {Maslov}},
  \bibinfo {author} {\bibfnamefont {Y.}~\bibnamefont {Nam}}, \bibinfo {author}
  {\bibfnamefont {N.~J.}\ \bibnamefont {Ross}}, \ and\ \bibinfo {author}
  {\bibfnamefont {Y.}~\bibnamefont {Su}},\ }\href {\doibase
  10.1073/pnas.1801723115} {\bibfield  {journal} {\bibinfo  {journal} {PNAS}\
  }\textbf {\bibinfo {volume} {115}},\ \bibinfo {pages} {9456} (\bibinfo {year}
  {2018})}\BibitemShut {NoStop}%
\bibitem [{\citenamefont {Childs}\ and\ \citenamefont
  {Su}(2019)}]{childs2019nearly}%
  \BibitemOpen
  \bibfield  {author} {\bibinfo {author} {\bibfnamefont {A.~M.}\ \bibnamefont
  {Childs}}\ and\ \bibinfo {author} {\bibfnamefont {Y.}~\bibnamefont {Su}},\
  }\href {\doibase 10.1103/PhysRevLett.123.050503} {\bibfield  {journal}
  {\bibinfo  {journal} {Phys. Rev. Lett.}\ }\textbf {\bibinfo {volume} {123}},\
  \bibinfo {pages} {050503} (\bibinfo {year} {2019})}\BibitemShut {NoStop}%
\bibitem [{\citenamefont {Tran}\ \emph {et~al.}(2020)\citenamefont {Tran},
  \citenamefont {Chu}, \citenamefont {Su}, \citenamefont {Childs},\ and\
  \citenamefont {Gorshkov}}]{tran2020destructive}%
  \BibitemOpen
  \bibfield  {author} {\bibinfo {author} {\bibfnamefont {M.~C.}\ \bibnamefont
  {Tran}}, \bibinfo {author} {\bibfnamefont {S.-K.}\ \bibnamefont {Chu}},
  \bibinfo {author} {\bibfnamefont {Y.}~\bibnamefont {Su}}, \bibinfo {author}
  {\bibfnamefont {A.~M.}\ \bibnamefont {Childs}}, \ and\ \bibinfo {author}
  {\bibfnamefont {A.~V.}\ \bibnamefont {Gorshkov}},\ }\href {\doibase
  10.1103/PhysRevLett.124.220502} {\bibfield  {journal} {\bibinfo  {journal}
  {Phys. Rev. Lett.}\ }\textbf {\bibinfo {volume} {124}},\ \bibinfo {pages}
  {220502} (\bibinfo {year} {2020})}\BibitemShut {NoStop}%
\bibitem [{\citenamefont {Childs}\ \emph {et~al.}(2021)\citenamefont {Childs},
  \citenamefont {Su}, \citenamefont {Tran}, \citenamefont {Wiebe},\ and\
  \citenamefont {Zhu}}]{childs2021theory}%
  \BibitemOpen
  \bibfield  {author} {\bibinfo {author} {\bibfnamefont {A.~M.}\ \bibnamefont
  {Childs}}, \bibinfo {author} {\bibfnamefont {Y.}~\bibnamefont {Su}}, \bibinfo
  {author} {\bibfnamefont {M.~C.}\ \bibnamefont {Tran}}, \bibinfo {author}
  {\bibfnamefont {N.}~\bibnamefont {Wiebe}}, \ and\ \bibinfo {author}
  {\bibfnamefont {S.}~\bibnamefont {Zhu}},\ }\href {\doibase
  10.1103/PhysRevX.11.011020} {\bibfield  {journal} {\bibinfo  {journal} {Phys.
  Rev. X}\ }\textbf {\bibinfo {volume} {11}},\ \bibinfo {pages} {011020}
  (\bibinfo {year} {2021})}\BibitemShut {NoStop}%
\bibitem [{\citenamefont {Layden}(2022)}]{layden2022first}%
  \BibitemOpen
  \bibfield  {author} {\bibinfo {author} {\bibfnamefont {D.}~\bibnamefont
  {Layden}},\ }\href {\doibase 10.1103/PhysRevLett.128.210501} {\bibfield
  {journal} {\bibinfo  {journal} {Phys. Rev. Lett.}\ }\textbf {\bibinfo
  {volume} {128}},\ \bibinfo {pages} {210501} (\bibinfo {year}
  {2022})}\BibitemShut {NoStop}%
\bibitem [{\citenamefont {Heyl}\ \emph {et~al.}(2019)\citenamefont {Heyl},
  \citenamefont {Hauke},\ and\ \citenamefont {Zoller}}]{heyl2019quantum}%
  \BibitemOpen
  \bibfield  {author} {\bibinfo {author} {\bibfnamefont {M.}~\bibnamefont
  {Heyl}}, \bibinfo {author} {\bibfnamefont {P.}~\bibnamefont {Hauke}}, \ and\
  \bibinfo {author} {\bibfnamefont {P.}~\bibnamefont {Zoller}},\ }\href
  {\doibase 10.1126/sciadv.aau8342} {\bibfield  {journal} {\bibinfo  {journal}
  {Science Adv.}\ }\textbf {\bibinfo {volume} {5}},\ \bibinfo {pages}
  {eaau8342} (\bibinfo {year} {2019})}\BibitemShut {NoStop}%
\bibitem [{\citenamefont {Ishii}\ \emph {et~al.}(2018)\citenamefont {Ishii},
  \citenamefont {Kuwahara}, \citenamefont {Mori},\ and\ \citenamefont
  {Hatano}}]{ishii2018heating}%
  \BibitemOpen
  \bibfield  {author} {\bibinfo {author} {\bibfnamefont {T.}~\bibnamefont
  {Ishii}}, \bibinfo {author} {\bibfnamefont {T.}~\bibnamefont {Kuwahara}},
  \bibinfo {author} {\bibfnamefont {T.}~\bibnamefont {Mori}}, \ and\ \bibinfo
  {author} {\bibfnamefont {N.}~\bibnamefont {Hatano}},\ }\href {\doibase
  10.1103/PhysRevLett.120.220602} {\bibfield  {journal} {\bibinfo  {journal}
  {Phys. Rev. Lett.}\ }\textbf {\bibinfo {volume} {120}},\ \bibinfo {pages}
  {220602} (\bibinfo {year} {2018})}\BibitemShut {NoStop}%
\bibitem [{\citenamefont {Sieberer}\ \emph {et~al.}(2019)\citenamefont
  {Sieberer}, \citenamefont {Olsacher}, \citenamefont {Elben}, \citenamefont
  {Heyl}, \citenamefont {Hauke}, \citenamefont {Haake},\ and\ \citenamefont
  {Zoller}}]{sieberer2019digital}%
  \BibitemOpen
  \bibfield  {author} {\bibinfo {author} {\bibfnamefont {L.~M.}\ \bibnamefont
  {Sieberer}}, \bibinfo {author} {\bibfnamefont {T.}~\bibnamefont {Olsacher}},
  \bibinfo {author} {\bibfnamefont {A.}~\bibnamefont {Elben}}, \bibinfo
  {author} {\bibfnamefont {M.}~\bibnamefont {Heyl}}, \bibinfo {author}
  {\bibfnamefont {P.}~\bibnamefont {Hauke}}, \bibinfo {author} {\bibfnamefont
  {F.}~\bibnamefont {Haake}}, \ and\ \bibinfo {author} {\bibfnamefont
  {P.}~\bibnamefont {Zoller}},\ }\href {\doibase 10.1038/s41534-019-0192-5}
  {\bibfield  {journal} {\bibinfo  {journal} {npj Quantum Inf.}\ }\textbf
  {\bibinfo {volume} {5}},\ \bibinfo {pages} {1} (\bibinfo {year}
  {2019})}\BibitemShut {NoStop}%
\bibitem [{\citenamefont {Kargi}\ \emph {et~al.}(2021)\citenamefont {Kargi},
  \citenamefont {Dehollain}, \citenamefont {Henriques}, \citenamefont
  {Sieberer}, \citenamefont {Olsacher}, \citenamefont {Hauke}, \citenamefont
  {Heyl}, \citenamefont {Zoller},\ and\ \citenamefont
  {Langford}}]{kargi2021quantum}%
  \BibitemOpen
  \bibfield  {author} {\bibinfo {author} {\bibfnamefont {C.}~\bibnamefont
  {Kargi}}, \bibinfo {author} {\bibfnamefont {J.~P.}\ \bibnamefont
  {Dehollain}}, \bibinfo {author} {\bibfnamefont {F.}~\bibnamefont
  {Henriques}}, \bibinfo {author} {\bibfnamefont {L.~M.}\ \bibnamefont
  {Sieberer}}, \bibinfo {author} {\bibfnamefont {T.}~\bibnamefont {Olsacher}},
  \bibinfo {author} {\bibfnamefont {P.}~\bibnamefont {Hauke}}, \bibinfo
  {author} {\bibfnamefont {M.}~\bibnamefont {Heyl}}, \bibinfo {author}
  {\bibfnamefont {P.}~\bibnamefont {Zoller}}, \ and\ \bibinfo {author}
  {\bibfnamefont {N.~K.}\ \bibnamefont {Langford}},\ }\href
  {https://arxiv.org/abs/2110.11113} {\bibfield  {journal} {\bibinfo  {journal}
  {arXiv:2110.11113}\ } (\bibinfo {year} {2021})}\BibitemShut {NoStop}%
\bibitem [{\citenamefont {Chinni}\ \emph {et~al.}(2022)\citenamefont {Chinni},
  \citenamefont {Mu\~noz Arias}, \citenamefont {Deutsch},\ and\ \citenamefont
  {Poggi}}]{chinni2022trotter}%
  \BibitemOpen
  \bibfield  {author} {\bibinfo {author} {\bibfnamefont {K.}~\bibnamefont
  {Chinni}}, \bibinfo {author} {\bibfnamefont {M.~H.}\ \bibnamefont {Mu\~noz
  Arias}}, \bibinfo {author} {\bibfnamefont {I.~H.}\ \bibnamefont {Deutsch}}, \
  and\ \bibinfo {author} {\bibfnamefont {P.~M.}\ \bibnamefont {Poggi}},\ }\href
  {\doibase 10.1103/PRXQuantum.3.010351} {\bibfield  {journal} {\bibinfo
  {journal} {PRX Quantum}\ }\textbf {\bibinfo {volume} {3}},\ \bibinfo {pages}
  {010351} (\bibinfo {year} {2022})}\BibitemShut {NoStop}%
\bibitem [{\citenamefont {Abanin}\ \emph {et~al.}(2015)\citenamefont {Abanin},
  \citenamefont {De~Roeck},\ and\ \citenamefont
  {Huveneers}}]{abanin2015exponentially}%
  \BibitemOpen
  \bibfield  {author} {\bibinfo {author} {\bibfnamefont {D.~A.}\ \bibnamefont
  {Abanin}}, \bibinfo {author} {\bibfnamefont {W.}~\bibnamefont {De~Roeck}}, \
  and\ \bibinfo {author} {\bibfnamefont {F.}~\bibnamefont {Huveneers}},\ }\href
  {\doibase 10.1103/PhysRevLett.115.256803} {\bibfield  {journal} {\bibinfo
  {journal} {Phys. Rev. Lett.}\ }\textbf {\bibinfo {volume} {115}},\ \bibinfo
  {pages} {256803} (\bibinfo {year} {2015})}\BibitemShut {NoStop}%
\bibitem [{\citenamefont {Abanin}\ \emph
  {et~al.}(2017{\natexlab{a}})\citenamefont {Abanin}, \citenamefont {De~Roeck},
  \citenamefont {Ho},\ and\ \citenamefont {Huveneers}}]{abanin2017effective}%
  \BibitemOpen
  \bibfield  {author} {\bibinfo {author} {\bibfnamefont {D.~A.}\ \bibnamefont
  {Abanin}}, \bibinfo {author} {\bibfnamefont {W.}~\bibnamefont {De~Roeck}},
  \bibinfo {author} {\bibfnamefont {W.~W.}\ \bibnamefont {Ho}}, \ and\ \bibinfo
  {author} {\bibfnamefont {F.}~\bibnamefont {Huveneers}},\ }\href {\doibase
  10.1103/PhysRevB.95.014112} {\bibfield  {journal} {\bibinfo  {journal} {Phys.
  Rev. B}\ }\textbf {\bibinfo {volume} {95}},\ \bibinfo {pages} {014112}
  (\bibinfo {year} {2017}{\natexlab{a}})}\BibitemShut {NoStop}%
\bibitem [{\citenamefont {Abanin}\ \emph
  {et~al.}(2017{\natexlab{b}})\citenamefont {Abanin}, \citenamefont {De~Roeck},
  \citenamefont {Ho},\ and\ \citenamefont {Huveneers}}]{abanin2017rigorous}%
  \BibitemOpen
  \bibfield  {author} {\bibinfo {author} {\bibfnamefont {D.}~\bibnamefont
  {Abanin}}, \bibinfo {author} {\bibfnamefont {W.}~\bibnamefont {De~Roeck}},
  \bibinfo {author} {\bibfnamefont {W.~W.}\ \bibnamefont {Ho}}, \ and\ \bibinfo
  {author} {\bibfnamefont {F.}~\bibnamefont {Huveneers}},\ }\href {\doibase
  10.1007/s00220-017-2930-x} {\bibfield  {journal} {\bibinfo  {journal} {Comm.
  Math. Phys.}\ }\textbf {\bibinfo {volume} {354}},\ \bibinfo {pages} {809}
  (\bibinfo {year} {2017}{\natexlab{b}})}\BibitemShut {NoStop}%
\bibitem [{\citenamefont {Mori}\ \emph {et~al.}(2016)\citenamefont {Mori},
  \citenamefont {Kuwahara},\ and\ \citenamefont {Saito}}]{mori2016rigorous}%
  \BibitemOpen
  \bibfield  {author} {\bibinfo {author} {\bibfnamefont {T.}~\bibnamefont
  {Mori}}, \bibinfo {author} {\bibfnamefont {T.}~\bibnamefont {Kuwahara}}, \
  and\ \bibinfo {author} {\bibfnamefont {K.}~\bibnamefont {Saito}},\ }\href
  {\doibase 10.1103/PhysRevLett.116.120401} {\bibfield  {journal} {\bibinfo
  {journal} {Phys. Rev. Lett.}\ }\textbf {\bibinfo {volume} {116}},\ \bibinfo
  {pages} {120401} (\bibinfo {year} {2016})}\BibitemShut {NoStop}%
\bibitem [{\citenamefont {Kuwahara}\ \emph {et~al.}(2016)\citenamefont
  {Kuwahara}, \citenamefont {Mori},\ and\ \citenamefont
  {Saito}}]{kuwahara2016floquet}%
  \BibitemOpen
  \bibfield  {author} {\bibinfo {author} {\bibfnamefont {T.}~\bibnamefont
  {Kuwahara}}, \bibinfo {author} {\bibfnamefont {T.}~\bibnamefont {Mori}}, \
  and\ \bibinfo {author} {\bibfnamefont {K.}~\bibnamefont {Saito}},\ }\href
  {\doibase https://doi.org/10.1016/j.aop.2016.01.012} {\bibfield  {journal}
  {\bibinfo  {journal} {Ann. Phys.}\ }\textbf {\bibinfo {volume} {367}},\
  \bibinfo {pages} {96} (\bibinfo {year} {2016})}\BibitemShut {NoStop}%
\bibitem [{\citenamefont {Calabrese}\ and\ \citenamefont
  {Cardy}(2006)}]{calabrese2006time}%
  \BibitemOpen
  \bibfield  {author} {\bibinfo {author} {\bibfnamefont {P.}~\bibnamefont
  {Calabrese}}\ and\ \bibinfo {author} {\bibfnamefont {J.}~\bibnamefont
  {Cardy}},\ }\href {\doibase 10.1103/PhysRevLett.96.136801} {\bibfield
  {journal} {\bibinfo  {journal} {Phys. Rev. Lett.}\ }\textbf {\bibinfo
  {volume} {96}},\ \bibinfo {pages} {136801} (\bibinfo {year}
  {2006})}\BibitemShut {NoStop}%
\bibitem [{\citenamefont {Calabrese}\ and\ \citenamefont
  {Cardy}(2007)}]{calabrese2007quantum}%
  \BibitemOpen
  \bibfield  {author} {\bibinfo {author} {\bibfnamefont {P.}~\bibnamefont
  {Calabrese}}\ and\ \bibinfo {author} {\bibfnamefont {J.}~\bibnamefont
  {Cardy}},\ }\href
  {https://iopscience.iop.org/article/10.1088/1742-5468/2007/06/P06008}
  {\bibfield  {journal} {\bibinfo  {journal} {J. Stat. Mech.}\ }\textbf
  {\bibinfo {volume} {2007}},\ \bibinfo {pages} {P06008} (\bibinfo {year}
  {2007})}\BibitemShut {NoStop}%
\bibitem [{\citenamefont {Korepin}\ \emph {et~al.}(1997)\citenamefont
  {Korepin}, \citenamefont {Bogoliubov},\ and\ \citenamefont
  {Izergin}}]{korepin1997quantum}%
  \BibitemOpen
  \bibfield  {author} {\bibinfo {author} {\bibfnamefont {V.~E.}\ \bibnamefont
  {Korepin}}, \bibinfo {author} {\bibfnamefont {N.~M.}\ \bibnamefont
  {Bogoliubov}}, \ and\ \bibinfo {author} {\bibfnamefont {A.~G.}\ \bibnamefont
  {Izergin}},\ }\href@noop {} {\emph {\bibinfo {title} {Quantum inverse
  scattering method and correlation functions}}},\ Vol.~\bibinfo {volume} {3}\
  (\bibinfo  {publisher} {Cambridge university press},\ \bibinfo {year}
  {1997})\BibitemShut {NoStop}%
\bibitem [{\citenamefont {Vanicat}\ \emph {et~al.}(2018)\citenamefont
  {Vanicat}, \citenamefont {Zadnik},\ and\ \citenamefont
  {Prosen}}]{vanicat2018integrable}%
  \BibitemOpen
  \bibfield  {author} {\bibinfo {author} {\bibfnamefont {M.}~\bibnamefont
  {Vanicat}}, \bibinfo {author} {\bibfnamefont {L.}~\bibnamefont {Zadnik}}, \
  and\ \bibinfo {author} {\bibfnamefont {T.}~\bibnamefont {Prosen}},\ }\href
  {\doibase 10.1103/PhysRevLett.121.030606} {\bibfield  {journal} {\bibinfo
  {journal} {Phys. Rev. Lett.}\ }\textbf {\bibinfo {volume} {121}},\ \bibinfo
  {pages} {030606} (\bibinfo {year} {2018})}\BibitemShut {NoStop}%
\bibitem [{\citenamefont {Ljubotina}\ \emph {et~al.}(2019)\citenamefont
  {Ljubotina}, \citenamefont {Zadnik},\ and\ \citenamefont
  {Prosen}}]{ljubotina2019ballistic}%
  \BibitemOpen
  \bibfield  {author} {\bibinfo {author} {\bibfnamefont {M.}~\bibnamefont
  {Ljubotina}}, \bibinfo {author} {\bibfnamefont {L.}~\bibnamefont {Zadnik}}, \
  and\ \bibinfo {author} {\bibfnamefont {T.}~\bibnamefont {Prosen}},\ }\href
  {\doibase 10.1103/PhysRevLett.122.150605} {\bibfield  {journal} {\bibinfo
  {journal} {Phys. Rev. Lett.}\ }\textbf {\bibinfo {volume} {122}},\ \bibinfo
  {pages} {150605} (\bibinfo {year} {2019})}\BibitemShut {NoStop}%
\bibitem [{\citenamefont {Medenjak}\ \emph {et~al.}(2020)\citenamefont
  {Medenjak}, \citenamefont {Prosen},\ and\ \citenamefont
  {Zadnik}}]{medenjak2020rigorous}%
  \BibitemOpen
  \bibfield  {author} {\bibinfo {author} {\bibfnamefont {M.}~\bibnamefont
  {Medenjak}}, \bibinfo {author} {\bibfnamefont {T.}~\bibnamefont {Prosen}}, \
  and\ \bibinfo {author} {\bibfnamefont {L.}~\bibnamefont {Zadnik}},\ }\href
  {\doibase 10.21468/SciPostPhys.9.1.003} {\bibfield  {journal} {\bibinfo
  {journal} {SciPost Phys.}\ }\textbf {\bibinfo {volume} {9}},\ \bibinfo
  {pages} {003} (\bibinfo {year} {2020})}\BibitemShut {NoStop}%
\bibitem [{\citenamefont {Rigol}\ \emph {et~al.}(2007)\citenamefont {Rigol},
  \citenamefont {Dunjko}, \citenamefont {Yurovsky},\ and\ \citenamefont
  {Olshanii}}]{rigol2007relaxation}%
  \BibitemOpen
  \bibfield  {author} {\bibinfo {author} {\bibfnamefont {M.}~\bibnamefont
  {Rigol}}, \bibinfo {author} {\bibfnamefont {V.}~\bibnamefont {Dunjko}},
  \bibinfo {author} {\bibfnamefont {V.}~\bibnamefont {Yurovsky}}, \ and\
  \bibinfo {author} {\bibfnamefont {M.}~\bibnamefont {Olshanii}},\ }\href
  {\doibase 10.1103/PhysRevLett.98.050405} {\bibfield  {journal} {\bibinfo
  {journal} {Phys. Rev. Lett.}\ }\textbf {\bibinfo {volume} {98}},\ \bibinfo
  {pages} {050405} (\bibinfo {year} {2007})}\BibitemShut {NoStop}%
\bibitem [{\citenamefont {Vidmar}\ and\ \citenamefont
  {Rigol}(2016)}]{vidmar2016generalized}%
  \BibitemOpen
  \bibfield  {author} {\bibinfo {author} {\bibfnamefont {L.}~\bibnamefont
  {Vidmar}}\ and\ \bibinfo {author} {\bibfnamefont {M.}~\bibnamefont {Rigol}},\
  }\href {https://iopscience.iop.org/article/10.1088/1742-5468/2016/06/064007}
  {\bibfield  {journal} {\bibinfo  {journal} {J. Stat. Mech.}\ }\textbf
  {\bibinfo {volume} {2016}},\ \bibinfo {pages} {064007} (\bibinfo {year}
  {2016})}\BibitemShut {NoStop}%
\bibitem [{\citenamefont {Essler}\ and\ \citenamefont
  {Fagotti}(2016)}]{essler2016quench}%
  \BibitemOpen
  \bibfield  {author} {\bibinfo {author} {\bibfnamefont {F.~H.}\ \bibnamefont
  {Essler}}\ and\ \bibinfo {author} {\bibfnamefont {M.}~\bibnamefont
  {Fagotti}},\ }\href
  {https://iopscience.iop.org/article/10.1088/1742-5468/2016/06/064002}
  {\bibfield  {journal} {\bibinfo  {journal} {J. Stat. Mech.}\ }\textbf
  {\bibinfo {volume} {2016}},\ \bibinfo {pages} {064002} (\bibinfo {year}
  {2016})}\BibitemShut {NoStop}%
\bibitem [{\citenamefont {Morvan}\ \emph {et~al.}(2022)\citenamefont {Morvan},
  \citenamefont {Andersen}, \citenamefont {Mi}, \citenamefont {Neill},
  \citenamefont {Petukhov}, \citenamefont {Kechedzhi}, \citenamefont {Abanin},
  \citenamefont {Acharya}, \citenamefont {Arute}, \citenamefont {Arya} \emph
  {et~al.}}]{morvan2022formation}%
  \BibitemOpen
  \bibfield  {author} {\bibinfo {author} {\bibfnamefont {A.}~\bibnamefont
  {Morvan}}, \bibinfo {author} {\bibfnamefont {T.~I.}\ \bibnamefont
  {Andersen}}, \bibinfo {author} {\bibfnamefont {X.}~\bibnamefont {Mi}},
  \bibinfo {author} {\bibfnamefont {C.}~\bibnamefont {Neill}}, \bibinfo
  {author} {\bibfnamefont {A.}~\bibnamefont {Petukhov}}, \bibinfo {author}
  {\bibfnamefont {K.}~\bibnamefont {Kechedzhi}}, \bibinfo {author}
  {\bibfnamefont {D.}~\bibnamefont {Abanin}}, \bibinfo {author} {\bibfnamefont
  {R.}~\bibnamefont {Acharya}}, \bibinfo {author} {\bibfnamefont
  {F.}~\bibnamefont {Arute}}, \bibinfo {author} {\bibfnamefont
  {K.}~\bibnamefont {Arya}},  \emph {et~al.},\ }\href
  {https://arxiv.org/abs/2206.05254} {\bibfield  {journal} {\bibinfo  {journal}
  {arXiv:2206.05254}\ } (\bibinfo {year} {2022})}\BibitemShut {NoStop}%
\bibitem [{\citenamefont {Maruyoshi}\ \emph {et~al.}(2022)\citenamefont
  {Maruyoshi}, \citenamefont {Okuda}, \citenamefont {Pedersen}, \citenamefont
  {Suzuki}, \citenamefont {Yamazaki},\ and\ \citenamefont
  {Yoshida}}]{maruyoshi2022conserved}%
  \BibitemOpen
  \bibfield  {author} {\bibinfo {author} {\bibfnamefont {K.}~\bibnamefont
  {Maruyoshi}}, \bibinfo {author} {\bibfnamefont {T.}~\bibnamefont {Okuda}},
  \bibinfo {author} {\bibfnamefont {J.~W.}\ \bibnamefont {Pedersen}}, \bibinfo
  {author} {\bibfnamefont {R.}~\bibnamefont {Suzuki}}, \bibinfo {author}
  {\bibfnamefont {M.}~\bibnamefont {Yamazaki}}, \ and\ \bibinfo {author}
  {\bibfnamefont {Y.}~\bibnamefont {Yoshida}},\ }\href
  {https://arxiv.org/abs/2208.00576} {\bibfield  {journal} {\bibinfo  {journal}
  {arXiv preprint arXiv:2208.00576}\ } (\bibinfo {year} {2022})}\BibitemShut
  {NoStop}%
\bibitem [{\citenamefont {Keenan}\ \emph {et~al.}(2022)\citenamefont {Keenan},
  \citenamefont {Robertson}, \citenamefont {Murphy}, \citenamefont {Zhuk},\
  and\ \citenamefont {Goold}}]{keenan2022evidence}%
  \BibitemOpen
  \bibfield  {author} {\bibinfo {author} {\bibfnamefont {N.}~\bibnamefont
  {Keenan}}, \bibinfo {author} {\bibfnamefont {N.}~\bibnamefont {Robertson}},
  \bibinfo {author} {\bibfnamefont {T.}~\bibnamefont {Murphy}}, \bibinfo
  {author} {\bibfnamefont {S.}~\bibnamefont {Zhuk}}, \ and\ \bibinfo {author}
  {\bibfnamefont {J.}~\bibnamefont {Goold}},\ }\href
  {https://arxiv.org/abs/2208.12243} {\bibfield  {journal} {\bibinfo  {journal}
  {arXiv:2208.12243}\ } (\bibinfo {year} {2022})}\BibitemShut {NoStop}%
\bibitem [{\citenamefont {Ilievski}\ \emph {et~al.}(2016)\citenamefont
  {Ilievski}, \citenamefont {Medenjak}, \citenamefont {Prosen},\ and\
  \citenamefont {Zadnik}}]{ilievski2016quasilocal}%
  \BibitemOpen
  \bibfield  {author} {\bibinfo {author} {\bibfnamefont {E.}~\bibnamefont
  {Ilievski}}, \bibinfo {author} {\bibfnamefont {M.}~\bibnamefont {Medenjak}},
  \bibinfo {author} {\bibfnamefont {T.}~\bibnamefont {Prosen}}, \ and\ \bibinfo
  {author} {\bibfnamefont {L.}~\bibnamefont {Zadnik}},\ }\href
  {https://iopscience.iop.org/article/10.1088/1742-5468/2016/06/064008}
  {\bibfield  {journal} {\bibinfo  {journal} {J. Stat. Mech.}\ }\textbf
  {\bibinfo {volume} {2016}},\ \bibinfo {pages} {064008} (\bibinfo {year}
  {2016})}\BibitemShut {NoStop}%
\bibitem [{SM()}]{SM}%
  \BibitemOpen
  \href@noop {} {}\bibinfo {note} {See Supplemental Material, which includes
  Refs.~\cite{Destri_de_Vega_1995,bertini2019exact,piroli2019integrable_su3_1,piroli2019integrable_su3_2,kuniba2011t,Kuniba_Sakai_Suzuki_1998,Murgan_Nepomechie_Shi_2006,borsi2021current},
  for details.}\BibitemShut {Stop}%
\bibitem [{\citenamefont {Giudice}\ \emph {et~al.}(2022)\citenamefont
  {Giudice}, \citenamefont {Giudici}, \citenamefont {Sonner}, \citenamefont
  {Thoenniss}, \citenamefont {Lerose}, \citenamefont {Abanin},\ and\
  \citenamefont {Piroli}}]{giudice2022temporal}%
  \BibitemOpen
  \bibfield  {author} {\bibinfo {author} {\bibfnamefont {G.}~\bibnamefont
  {Giudice}}, \bibinfo {author} {\bibfnamefont {G.}~\bibnamefont {Giudici}},
  \bibinfo {author} {\bibfnamefont {M.}~\bibnamefont {Sonner}}, \bibinfo
  {author} {\bibfnamefont {J.}~\bibnamefont {Thoenniss}}, \bibinfo {author}
  {\bibfnamefont {A.}~\bibnamefont {Lerose}}, \bibinfo {author} {\bibfnamefont
  {D.~A.}\ \bibnamefont {Abanin}}, \ and\ \bibinfo {author} {\bibfnamefont
  {L.}~\bibnamefont {Piroli}},\ }\href {\doibase
  10.1103/PhysRevLett.128.220401} {\bibfield  {journal} {\bibinfo  {journal}
  {Phys. Rev. Lett.}\ }\textbf {\bibinfo {volume} {128}},\ \bibinfo {pages}
  {220401} (\bibinfo {year} {2022})}\BibitemShut {NoStop}%
\bibitem [{\citenamefont {Caux}(2016)}]{caux2016quench}%
  \BibitemOpen
  \bibfield  {author} {\bibinfo {author} {\bibfnamefont {J.-S.}\ \bibnamefont
  {Caux}},\ }\href
  {https://iopscience.iop.org/article/10.1088/1742-5468/2016/06/064006}
  {\bibfield  {journal} {\bibinfo  {journal} {J. Stat. Mech.}\ }\textbf
  {\bibinfo {volume} {2016}},\ \bibinfo {pages} {064006} (\bibinfo {year}
  {2016})}\BibitemShut {NoStop}%
\bibitem [{\citenamefont {Calabrese}\ \emph {et~al.}(2016)\citenamefont
  {Calabrese}, \citenamefont {Essler},\ and\ \citenamefont
  {Mussardo}}]{calabrese2016introduction}%
  \BibitemOpen
  \bibfield  {author} {\bibinfo {author} {\bibfnamefont {P.}~\bibnamefont
  {Calabrese}}, \bibinfo {author} {\bibfnamefont {F.~H.}\ \bibnamefont
  {Essler}}, \ and\ \bibinfo {author} {\bibfnamefont {G.}~\bibnamefont
  {Mussardo}},\ }\href
  {https://iopscience.iop.org/article/10.1088/1742-5468/2016/06/064001}
  {\bibfield  {journal} {\bibinfo  {journal} {J. Stat. Mech.}\ }\textbf
  {\bibinfo {volume} {2016}},\ \bibinfo {pages} {064001} (\bibinfo {year}
  {2016})}\BibitemShut {NoStop}%
\bibitem [{\citenamefont {Piroli}\ \emph
  {et~al.}(2017{\natexlab{a}})\citenamefont {Piroli}, \citenamefont {Pozsgay},\
  and\ \citenamefont {Vernier}}]{piroli2017integrable}%
  \BibitemOpen
  \bibfield  {author} {\bibinfo {author} {\bibfnamefont {L.}~\bibnamefont
  {Piroli}}, \bibinfo {author} {\bibfnamefont {B.}~\bibnamefont {Pozsgay}}, \
  and\ \bibinfo {author} {\bibfnamefont {E.}~\bibnamefont {Vernier}},\ }\href
  {\doibase 10.1016/j.nuclphysb.2017.10.012} {\bibfield  {journal} {\bibinfo
  {journal} {Nucl. Phys. B}\ }\textbf {\bibinfo {volume} {925}},\ \bibinfo
  {pages} {362} (\bibinfo {year} {2017}{\natexlab{a}})}\BibitemShut {NoStop}%
\bibitem [{\citenamefont {Aleiner}(2021)}]{aleiner2021bethe}%
  \BibitemOpen
  \bibfield  {author} {\bibinfo {author} {\bibfnamefont {I.~L.}\ \bibnamefont
  {Aleiner}},\ }\href {\doibase 10.1016/j.aop.2021.168593} {\bibfield
  {journal} {\bibinfo  {journal} {Ann. Phys.}\ }\textbf {\bibinfo {volume}
  {433}},\ \bibinfo {pages} {168593} (\bibinfo {year} {2021})}\BibitemShut
  {NoStop}%
\bibitem [{\citenamefont {Claeys}\ \emph {et~al.}(2022)\citenamefont {Claeys},
  \citenamefont {Herzog-Arbeitman},\ and\ \citenamefont
  {Lamacraft}}]{pieter2022correlations}%
  \BibitemOpen
  \bibfield  {author} {\bibinfo {author} {\bibfnamefont {P.~W.}\ \bibnamefont
  {Claeys}}, \bibinfo {author} {\bibfnamefont {J.}~\bibnamefont
  {Herzog-Arbeitman}}, \ and\ \bibinfo {author} {\bibfnamefont
  {A.}~\bibnamefont {Lamacraft}},\ }\href {\doibase
  10.21468/SciPostPhys.12.1.007} {\bibfield  {journal} {\bibinfo  {journal}
  {SciPost Phys.}\ }\textbf {\bibinfo {volume} {12}},\ \bibinfo {pages} {007}
  (\bibinfo {year} {2022})}\BibitemShut {NoStop}%
\bibitem [{\citenamefont {Miao}\ \emph {et~al.}(2022)\citenamefont {Miao},
  \citenamefont {Gritsev},\ and\ \citenamefont {Kurlov}}]{miao2022floquet}%
  \BibitemOpen
  \bibfield  {author} {\bibinfo {author} {\bibfnamefont {Y.}~\bibnamefont
  {Miao}}, \bibinfo {author} {\bibfnamefont {V.}~\bibnamefont {Gritsev}}, \
  and\ \bibinfo {author} {\bibfnamefont {D.~V.}\ \bibnamefont {Kurlov}},\
  }\href {https://arxiv.org/abs/2206.15142} {\bibfield  {journal} {\bibinfo
  {journal} {arXiv:2206.15142}\ } (\bibinfo {year} {2022})}\BibitemShut
  {NoStop}%
\bibitem [{\citenamefont {Takahashi}(2005)}]{takahashi2005thermodynamics}%
  \BibitemOpen
  \bibfield  {author} {\bibinfo {author} {\bibfnamefont {M.}~\bibnamefont
  {Takahashi}},\ }\href@noop {} {\bibfield  {journal} {\bibinfo  {journal}
  {Thermodynamics of One-Dimensional Solvable Models}\ } (\bibinfo {year}
  {2005})}\BibitemShut {NoStop}%
\bibitem [{Note1()}]{Note1}%
  \BibitemOpen
  \bibinfo {note} {The validity of this assumption is verified a posteriori,
  based on the agreement of our predictions with numerical
  computations}\BibitemShut {NoStop}%
\bibitem [{\citenamefont {Wouters}\ \emph {et~al.}(2014)\citenamefont
  {Wouters}, \citenamefont {De~Nardis}, \citenamefont {Brockmann},
  \citenamefont {Fioretto}, \citenamefont {Rigol},\ and\ \citenamefont
  {Caux}}]{wouters2014quenching}%
  \BibitemOpen
  \bibfield  {author} {\bibinfo {author} {\bibfnamefont {B.}~\bibnamefont
  {Wouters}}, \bibinfo {author} {\bibfnamefont {J.}~\bibnamefont {De~Nardis}},
  \bibinfo {author} {\bibfnamefont {M.}~\bibnamefont {Brockmann}}, \bibinfo
  {author} {\bibfnamefont {D.}~\bibnamefont {Fioretto}}, \bibinfo {author}
  {\bibfnamefont {M.}~\bibnamefont {Rigol}}, \ and\ \bibinfo {author}
  {\bibfnamefont {J.-S.}\ \bibnamefont {Caux}},\ }\href {\doibase
  10.1103/PhysRevLett.113.117202} {\bibfield  {journal} {\bibinfo  {journal}
  {Phys. Rev. Lett.}\ }\textbf {\bibinfo {volume} {113}},\ \bibinfo {pages}
  {117202} (\bibinfo {year} {2014})}\BibitemShut {NoStop}%
\bibitem [{\citenamefont {Pozsgay}\ \emph {et~al.}(2014)\citenamefont
  {Pozsgay}, \citenamefont {Mesty\'an}, \citenamefont {Werner}, \citenamefont
  {Kormos}, \citenamefont {Zar\'and},\ and\ \citenamefont
  {Tak\'acs}}]{pozsgay2014correlations}%
  \BibitemOpen
  \bibfield  {author} {\bibinfo {author} {\bibfnamefont {B.}~\bibnamefont
  {Pozsgay}}, \bibinfo {author} {\bibfnamefont {M.}~\bibnamefont {Mesty\'an}},
  \bibinfo {author} {\bibfnamefont {M.~A.}\ \bibnamefont {Werner}}, \bibinfo
  {author} {\bibfnamefont {M.}~\bibnamefont {Kormos}}, \bibinfo {author}
  {\bibfnamefont {G.}~\bibnamefont {Zar\'and}}, \ and\ \bibinfo {author}
  {\bibfnamefont {G.}~\bibnamefont {Tak\'acs}},\ }\href {\doibase
  10.1103/PhysRevLett.113.117203} {\bibfield  {journal} {\bibinfo  {journal}
  {Phys. Rev. Lett.}\ }\textbf {\bibinfo {volume} {113}},\ \bibinfo {pages}
  {117203} (\bibinfo {year} {2014})}\BibitemShut {NoStop}%
\bibitem [{\citenamefont {Brockmann}\ \emph {et~al.}(2014)\citenamefont
  {Brockmann}, \citenamefont {Wouters}, \citenamefont {Fioretto}, \citenamefont
  {De~Nardis}, \citenamefont {Vlijm},\ and\ \citenamefont
  {Caux}}]{brockmann2014quench}%
  \BibitemOpen
  \bibfield  {author} {\bibinfo {author} {\bibfnamefont {M.}~\bibnamefont
  {Brockmann}}, \bibinfo {author} {\bibfnamefont {B.}~\bibnamefont {Wouters}},
  \bibinfo {author} {\bibfnamefont {D.}~\bibnamefont {Fioretto}}, \bibinfo
  {author} {\bibfnamefont {J.}~\bibnamefont {De~Nardis}}, \bibinfo {author}
  {\bibfnamefont {R.}~\bibnamefont {Vlijm}}, \ and\ \bibinfo {author}
  {\bibfnamefont {J.-S.}\ \bibnamefont {Caux}},\ }\href {\doibase
  10.1088/1742-5468/2014/12/P12009} {\bibfield  {journal} {\bibinfo  {journal}
  {J. Stat. Mech.}\ }\textbf {\bibinfo {volume} {2014}},\ \bibinfo {pages}
  {P12009} (\bibinfo {year} {2014})}\BibitemShut {NoStop}%
\bibitem [{\citenamefont {Mesty{\'a}n}\ \emph {et~al.}(2015)\citenamefont
  {Mesty{\'a}n}, \citenamefont {Pozsgay}, \citenamefont {Tak{\'a}cs},\ and\
  \citenamefont {Werner}}]{mestyan2015quenching}%
  \BibitemOpen
  \bibfield  {author} {\bibinfo {author} {\bibfnamefont {M.}~\bibnamefont
  {Mesty{\'a}n}}, \bibinfo {author} {\bibfnamefont {B.}~\bibnamefont
  {Pozsgay}}, \bibinfo {author} {\bibfnamefont {G.}~\bibnamefont {Tak{\'a}cs}},
  \ and\ \bibinfo {author} {\bibfnamefont {M.}~\bibnamefont {Werner}},\ }\href
  {\doibase 10.1088/1742-5468/2014/12/P12009} {\bibfield  {journal} {\bibinfo
  {journal} {J. Stat. Mech.}\ }\textbf {\bibinfo {volume} {2015}},\ \bibinfo
  {pages} {P04001} (\bibinfo {year} {2015})}\BibitemShut {NoStop}%
\bibitem [{\citenamefont {Caux}\ and\ \citenamefont
  {Essler}(2013)}]{caux2013time}%
  \BibitemOpen
  \bibfield  {author} {\bibinfo {author} {\bibfnamefont {J.-S.}\ \bibnamefont
  {Caux}}\ and\ \bibinfo {author} {\bibfnamefont {F.~H.~L.}\ \bibnamefont
  {Essler}},\ }\href {\doibase 10.1103/PhysRevLett.110.257203} {\bibfield
  {journal} {\bibinfo  {journal} {Phys. Rev. Lett.}\ }\textbf {\bibinfo
  {volume} {110}},\ \bibinfo {pages} {257203} (\bibinfo {year}
  {2013})}\BibitemShut {NoStop}%
\bibitem [{\citenamefont {Pozsgay}(2013)}]{Pozsgay_2013}%
  \BibitemOpen
  \bibfield  {author} {\bibinfo {author} {\bibfnamefont {B.}~\bibnamefont
  {Pozsgay}},\ }\href {\doibase 10.1088/1742-5468/2013/10/P10028} {\bibfield
  {journal} {\bibinfo  {journal} {J. Stat. Mech.}\ }\textbf {\bibinfo {volume}
  {2013}},\ \bibinfo {pages} {P10028} (\bibinfo {year} {2013})}\BibitemShut
  {NoStop}%
\bibitem [{\citenamefont {Piroli}\ \emph
  {et~al.}(2017{\natexlab{b}})\citenamefont {Piroli}, \citenamefont {Pozsgay},\
  and\ \citenamefont {Vernier}}]{Piroli_Pozsgay_Vernier_2017}%
  \BibitemOpen
  \bibfield  {author} {\bibinfo {author} {\bibfnamefont {L.}~\bibnamefont
  {Piroli}}, \bibinfo {author} {\bibfnamefont {B.}~\bibnamefont {Pozsgay}}, \
  and\ \bibinfo {author} {\bibfnamefont {E.}~\bibnamefont {Vernier}},\ }\href
  {\doibase 10.1088/1742-5468/aa5d1e} {\bibfield  {journal} {\bibinfo
  {journal} {J. Stat. Mech.}\ }\textbf {\bibinfo {volume} {2017}},\ \bibinfo
  {pages} {023106} (\bibinfo {year} {2017}{\natexlab{b}})}\BibitemShut
  {NoStop}%
\bibitem [{\citenamefont {Piroli}\ \emph {et~al.}(2018)\citenamefont {Piroli},
  \citenamefont {Pozsgay},\ and\ \citenamefont
  {Vernier}}]{Piroli_Pozsgay_Vernier_2018}%
  \BibitemOpen
  \bibfield  {author} {\bibinfo {author} {\bibfnamefont {L.}~\bibnamefont
  {Piroli}}, \bibinfo {author} {\bibfnamefont {B.}~\bibnamefont {Pozsgay}}, \
  and\ \bibinfo {author} {\bibfnamefont {E.}~\bibnamefont {Vernier}},\ }\href
  {\doibase 10.1016/j.nuclphysb.2018.06.015} {\bibfield  {journal} {\bibinfo
  {journal} {Nucl. Phys. B}\ }\textbf {\bibinfo {volume} {933}},\ \bibinfo
  {pages} {454} (\bibinfo {year} {2018})}\BibitemShut {NoStop}%
\bibitem [{\citenamefont {Pozsgay}\ \emph {et~al.}(2019)\citenamefont
  {Pozsgay}, \citenamefont {Piroli},\ and\ \citenamefont
  {Vernier}}]{pozsgay2019integrable}%
  \BibitemOpen
  \bibfield  {author} {\bibinfo {author} {\bibfnamefont {B.}~\bibnamefont
  {Pozsgay}}, \bibinfo {author} {\bibfnamefont {L.}~\bibnamefont {Piroli}}, \
  and\ \bibinfo {author} {\bibfnamefont {E.}~\bibnamefont {Vernier}},\ }\href
  {\doibase 10.21468/SciPostPhys.6.5.062} {\bibfield  {journal} {\bibinfo
  {journal} {SciPost Phys.}\ }\textbf {\bibinfo {volume} {6}},\ \bibinfo
  {pages} {062} (\bibinfo {year} {2019})}\BibitemShut {NoStop}%
\bibitem [{\citenamefont {Klümper}(2004)}]{klumper_integrability_2004}%
  \BibitemOpen
  \bibfield  {author} {\bibinfo {author} {\bibfnamefont {A.}~\bibnamefont
  {Klümper}},\ }in\ \href {\doibase 10.1007/BFb0119598} {\emph {\bibinfo
  {booktitle} {Quantum {Magnetism}}}},\ \bibinfo {series and number} {\bibinfo
  {series} {Lecture {Notes} in {Physics}}\ No.\ \bibinfo {number} {645}},\
  \bibinfo {editor} {edited by\ \bibinfo {editor} {\bibfnamefont
  {U.}~\bibnamefont {Schollwöck}}, \bibinfo {editor} {\bibfnamefont
  {J.}~\bibnamefont {Richter}}, \bibinfo {editor} {\bibfnamefont {D.~J.~J.}\
  \bibnamefont {Farnell}}, \ and\ \bibinfo {editor} {\bibfnamefont {R.~F.}\
  \bibnamefont {Bishop}}}\ (\bibinfo  {publisher} {Springer Berlin
  Heidelberg},\ \bibinfo {year} {2004})\ pp.\ \bibinfo {pages}
  {349--379}\BibitemShut {NoStop}%
\bibitem [{InP()}]{InPrep}%
  \BibitemOpen
  \href@noop {} {}\bibinfo {note} {Bruno Bertini, Giuliano Giudici, Lorenzo
  Piroli, Eric Vernier, \emph{In preparation}.}\BibitemShut {Stop}%
\bibitem [{Note2()}]{Note2}%
  \BibitemOpen
  \bibinfo {note} {Interestingly, for $\tau $ approaching $\tau _{\protect \rm
  th}(\Delta )$ from below, the stationary state described by Eqs.~\protect
  \textup {\hbox {\mathsurround \z@ \protect \normalfont (\ignorespaces \ref
  {eq:eta1gapped}\unskip \@@italiccorr )}}--\protect \textup {\hbox
  {\mathsurround \z@ \protect \normalfont (\ignorespaces \ref
  {eq:higher_n}\unskip \@@italiccorr )}} approaches the one of the isotropic
  Heisenberg chain in continuous time}\BibitemShut {NoStop}%
\bibitem [{\citenamefont {Mesty{\'a}n}\ and\ \citenamefont
  {Pozsgay}(2014)}]{mestyan2014short}%
  \BibitemOpen
  \bibfield  {author} {\bibinfo {author} {\bibfnamefont {M.}~\bibnamefont
  {Mesty{\'a}n}}\ and\ \bibinfo {author} {\bibfnamefont {B.}~\bibnamefont
  {Pozsgay}},\ }\href {\doibase 10.1088/1742-5468/2014/09/P09020} {\bibfield
  {journal} {\bibinfo  {journal} {J. Stat. Mech.}\ }\textbf {\bibinfo {volume}
  {2014}},\ \bibinfo {pages} {P09020} (\bibinfo {year} {2014})}\BibitemShut
  {NoStop}%
\bibitem [{\citenamefont {Piroli}\ \emph {et~al.}(2016)\citenamefont {Piroli},
  \citenamefont {Vernier},\ and\ \citenamefont {Calabrese}}]{exact2016piroli}%
  \BibitemOpen
  \bibfield  {author} {\bibinfo {author} {\bibfnamefont {L.}~\bibnamefont
  {Piroli}}, \bibinfo {author} {\bibfnamefont {E.}~\bibnamefont {Vernier}}, \
  and\ \bibinfo {author} {\bibfnamefont {P.}~\bibnamefont {Calabrese}},\ }\href
  {\doibase 10.1103/PhysRevB.94.054313} {\bibfield  {journal} {\bibinfo
  {journal} {Phys. Rev. B}\ }\textbf {\bibinfo {volume} {94}},\ \bibinfo
  {pages} {054313} (\bibinfo {year} {2016})}\BibitemShut {NoStop}%
\bibitem [{\citenamefont {Piroli}\ \emph
  {et~al.}(2017{\natexlab{c}})\citenamefont {Piroli}, \citenamefont {Vernier},
  \citenamefont {Calabrese},\ and\ \citenamefont
  {Rigol}}]{correlation2016piroli}%
  \BibitemOpen
  \bibfield  {author} {\bibinfo {author} {\bibfnamefont {L.}~\bibnamefont
  {Piroli}}, \bibinfo {author} {\bibfnamefont {E.}~\bibnamefont {Vernier}},
  \bibinfo {author} {\bibfnamefont {P.}~\bibnamefont {Calabrese}}, \ and\
  \bibinfo {author} {\bibfnamefont {M.}~\bibnamefont {Rigol}},\ }\href
  {\doibase 10.1103/PhysRevB.95.054308} {\bibfield  {journal} {\bibinfo
  {journal} {Phys. Rev. B}\ }\textbf {\bibinfo {volume} {95}},\ \bibinfo
  {pages} {054308} (\bibinfo {year} {2017}{\natexlab{c}})}\BibitemShut
  {NoStop}%
\bibitem [{\citenamefont {Bastianello}\ \emph {et~al.}(2018)\citenamefont
  {Bastianello}, \citenamefont {Piroli},\ and\ \citenamefont
  {Calabrese}}]{bastianello2018exact}%
  \BibitemOpen
  \bibfield  {author} {\bibinfo {author} {\bibfnamefont {A.}~\bibnamefont
  {Bastianello}}, \bibinfo {author} {\bibfnamefont {L.}~\bibnamefont {Piroli}},
  \ and\ \bibinfo {author} {\bibfnamefont {P.}~\bibnamefont {Calabrese}},\
  }\href {\doibase 10.1103/PhysRevLett.120.190601} {\bibfield  {journal}
  {\bibinfo  {journal} {Phys. Rev. Lett.}\ }\textbf {\bibinfo {volume} {120}},\
  \bibinfo {pages} {190601} (\bibinfo {year} {2018})}\BibitemShut {NoStop}%
\bibitem [{\citenamefont {Kitanine}\ \emph {et~al.}(1999)\citenamefont
  {Kitanine}, \citenamefont {Maillet},\ and\ \citenamefont
  {Terras}}]{kitanine1999form}%
  \BibitemOpen
  \bibfield  {author} {\bibinfo {author} {\bibfnamefont {N.}~\bibnamefont
  {Kitanine}}, \bibinfo {author} {\bibfnamefont {J.}~\bibnamefont {Maillet}}, \
  and\ \bibinfo {author} {\bibfnamefont {V.}~\bibnamefont {Terras}},\ }\href
  {\doibase 10.1016/S0550-3213(99)00295-3} {\bibfield  {journal} {\bibinfo
  {journal} {Nucl. Phys. B}\ }\textbf {\bibinfo {volume} {554}},\ \bibinfo
  {pages} {647} (\bibinfo {year} {1999})}\BibitemShut {NoStop}%
\bibitem [{\citenamefont {Kitanine}\ \emph {et~al.}(2000)\citenamefont
  {Kitanine}, \citenamefont {Maillet},\ and\ \citenamefont
  {Terras}}]{kitanine2000correlation}%
  \BibitemOpen
  \bibfield  {author} {\bibinfo {author} {\bibfnamefont {N.}~\bibnamefont
  {Kitanine}}, \bibinfo {author} {\bibfnamefont {J.~M.}\ \bibnamefont
  {Maillet}}, \ and\ \bibinfo {author} {\bibfnamefont {V.}~\bibnamefont
  {Terras}},\ }\href {\doibase 10.1016/S0550-3213(99)00619-7} {\bibfield
  {journal} {\bibinfo  {journal} {Nucl. Phys. B}\ }\textbf {\bibinfo {volume}
  {567}},\ \bibinfo {pages} {554} (\bibinfo {year} {2000})}\BibitemShut
  {NoStop}%
\bibitem [{\citenamefont {Fagotti}\ \emph {et~al.}(2014)\citenamefont
  {Fagotti}, \citenamefont {Collura}, \citenamefont {Essler},\ and\
  \citenamefont {Calabrese}}]{fagotti2014relaxation}%
  \BibitemOpen
  \bibfield  {author} {\bibinfo {author} {\bibfnamefont {M.}~\bibnamefont
  {Fagotti}}, \bibinfo {author} {\bibfnamefont {M.}~\bibnamefont {Collura}},
  \bibinfo {author} {\bibfnamefont {F.~H.~L.}\ \bibnamefont {Essler}}, \ and\
  \bibinfo {author} {\bibfnamefont {P.}~\bibnamefont {Calabrese}},\ }\href
  {\doibase 10.1103/PhysRevB.89.125101} {\bibfield  {journal} {\bibinfo
  {journal} {Phys. Rev. B}\ }\textbf {\bibinfo {volume} {89}},\ \bibinfo
  {pages} {125101} (\bibinfo {year} {2014})}\BibitemShut {NoStop}%
\bibitem [{\citenamefont {Schollw{\"o}ck}(2011)}]{schollwock2011density}%
  \BibitemOpen
  \bibfield  {author} {\bibinfo {author} {\bibfnamefont {U.}~\bibnamefont
  {Schollw{\"o}ck}},\ }\href {\doibase 10.1103/PhysRevLett.93.040502}
  {\bibfield  {journal} {\bibinfo  {journal} {Ann. Phys.}\ }\textbf {\bibinfo
  {volume} {326}},\ \bibinfo {pages} {96} (\bibinfo {year} {2011})}\BibitemShut
  {NoStop}%
\bibitem [{\citenamefont {Bulchandani}\ \emph {et~al.}(2021)\citenamefont
  {Bulchandani}, \citenamefont {Gopalakrishnan},\ and\ \citenamefont
  {Ilievski}}]{bulchandani2021superdiffusion}%
  \BibitemOpen
  \bibfield  {author} {\bibinfo {author} {\bibfnamefont {V.~B.}\ \bibnamefont
  {Bulchandani}}, \bibinfo {author} {\bibfnamefont {S.}~\bibnamefont
  {Gopalakrishnan}}, \ and\ \bibinfo {author} {\bibfnamefont {E.}~\bibnamefont
  {Ilievski}},\ }\href {\doibase 10.1088/1742-5468/ac12c7} {\bibfield
  {journal} {\bibinfo  {journal} {J. Stat. Mech}\ }\textbf {\bibinfo {volume}
  {2021}},\ \bibinfo {pages} {084001} (\bibinfo {year} {2021})}\BibitemShut
  {NoStop}%
\bibitem [{\citenamefont {Bertini}\ \emph {et~al.}(2021)\citenamefont
  {Bertini}, \citenamefont {Heidrich-Meisner}, \citenamefont {Karrasch},
  \citenamefont {Prosen}, \citenamefont {Steinigeweg},\ and\ \citenamefont
  {\ifmmode \check{Z}\else \v{Z}\fi{}nidari\ifmmode~\check{c}\else
  \v{c}\fi{}}}]{bertini2021finite}%
  \BibitemOpen
  \bibfield  {author} {\bibinfo {author} {\bibfnamefont {B.}~\bibnamefont
  {Bertini}}, \bibinfo {author} {\bibfnamefont {F.}~\bibnamefont
  {Heidrich-Meisner}}, \bibinfo {author} {\bibfnamefont {C.}~\bibnamefont
  {Karrasch}}, \bibinfo {author} {\bibfnamefont {T.}~\bibnamefont {Prosen}},
  \bibinfo {author} {\bibfnamefont {R.}~\bibnamefont {Steinigeweg}}, \ and\
  \bibinfo {author} {\bibfnamefont {M.}~\bibnamefont {\ifmmode \check{Z}\else
  \v{Z}\fi{}nidari\ifmmode~\check{c}\else \v{c}\fi{}}},\ }\href {\doibase
  10.1103/RevModPhys.93.025003} {\bibfield  {journal} {\bibinfo  {journal}
  {Rev. Mod. Phys.}\ }\textbf {\bibinfo {volume} {93}},\ \bibinfo {pages}
  {025003} (\bibinfo {year} {2021})}\BibitemShut {NoStop}%
\bibitem [{\citenamefont {Prosen}(2011)}]{prosen2011open}%
  \BibitemOpen
  \bibfield  {author} {\bibinfo {author} {\bibfnamefont {T.}~\bibnamefont
  {Prosen}},\ }\href {\doibase 10.1103/PhysRevLett.106.217206} {\bibfield
  {journal} {\bibinfo  {journal} {Phys. Rev. Lett.}\ }\textbf {\bibinfo
  {volume} {106}},\ \bibinfo {pages} {217206} (\bibinfo {year}
  {2011})}\BibitemShut {NoStop}%
\bibitem [{\citenamefont {Prosen}\ and\ \citenamefont
  {Ilievski}(2013)}]{prosen2013families}%
  \BibitemOpen
  \bibfield  {author} {\bibinfo {author} {\bibfnamefont {T.}~\bibnamefont
  {Prosen}}\ and\ \bibinfo {author} {\bibfnamefont {E.}~\bibnamefont
  {Ilievski}},\ }\href {\doibase 10.1103/PhysRevLett.111.057203} {\bibfield
  {journal} {\bibinfo  {journal} {Phys. Rev. Lett.}\ }\textbf {\bibinfo
  {volume} {111}},\ \bibinfo {pages} {057203} (\bibinfo {year}
  {2013})}\BibitemShut {NoStop}%
\bibitem [{\citenamefont {Collura}\ \emph {et~al.}(2018)\citenamefont
  {Collura}, \citenamefont {De~Luca},\ and\ \citenamefont
  {Viti}}]{collura2018analytic}%
  \BibitemOpen
  \bibfield  {author} {\bibinfo {author} {\bibfnamefont {M.}~\bibnamefont
  {Collura}}, \bibinfo {author} {\bibfnamefont {A.}~\bibnamefont {De~Luca}}, \
  and\ \bibinfo {author} {\bibfnamefont {J.}~\bibnamefont {Viti}},\ }\href
  {\doibase 10.1103/PhysRevB.97.081111} {\bibfield  {journal} {\bibinfo
  {journal} {Phys. Rev. B}\ }\textbf {\bibinfo {volume} {97}},\ \bibinfo
  {pages} {081111} (\bibinfo {year} {2018})}\BibitemShut {NoStop}%
\bibitem [{\citenamefont {Alba}\ and\ \citenamefont
  {Calabrese}(2017)}]{alba2017entanglement}%
  \BibitemOpen
  \bibfield  {author} {\bibinfo {author} {\bibfnamefont {V.}~\bibnamefont
  {Alba}}\ and\ \bibinfo {author} {\bibfnamefont {P.}~\bibnamefont
  {Calabrese}},\ }\href {\doibase 10.1073/pnas.1703516114} {\bibfield
  {journal} {\bibinfo  {journal} {PNAS}\ }\textbf {\bibinfo {volume} {114}},\
  \bibinfo {pages} {7947} (\bibinfo {year} {2017})}\BibitemShut {NoStop}%
\bibitem [{\citenamefont {Bertini}\ \emph {et~al.}(2022)\citenamefont
  {Bertini}, \citenamefont {Klobas}, \citenamefont {Alba}, \citenamefont
  {Lagnese},\ and\ \citenamefont {Calabrese}}]{bertini2022growth}%
  \BibitemOpen
  \bibfield  {author} {\bibinfo {author} {\bibfnamefont {B.}~\bibnamefont
  {Bertini}}, \bibinfo {author} {\bibfnamefont {K.}~\bibnamefont {Klobas}},
  \bibinfo {author} {\bibfnamefont {V.}~\bibnamefont {Alba}}, \bibinfo {author}
  {\bibfnamefont {G.}~\bibnamefont {Lagnese}}, \ and\ \bibinfo {author}
  {\bibfnamefont {P.}~\bibnamefont {Calabrese}},\ }\href {\doibase
  10.1103/PhysRevX.12.031016} {\bibfield  {journal} {\bibinfo  {journal} {Phys.
  Rev. X}\ }\textbf {\bibinfo {volume} {12}},\ \bibinfo {pages} {031016}
  (\bibinfo {year} {2022})}\BibitemShut {NoStop}%
\bibitem [{\citenamefont {Bertini}\ \emph {et~al.}(2016)\citenamefont
  {Bertini}, \citenamefont {Collura}, \citenamefont {De~Nardis},\ and\
  \citenamefont {Fagotti}}]{bertini2016transport}%
  \BibitemOpen
  \bibfield  {author} {\bibinfo {author} {\bibfnamefont {B.}~\bibnamefont
  {Bertini}}, \bibinfo {author} {\bibfnamefont {M.}~\bibnamefont {Collura}},
  \bibinfo {author} {\bibfnamefont {J.}~\bibnamefont {De~Nardis}}, \ and\
  \bibinfo {author} {\bibfnamefont {M.}~\bibnamefont {Fagotti}},\ }\href
  {\doibase 10.1103/PhysRevLett.117.207201} {\bibfield  {journal} {\bibinfo
  {journal} {Phys. Rev. Lett.}\ }\textbf {\bibinfo {volume} {117}},\ \bibinfo
  {pages} {207201} (\bibinfo {year} {2016})}\BibitemShut {NoStop}%
\bibitem [{\citenamefont {Castro-Alvaredo}\ \emph {et~al.}(2016)\citenamefont
  {Castro-Alvaredo}, \citenamefont {Doyon},\ and\ \citenamefont
  {Yoshimura}}]{castro2016emergent}%
  \BibitemOpen
  \bibfield  {author} {\bibinfo {author} {\bibfnamefont {O.~A.}\ \bibnamefont
  {Castro-Alvaredo}}, \bibinfo {author} {\bibfnamefont {B.}~\bibnamefont
  {Doyon}}, \ and\ \bibinfo {author} {\bibfnamefont {T.}~\bibnamefont
  {Yoshimura}},\ }\href {\doibase 10.1103/PhysRevX.6.041065} {\bibfield
  {journal} {\bibinfo  {journal} {Phys. Rev. X}\ }\textbf {\bibinfo {volume}
  {6}},\ \bibinfo {pages} {041065} (\bibinfo {year} {2016})}\BibitemShut
  {NoStop}%
\bibitem [{\citenamefont {Destri}\ and\ \citenamefont
  {de~Vega}(1995)}]{Destri_de_Vega_1995}%
  \BibitemOpen
  \bibfield  {author} {\bibinfo {author} {\bibfnamefont {C.}~\bibnamefont
  {Destri}}\ and\ \bibinfo {author} {\bibfnamefont {H.~J.}\ \bibnamefont
  {de~Vega}},\ }\href {\doibase 10.1016/0550-3213(94)00547-R} {\bibfield
  {journal} {\bibinfo  {journal} {Nucl. Phys. B}\ }\textbf {\bibinfo {volume}
  {438}},\ \bibinfo {pages} {413–454} (\bibinfo {year} {1995})},\ \bibinfo
  {note} {arXiv: hep-th/9407117}\BibitemShut {NoStop}%
\bibitem [{\citenamefont {Bertini}\ \emph {et~al.}(2019)\citenamefont
  {Bertini}, \citenamefont {Kos},\ and\ \citenamefont
  {Prosen}}]{bertini2019exact}%
  \BibitemOpen
  \bibfield  {author} {\bibinfo {author} {\bibfnamefont {B.}~\bibnamefont
  {Bertini}}, \bibinfo {author} {\bibfnamefont {P.}~\bibnamefont {Kos}}, \ and\
  \bibinfo {author} {\bibfnamefont {T.}~\bibnamefont {Prosen}},\ }\href
  {\doibase 10.1103/PhysRevLett.123.210601} {\bibfield  {journal} {\bibinfo
  {journal} {Phys. Rev. Lett.}\ }\textbf {\bibinfo {volume} {123}},\ \bibinfo
  {pages} {210601} (\bibinfo {year} {2019})}\BibitemShut {NoStop}%
\bibitem [{\citenamefont {Piroli}\ \emph
  {et~al.}(2019{\natexlab{a}})\citenamefont {Piroli}, \citenamefont {Vernier},
  \citenamefont {Calabrese},\ and\ \citenamefont
  {Pozsgay}}]{piroli2019integrable_su3_1}%
  \BibitemOpen
  \bibfield  {author} {\bibinfo {author} {\bibfnamefont {L.}~\bibnamefont
  {Piroli}}, \bibinfo {author} {\bibfnamefont {E.}~\bibnamefont {Vernier}},
  \bibinfo {author} {\bibfnamefont {P.}~\bibnamefont {Calabrese}}, \ and\
  \bibinfo {author} {\bibfnamefont {B.}~\bibnamefont {Pozsgay}},\ }\href
  {\doibase 10.1088/1742-5468/ab1c51} {\bibfield  {journal} {\bibinfo
  {journal} {J. Stat. Mech.}\ }\textbf {\bibinfo {volume} {2019}},\ \bibinfo
  {pages} {063103} (\bibinfo {year} {2019}{\natexlab{a}})}\BibitemShut
  {NoStop}%
\bibitem [{\citenamefont {Piroli}\ \emph
  {et~al.}(2019{\natexlab{b}})\citenamefont {Piroli}, \citenamefont {Vernier},
  \citenamefont {Calabrese},\ and\ \citenamefont
  {Pozsgay}}]{piroli2019integrable_su3_2}%
  \BibitemOpen
  \bibfield  {author} {\bibinfo {author} {\bibfnamefont {L.}~\bibnamefont
  {Piroli}}, \bibinfo {author} {\bibfnamefont {E.}~\bibnamefont {Vernier}},
  \bibinfo {author} {\bibfnamefont {P.}~\bibnamefont {Calabrese}}, \ and\
  \bibinfo {author} {\bibfnamefont {B.}~\bibnamefont {Pozsgay}},\ }\href
  {\doibase 10.1088/1742-5468/ab1c52} {\bibfield  {journal} {\bibinfo
  {journal} {J. Stat. Mech.}\ }\textbf {\bibinfo {volume} {2019}},\ \bibinfo
  {pages} {063104} (\bibinfo {year} {2019}{\natexlab{b}})}\BibitemShut
  {NoStop}%
\bibitem [{\citenamefont {Kuniba}\ \emph {et~al.}(2011)\citenamefont {Kuniba},
  \citenamefont {Nakanishi},\ and\ \citenamefont {Suzuki}}]{kuniba2011t}%
  \BibitemOpen
  \bibfield  {author} {\bibinfo {author} {\bibfnamefont {A.}~\bibnamefont
  {Kuniba}}, \bibinfo {author} {\bibfnamefont {T.}~\bibnamefont {Nakanishi}}, \
  and\ \bibinfo {author} {\bibfnamefont {J.}~\bibnamefont {Suzuki}},\ }\href
  {\doibase 10.1088/1742-5468/aa5d1e} {\bibfield  {journal} {\bibinfo
  {journal} {J. Phys. A: Math. Theor.}\ }\textbf {\bibinfo {volume} {44}},\
  \bibinfo {pages} {103001} (\bibinfo {year} {2011})}\BibitemShut {NoStop}%
\bibitem [{\citenamefont {Kuniba}\ \emph {et~al.}(1998)\citenamefont {Kuniba},
  \citenamefont {Sakai},\ and\ \citenamefont
  {Suzuki}}]{Kuniba_Sakai_Suzuki_1998}%
  \BibitemOpen
  \bibfield  {author} {\bibinfo {author} {\bibfnamefont {A.}~\bibnamefont
  {Kuniba}}, \bibinfo {author} {\bibfnamefont {K.}~\bibnamefont {Sakai}}, \
  and\ \bibinfo {author} {\bibfnamefont {J.}~\bibnamefont {Suzuki}},\ }\href
  {\doibase 10.1016/S0550-3213(98)00300-9} {\bibfield  {journal} {\bibinfo
  {journal} {Nucl. Phys. B}\ }\textbf {\bibinfo {volume} {525}},\ \bibinfo
  {pages} {597} (\bibinfo {year} {1998})}\BibitemShut {NoStop}%
\bibitem [{\citenamefont {Murgan}\ \emph {et~al.}(2006)\citenamefont {Murgan},
  \citenamefont {Nepomechie},\ and\ \citenamefont
  {Shi}}]{Murgan_Nepomechie_Shi_2006}%
  \BibitemOpen
  \bibfield  {author} {\bibinfo {author} {\bibfnamefont {R.}~\bibnamefont
  {Murgan}}, \bibinfo {author} {\bibfnamefont {R.~I.}\ \bibnamefont
  {Nepomechie}}, \ and\ \bibinfo {author} {\bibfnamefont {C.}~\bibnamefont
  {Shi}},\ }\href {\doibase 10.1088/1742-5468/2006/08/P08006} {\bibfield
  {journal} {\bibinfo  {journal} {J. Stat. Mech.}\ }\textbf {\bibinfo {volume}
  {2006}},\ \bibinfo {pages} {P08006} (\bibinfo {year} {2006})}\BibitemShut
  {NoStop}%
\bibitem [{\citenamefont {Borsi}\ \emph {et~al.}(2021)\citenamefont {Borsi},
  \citenamefont {Pozsgay},\ and\ \citenamefont
  {Pristy{\'{a}}k}}]{borsi2021current}%
  \BibitemOpen
  \bibfield  {author} {\bibinfo {author} {\bibfnamefont {M.}~\bibnamefont
  {Borsi}}, \bibinfo {author} {\bibfnamefont {B.}~\bibnamefont {Pozsgay}}, \
  and\ \bibinfo {author} {\bibfnamefont {L.}~\bibnamefont {Pristy{\'{a}}k}},\
  }\href {\doibase 10.1088/1742-5468/ac0f6b} {\bibfield  {journal} {\bibinfo
  {journal} {J. Stat. Mech.}\ }\textbf {\bibinfo {volume} {2021}},\ \bibinfo
  {pages} {094001} (\bibinfo {year} {2021})}\BibitemShut {NoStop}%
\end{thebibliography}%
	\let\addcontentsline\oldaddcontentsline

	\onecolumngrid
	\newpage

	\appendix
	\setcounter{equation}{0}
	\setcounter{figure}{0}
	\renewcommand{\thetable}{S\arabic{table}}
	\renewcommand{\theequation}{S\thesection.\arabic{equation}}
	\renewcommand{\thefigure}{S\arabic{figure}}
	\setcounter{secnumdepth}{2}

	\begin{center}
		{\Large Supplementary Material \\ 
			\titleinfo
		}
	\end{center}
	\tableofcontents

\section{Conserved charges and the Bethe ansatz}
\label{sec:appendixbetheansatz}

\begin{figure}
	\centering
		\includegraphics[scale=0.8]{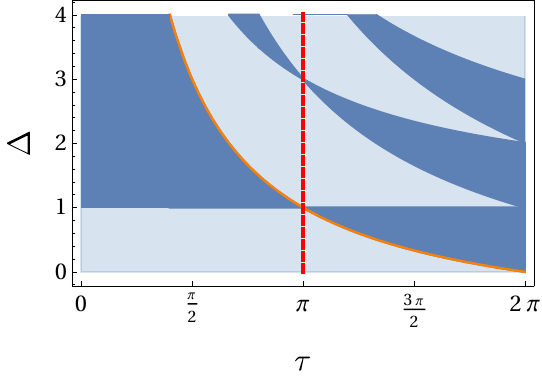}
		\caption{Phase diagram of the integrable Trotterized dynamics as a function of $\tau,\Delta$. The dark and light regions correspond to the gapped ($|\Delta'|>1$) and gapless ($|\Delta'|\leq 1$) regimes respectively, and we have indicated in red the dual-unitary line, corresponding to $\tau=\pi$~\cite{bertini2019exact}. The orange line indicates the Trotter transition \eqref{eq:transition}, and further transitions are visible when $\tau$ keeps increasing for a given value of $\Delta$.
        }
		\label{fig:phasediagram}
	\end{figure}

Integrability of the Trotterized dynamics considered in the main text follows from the fact that $U(\tau)$ can be written as a product of integrable transfer matrices analogous to those generating the XXZ Heisenberg chain, albeit with spatial inhomogeneities depending on $\tau$ \cite{vanicat2018integrable,Destri_de_Vega_1995}.  
We introduce the row-to-row transfer matrices
\begin{equation}
T(u) =  \mathrm{tr}_a \left[ R_{a, L}(u - x/2) R_{a, L-1}(u + x/2)
\ldots
R_{a, 2}(u -  x/2) R_{a, 1}(u + x/2) \right] \,,
\label{eq:rowtorowTM}
\end{equation}
where $a$ refers to a two-dimensional auxilliary space, and $R_{a,i}(u)$ is a matrix acting non-trivially on $a$ and the $i$th spin with matrix elements $R(u)_{\epsilon,\epsilon}^{\epsilon,\epsilon} = \frac{\sin(u+\gamma)}{\sin \gamma}$, $R(u)_{\epsilon,1-\epsilon}^{1-\epsilon,\epsilon} = 1$, $R(u)_{\epsilon,1-\epsilon}^{\epsilon,1-\epsilon} = \frac{\sin(u)}{\sin \gamma}$ for all $\epsilon\in \{0,1\}$, and as the identity on the rest of the chain.
As a feature of integrability, transfer matrices for different values of $u$ and fixed $x$ form a commuting family.
Choosing the parameters $\gamma,x$ as in the main text, see eqs. \eqref{eq:def_x}, the correspondence with our Trotterized dynamics is
\begin{equation}
U(\tau)=T(x/2) T(-\gamma-x/2)\,,
\end{equation}  
as ensured by the fact that $R(x)_{a,b}^{c,d} = \frac{\sin (\gamma+ x )}{\sin \gamma}V_{a,b}^{d,c}$, where $V$ is the two-site gate of eq. \eqref{eq:twositegate}. Hence $U(\tau)$ commutes with the transfer matrices $T(u)$, and therefore with an extensive number of local charges $Q_n^\pm = \left. \frac{\mathrm{d}^n}{\mathrm{d}u^n} \log T(u) \right|_{u=\pm x/2}$, as well as with quasilocal charges built from higher-spin families of transfer matrices~\cite{vanicat2018integrable,ljubotina2019ballistic,medenjak2020rigorous}. 
The first local charges take the form
\begin{eqnarray}
Q_1^{\pm} &=& \sum_{j=1}^L \frac{\sin\gamma}{\cos(2x)-\cos(2\gamma)} \left(  \cos x ( \sigma_j^x \sigma_{j+1}^x + \sigma_j^y \sigma_{j+1}^y )  + \cos\gamma ( \sigma_j^z \sigma_{j+1}^z -1)  \right) 
+ \sum_{j~{\rm odd/even}} q_{j,j+1,j+2} \,,
\end{eqnarray} 
where 
\begin{eqnarray}
q_{j,j+1,j+2} &=& 
 -\frac{\cot\gamma  (\sin x)^2}{\cos(2x)-\cos(2\gamma)}
 \left(  \sigma_j^x \sigma_{j+2}^x  +  \sigma_j^y \sigma_{j+2}^y +  \sigma_j^z \sigma_{j+2}^z  \right)  
 +
\frac{i \sin x}{\cos(2x)-\cos(2\gamma)}
 \left[ \cos x \sigma_{j+1}^z  ( \sigma_j^x \sigma_{j+2}^y- \sigma_j^y \sigma_{j+2}^x)  \right.
\nonumber 
\\
& & \left.  + \cos\gamma \left(
\sigma_{j}^z  ( \sigma_{j+1}^x \sigma_{j+2}^y- \sigma_{j+1}^y \sigma_{j+2}^x)  
+ 
( \sigma_{j}^x \sigma_{j+1}^y- \sigma_{j}^y \sigma_{j+1}^x) \sigma_{j+2}^z
  \right)      \right]\,.
\end{eqnarray}
In the translationally-invariant case $x=0$, $Q_1^\pm$ coincide with the XXZ Hamiltonian \eqref{eq:xxz_hamiltonian}, with a value of $\Delta$ changed to $\Delta'=\cos\gamma$. 
For $x\neq 0$ the operators $Q_1^\pm$ are different, but nevertheless share some common features with the XXZ Hamiltonian, in particular they are gapless for $|\Delta'|\leq 1$ and gapped for $|\Delta'|>1$.
The phase diagram as a function of $(\tau,\Delta)$ is displayed on Fig. \ref{fig:phasediagram}. We see that the phase diagram has a complicated structure with multiple phases. In the figure, we only display it up to $\tau=2\pi$, but other points can be easily determined by inverting the relations~\eqref{eq:bethe_eq} (we recall that the gapped and gapless phases correspond to real and purely imaginary values of $x$, respectively). As anticipated in the main text, we see that, as $\Delta$ is fixed and $\tau$ is increased, multiple phase transitions appear. In this work, we have focused on the first one, but subsequent transitions can be analyzed in a similar way.

The simultaneous diagonalization of $U(\tau)$ and the matrices $T(u)$ can be performed using Bethe ansatz, see e.g.~\cite{aleiner2021bethe}. As reviewed in the main text, the eigenvectors are parametrized by a set of quasimomenta satisfying the quantization conditions \eqref{eq:bethe_eq}. 
In the thermodynamic limit we assume, as is commonly done in the homogeneous case, that these organize in regular patterns called strings \cite{takahashi2005thermodynamics}. A string of type $n$ is typically formed of $n$ quasimomenta and is fully characterized by the position of its center, which is a real variable in the gapless case and restricted to $[-\pi,\pi]$ in the gapped case. A given macrostate can then be described by a smooth density of string centers $\rho_n(\lambda)$ for each string species, as well as a density of holes $\rho_n^h(\lambda)$.

\subsection{Gapped case: $\gamma= i \eta\,, \eta \in \mathbb{R}$}

In the gapped regime, the string length $n$ can take any possible integer value and the quantization equations \eqref{eq:bethe_eq} results in the following set of coupled integral equations for the functions $\rho_n(\lambda)$, $\rho_n^h(\lambda)$ 	\begin{equation}
		\rho_n(\lambda)+\rho_n^h(\lambda)=a^{(x/2)}_n(\lambda)-\sum_{m=1}^{\infty}\left(a_{n m} \ast \rho_m\right)(\lambda)\,,
		\label{eq:thermo_bethe_SM}
	\end{equation}
where $(f\ast g)(\lambda):=\int_{-\pi/2}^{\pi/2}d\mu f(\mu-\lambda)g(\mu)$, while we introduced the notation $f^{(x)}(\lambda)=(f(\lambda+x)+f(\lambda-x))/2$, and defined
	\begin{align}
		a_{nm}(\lambda)=&(1-\delta_{nm})a_{|n-m|}(\lambda)+2a_{|n-m|+2}(\lambda)\nonumber\\
		+&\ldots +2a_{n+m-2}(\lambda)+a_{n+m}(\lambda)\,.
	\end{align}
	with  $a_n(\lambda)=\pi^{-1}\sinh\left( n\eta\right)/[\cosh (n \eta) - \cos( 2 \lambda)]$ 
For $x=0$, these recover the usual result for the homogeneous XXZ chain \cite{takahashi2005thermodynamics}. 
Introducing the ratios $\eta_n(\lambda) = \rho_n^h(\lambda)/\rho_n(\lambda)$, eq. \eqref{eq:thermo_bethe} of main text is recovered.

\subsection{Gapless case $\gamma \in \mathbb{R}$}
\label{sec:gaplessTBA}

In the gapless regime, the number and types of allowed strings depends on a complicated manner on $\gamma$, due to the presence of a dense set of ``root-of-unity points'' $\gamma / \pi \in \mathbb{Q}$. In the following we shall restrict to the simplest of such points, namely  
\be
\frac{\gamma}{\pi}= \frac{1}{\nu_1+1/\nu_2}, \qquad \nu_1,\nu_2\geq 1\,.
\label{eq:special_points}
\ee
There are then $N_b=\nu_1+\nu_2$ different types of strings, which we denote by an integer $j\in \{1,\ldots N_b\}$. Each type is characterized by a number of quasimomenta $n_j$ and a parity $\upsilon_j =\pm$, which we list as follows \cite{takahashi2005thermodynamics} 
\begin{align}
(&1,+), (2,+) \ldots (\nu_1 -1 , +) , (1,-), \nonumber \\
(&1+\nu_1,+), (1+2\nu_1, -), \ldots (1+(\nu_2-1) \nu_1, (-1)^{\nu_2-1}) , (\nu_1, (-1)^{\nu_2}) \nonumber  \,.
\end{align}
Following \cite{takahashi2005thermodynamics}, we also introduce the numbers $\{p_0,p_1,p_2\}=\{\nu_1+1/\nu_2,1,1/\nu_2\}$, $\{m_0,m_1,m_2\}=\{0,\nu_1,\nu_1+\nu_2\}$ as well as, for each string type, $q_j \equiv (-1)^i (p_i - (j-m_i)p_{i+1})$ for $m_i \leq j < m_{i+1}$.
The thermodynamic limit of the Bethe equations can be written as the following system of equations, which are once again a straightforward generalization of the homogeneous case \cite{takahashi2005thermodynamics} 
\begin{equation}
\mathrm{sign}(q_m) \left[\rho_m(\lambda)+\rho_m^h(\lambda)\right] = a_m^{(i x/2)}(\lambda)-\sum_{n=1}^{N_b} \left(a_{mn}  \ast \rho_m \right) (\lambda) \,,
\label{eq:tbagapless}
\end{equation}
where we have again made used of the notation $f^{(x)}(\lambda)=(f(\lambda+x)+f(\lambda-x))/2$, and where the convolution is now defined as $(f\ast g)(\lambda):=\int_{-\infty}^{\infty}d\mu f(\mu-\lambda)g(\mu)$. The various kernels involved in eq. \eqref{eq:tbagapless} are
\begin{align}
a_j  (\lambda) &=  \frac{\upsilon_j}{\pi} \frac{\sin(\gamma n_j)}{\cosh(2\lambda) - \upsilon_j \cos(\gamma n_j)}\equiv a^{\upsilon_j}_{n_j}  (\lambda)  \\
a_{jk}(\lambda)  &=(1-\delta_{n_j n_k}) a_{|n_j-n_k|}^{\upsilon_j \upsilon_k}(\lambda) + 2a_{|n_j-n_k|+2}^{\upsilon_j \upsilon_k}(\lambda) + ... + 
2a_{n_j+n_k-2}^{\upsilon_j \upsilon_k} (\lambda) + a_{n_j+n_k}^{\upsilon_j \upsilon_k} (\lambda)\,.
\end{align}

\section{Derivation of the dGGE rapidity distribution functions}

In this section we describe the derivation of the dGGE rapidity distribution functions for the quasiparticles and holes $\rho_n(\lambda)$, $\rho_n^h(\lambda)$. Our derivation is highly technical, and relies on a sophisticated mathematical theory developed in previous work. We follow in particular the Boundary Quantum Transfer Matrix (BQTM) approach explained in Ref.~\cite{piroli2017integrable} for the case of a continuous Hamiltonian evolution, see also~\cite{Pozsgay_2013,Piroli_Pozsgay_Vernier_2017,Piroli_Pozsgay_Vernier_2018,piroli2019integrable_su3_1,piroli2019integrable_su3_2}. There, the generator of the continuous time evolution $e^{i t H}$ was written as the Trotter limit of a family of mutually commuting transfer matrices. The functions $\eta_n(\lambda) = \frac{\rho_n^h(\lambda) }{ \rho_h(\lambda)}$ were shown to correspond to a set of auxilliary ``Y-functions'', derived from the leading eigenvalue of a BQTM generating a rotated space-time evolution. More precisely, the functions $\eta_n$ were obtained as the limit $t \to 0$ of the Y-functions. In this limit, $e^{i t H}$ can be written as a resolution of identity in terms of the Hamiltonian eigenstates. In the following, we sketch how these computations can be generalized to the present setting, but we refer the reader to Refs.~\cite{piroli2017integrable,Piroli_Pozsgay_Vernier_2017} for a thorough explanation of the method and the underlying mathematical structures, which our discussion heavily relies on.

\subsection{The Boundary Quantum Transfer Matrix construction}

\begin{figure}
	\centering
		\includegraphics[scale=0.55]{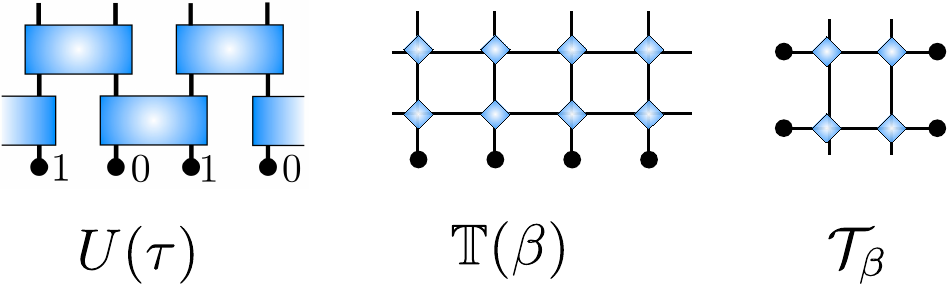}
		\caption{Left: the generator $U(\tau)$ of the discrete time evolution. Center the family of transfer matrices $\mathbb{T}(\beta)$ commuting with $U(\tau)$. Right: the Boundary Quantum Transfer Matrix acting on two auxilliary spins in the transverse channel.}
		\label{fig:QTM}
	\end{figure}

Following~\cite{piroli2017integrable,Piroli_Pozsgay_Vernier_2017}, we introduce a resolution of the identity as a limit of operators commuting with the time evolution $U(\tau)$. From the expression of $U(\tau)$ as a product of integrable row-to-row transfer matrices described in Appendix \ref{sec:appendixbetheansatz}, we see that the family of double row transfer matrices 
\begin{equation}
\mathbb{T}(\beta) =  \left( \frac{\sin^2 \gamma}{\sin(\gamma+x)\sin(\gamma-x)} \right)^{L/2}  T(x/2 + \beta) T(-\gamma- x/2-\beta)
\end{equation} 
commute with one another for different values of $\beta$ as well as with $U(\tau)$, and furthermore can be checked to obey $\mathbb{T}(0)=1$. 
Following the argument of \cite{Piroli_Pozsgay_Vernier_2017}, the dGGE rapidity distribution functions can be derived from the $L\to\infty$ limit of the partition function $\langle \Psi_0 | \mathbb{T}(\beta)|  \Psi_0 \rangle$, which can be recast as  
\begin{equation}
\langle \Psi_0 | \mathbb{T}(\beta)|  \Psi_0 \rangle 
= 
\mathrm{Tr}((\mathcal{T}_{\beta})^{L/2}) \,,
\label{eq:QTMdef}
\end{equation}
where $\mathcal{T}_{\beta}$ is a Quantum Transfer Matrix~\cite{klumper_integrability_2004} acting on the transverse direction, see Figure \ref{fig:QTM}. In contrast with the continuous Hamiltonian evolution, where a Trotter number $N$ was introduced and the QTM acted on $2N$ copies of the auxilliary space, the QTM here acts on only two sites.     
As in the continuous case, it is part of a family of mutually commuting open transfer matrices $\mathbb{T}_{\beta}(u)$ similar to those of \cite{Piroli_Pozsgay_Vernier_2017} after having replaced the Trotter parameter $N$ by $1$, and taken the inhomogeneity and boundary parameters to be
\begin{align}
\xi_1=\beta + \frac{x}{2} + \gamma,\qquad
\xi_2= \beta + \frac{x}{2}, \qquad
\xi_\pm = \frac{x}{2} \pm \frac{\gamma}{2}.
\end{align}

In the $L\to \infty$ limit, the partition function \eqref{eq:QTMdef} reduces to the computation of the single leading eigenvalue $\Lambda_{\beta}(u)$ of the transfer matrix $\mathcal{T}_{\beta}$, which can be computed using Bethe ansatz. Following once again~\cite{Piroli_Pozsgay_Vernier_2017}, we find
\begin{equation}
\Lambda_\beta(u) = \omega_1(u) \phi(u+\gamma/2) \frac{Q(u-\gamma)}{Q(u)} 
+\omega_2(u) \phi(u-\gamma/2) \frac{Q(u+\gamma)}{Q(u)} \,, 
\label{eq:TQ}
\end{equation}
where 
\begin{eqnarray}
Q(u) &=&  \sin(u -  (\beta + x/2)) \sin(u +  (\beta + x/2)) , \\
\phi(u) & = & \prod_{k=1}^{2}\sin\left(u-\gamma/2+\xi_k\right)\sin\left(u+\gamma/2-\xi_k\right)\,,\label{phi_function}\\
\omega_1(u)&=&\frac{\sin(2u+\gamma)\sin(u+\xi^+-\gamma/2)\sin(u+\xi^--\gamma/2)}{\sin(2u)}\,,\label{omega1_function}\\
\omega_2(u)&=&\frac{\sin(2u-\gamma)\sin(u-\xi^++\gamma/2)\sin(u-\xi^-+\gamma/2)}{\sin(2u)}\,.\label{omega2_function}
\end{eqnarray}

\subsection{The T--system}

The transfer matrices $\mathcal{T}_\beta(u)$ are part of a hierarchy of commuting transfer matrices $\mathcal{T}_i(u)$, satisfying the recursive relations~\cite{Piroli_Pozsgay_Vernier_2017} 
\begin{eqnarray}
\mathcal{T}_0(u) &=& 1, \nonumber\\
\mathcal{T}_1(u) &=& \mathcal{T}_\beta(u),  \nonumber\\
\mathcal{T}_j(u) &=& \mathcal{T}_{j-1}\left(u-\frac{\gamma}{2}\right)\mathcal{T}_{1}\left(u+(j-1)\frac{\gamma}{2}\right)  - f\left(u + (j-3)\frac{\gamma}{2}\right) \mathcal{T}_{j-2}(u-\gamma) \,, \qquad j\geq 2 ,
\end{eqnarray}
where $f(u - \gamma/2)=  \phi(u+\gamma) \phi(u-\gamma) \omega_1 (u+\gamma/2) \omega_2 (u-\gamma/2)$.
Since those are all mutually commuting, we use the same notations for the corresponding eigenvalues associated to the leading QTM eigenvector, which can be shown by induction starting from \eqref{eq:TQ} to have the form
\begin{equation}
\mathcal{T}_j(u) = \left(\prod_{l=1}^{j-1} \phi^{[2l-j]}\right)  \sum_{k=1}^{j+1} \phi^{[2k-j-2]}  \left( \prod_{l=1}^{k-1} \omega_1^{[2l-j-1]}  \prod_{l=k}^{j} \omega_2^{[2l-j-1]}   \right)  \frac{Q^{[j+1]} Q^{[-j-1]}}{Q^{[2k-j-3]} Q^{[2k-j-1]}}  \,, 
\end{equation}
where we have introduced the short-hand notation $f^{[k]}(u) \equiv f(u+i k \gamma/2)$. The transfer matrices $\mathcal{T}_j$, and hence their eigenvalues, satisfy a set of relations known as the T--system~\cite{kuniba2011t,Piroli_Pozsgay_Vernier_2017}, and which are best expressed in terms of the rescaled $t_0\equiv \phi$, $t_j = \frac{\mathcal{T}_j}{\prod_{l=1}^{j-1} \phi^{[2l-j]}}$ as :
\begin{eqnarray}  
t_j^{[m]}t_j^{[-m]} 
&=&
t_{j+m} t_{j-m} +  \Psi_{j-m+1}    t_{m-1}^{[j+1]} t_{m-1}^{[-j-1]} ,
\label{tsystemrescaled}
\end{eqnarray}
where we have introduced the function $\Psi_{j} = \prod_{l=1}^{j}  \omega_1^{[2k-j]}\omega_2^{[2k-j-2]}$.

The dGGE functions $\eta_m(\lambda)$ are then obtained from a set of ``Y--functions'' constructed from the T--system above, the precise definition of which depends on the value of $\gamma$ under consideration, in particular whether it belongs to the gapped ($\gamma \in i \mathbb{R}$) or gapless phase ($\gamma \in \mathbb{R}$). We will therefore treat these two cases separately.

\subsection{Gapped case: $\gamma = i\eta$, $\eta \in\mathbb{R}$}

For $\gamma = i\eta$, $\eta \in \mathbb{R}$, one introduces the family of Y--functions $Y_0 = 0$, and 
\begin{equation}
Y_j(\lambda)
=\frac{t_{j-1}(u)t_{j+1}(u)}{\Psi_{j}(u) \phi^{[j+1]}(u)\phi^{[-j-1]}(u)} \,, \qquad j \geq 1
\,,
\label{y_function}
\end{equation} 
The Y--functions obey the following set of relations, known as Y--system~\cite{kuniba2011t}
\begin{equation}
Y_j\left(u+\frac{\gamma}{2}\right)Y_j\left(u-\frac{\gamma}{2}\right)=\left[1+Y_{j+1}\left(u\right)\right]\left[1+Y_{j-1}\left(u\right)\right]\,,
\label{y_system}
\end{equation}
and can therefore be obtained recursively from the knowledge of $Y_1(\lambda)$, which is : 
\begin{equation} 
1+Y_1(u) = \frac{\Lambda_\beta(u+\gamma/2)\Lambda_\beta(u-\gamma/2)}{f(u-\gamma/2)} = \left(1 + \mathfrak{a}_\beta(u-\gamma/2) \right)
\left(1 + 1/\mathfrak{a}_\beta(u+\gamma/2) \right) \,,
\label{y1exact}
\end{equation} 
where the function $\mathfrak{a}_\beta$ is defined as 
$  \mathfrak{a}_\beta(u) = \frac{\omega_1(u)}{\omega_2(u)} \frac{\phi(u+\gamma/2)}{\phi(u-\gamma/2)} \frac{Q(u-\gamma)}{Q(u+\gamma)} \,.
$
We can now state the correspondence with the functions $\eta_j(\lambda)$: they are obtained as 
\begin{equation}
    \eta_m(\lambda) = \lim_{\beta \to 0} Y_m(i \lambda) \,,
\end{equation}
which can be seen by turning the Y--system \eqref{y_system} into a set of non-linear integral equations, and comparing those with the generic form of the thermodynamic Bethe ansatz equations obeyed by the functions $\eta_m$ \cite{takahashi2005thermodynamics}. 
In conclusion, the functions $\eta_m$ can be computed recursively from the Y-system, where for $\eta_1(\lambda)$ we use : 
\begin{equation}  
\lim_{\beta \to 0}  \mathfrak{a}_\beta(u) =  \frac{\sin(2u+\gamma)}{\sin(2u-\gamma)}
\frac{\sin(u-x/2-\gamma)}{\sin(u+x/2+\gamma)}
\frac{\sin(u-x/2)}{\sin(u+x/2)} \,.
\label{afrak}
\end{equation}   
This recovers Eqs.~\eqref{eq:eta1gapped},\eqref{eq:afrakgapped} of the main text.

\subsection{Gapless case: $\gamma \in \mathbb{R}$}

The gapless phase is densely spanned by the so-called root-of-unity points $\gamma/\pi \in \mathbb{Q}$ which have highly peculiar and $\gamma$-dependent features. 
At such points, the T-- and Y--systems truncate to a finite number of equations and one needs to study only a finite set of functions $\eta_m(\lambda)$, in accordance with the finite number of strings types introduced in Appendix \ref{sec:appendixbetheansatz}.

Restricting once again to the points \eqref{eq:special_points}, the truncation of the T--system takes the form of a linear relation between the functions $t_{\nu_1(\nu_2+1)}$, $t_{\nu_1(\nu_2-1)}$ and $t_{\nu_1-1}^{[\nu_1\nu_2+1]}$, namely
\begin{eqnarray} 
t_{\nu_1(\nu_2+1)}
 &=& 
\left(  \prod_{l=1}^{\nu_1} \omega_1^{[2l-\nu_1(\nu_2+1)-1]}\omega_2^{[2l-\nu_1(\nu_2+1)-3]} 
 \right)
  t_{\nu_1(\nu_2-1)}
\nonumber \\ 
& & + 
(-1)^{m \nu_2}
\left( 
\prod_{l=1}^{1+\nu_1 \nu_2} \omega_1^{[2l-\nu_1(\nu_2+1)-1]}
+
\prod_{l=1}^{1+\nu_1 \nu_2}  \omega_2^{[2l-\nu_1(\nu_2+1)-3]}
\right)
t_{\nu_1-1}^{[\nu_1\nu_2+1]}
\label{tsystemtruncationalpha2}
\end{eqnarray}

From there, we introduce the following Y--functions 
\begin{eqnarray}
Y_j &=&  \begin{cases}
=
\frac{t_{{j+1}} t_{j-1} }{\Psi_{j} t_{0}^{[j+1]} t_{0}^{[-j-1]} }    \,, \qquad {1\leq j \leq \nu_1-1} 
  \\ 
 \frac{t_{\nu_1(j+2-\nu_1)}^{[w_j p_0]} t_{\nu_1(j-\nu_1)}^{[w_j p_0]} }{\Psi_{\nu_1(j-\nu_1)+1}^{[w_j p_0]}  t_{\nu_1-1}^{[\nu_1(j+1-\nu_1)+1+w_j p_0]} t_{\nu_1-1}^{[-\nu_1(j+1-\nu_1)-1+w_j p_0]} }   
     \,, \quad {\nu_1 \leq j \leq \nu_1+\nu_2-1} 
\end{cases} 
\nonumber
\\
K&=&   \frac{t_{\nu_1 (\nu_2-1)}^{[(\nu_2-2)p_0]}}{t_{\nu_1-1}^{[1+\nu_1\nu_2+(\nu_2-2)p_0]}}
  \frac{\Psi_{\nu_1}^{[-(1+\nu_1 \nu_2)+(\nu_2-2)p_0]}}
{ \Omega^{[1+(\nu_2-2)p_0]}  \Omega_1^{[1+(\nu_2-2)p_0]}  \Omega_2^{[1+(\nu_2-2)p_0]}   } \,, 
\end{eqnarray}
where we have used the following definitions
\begin{eqnarray} \Omega(u)  &=& \frac{\sin((1+\nu_1 \nu_2)(u+\gamma/2))\sin((1+\nu_1 \nu_2)(u+\gamma/2+\pi/2))}{\sin((1+\nu_1 \nu_2)(u))\sin((1+\nu_1 \nu_2)(u+\pi/2))}  \,, 
\\ 
\Omega_{1,2}(u) &=& 2^{-\nu_1 \nu_2}  \sin\left((1+\nu_1 \nu_2) (u \pm  \frac{x}{2}) \right)
\end{eqnarray}
Using the T--system relation \eqref{tsystemrescaled} and the truncation identity \eqref{tsystemtruncationalpha2}, we check that these functions obey the following truncated Y--system  
\begin{eqnarray} 
Y_j^{[p_1]}Y_j^{[-p_1]} &=& (1+Y_{j+1})(1+Y_{j-1})^{1-2 \delta_{j,m_0}} \,,\qquad 1 \leq j \leq \nu_1 -2 
\label{Ysystemnu1nu2beginning}
\\
Y_{\nu_1-1}^{[p_1+p_2]}Y_{\nu_1-1}^{[-p_1-p_2]}Y_{\nu_1-1}^{[p_1-p_2]}Y_{\nu_1-1}^{[-p_1+p_2]} &=& \left((1+Y_{\nu_1-2}^{[p_2]})(1+Y_{\nu_1-2}^{[-p_2]})\right)^{1-2 \delta_{\nu_1,1}}
\nonumber \\ 
&\times &   (1+Y_{\nu_1}^{[p_1]})(1+Y_{\nu_1}^{[-p_1]})  (1+Y_{\nu_1-1}^{[p_1-p_2]})(1+Y_{\nu_1-1}^{[-p_1+p_2]})   \\
Y_j^{[p_2]}Y_j^{[-p_2]} &=& (1+Y_{j+1})(1+Y_{j-1})^{1-2 \delta_{j,m_1}} \,,\qquad \nu_1 \leq j \leq \nu_1+\nu_2 -2 \\
 1+ Y_{\nu_1 + \nu_2 -1}   &=& 1 + (\mathfrak{b} + \mathfrak{b}^{-1}) K + K^2 
\\
K^{[p_2]}K^{[-p_2]} &=& 1+ Y_{\nu_1 + \nu_2 -2}  \,,
\label{Ysystemnu1nu2}
\end{eqnarray} 
where 
\begin{equation}  
\mathfrak{b}(u) =
(-1)^{\nu_2}  \frac{\sin\left((1+\nu_1 \nu_2) \left( u+ \frac{x}{2}   \right) + \nu_1 \nu_2\frac{\pi}{2} \right)  }{\sin\left((1+\nu_1 \nu_2) \left( u- \frac{x}{2}   \right) + \nu_1 \nu_2\frac{\pi}{2} \right) }  \,. 
\label{betanu1nu2}
\end{equation} 
While the truncation of the T-- and Y--systems was derived for arbitrary roots of unity in the case of periodic transfer matrices in \cite{Kuniba_Sakai_Suzuki_1998}, and for open transfer matrices in the case of principal roots of unity in \cite{Murgan_Nepomechie_Shi_2006} (see also \cite{Piroli_Pozsgay_Vernier_2018}), the above result is new, to the best of our knowledge.

As in the gapped regime, and following the argument of Ref.~\cite{Piroli_Pozsgay_Vernier_2017}, we can read off $\eta_m$ from the Y-functions, leading to the identification: 
\begin{eqnarray} 
\eta_j(\lambda) &\equiv &  \lim_{\beta \to 0} Y_j(i\lambda)  \,, \qquad j=1, \ldots \nu_1+\nu_2 -2
\\
\eta_{\nu_1 + \nu_2 - 1}(\lambda) &\equiv & \lim_{\beta \to 0} \mathfrak{b}(i\lambda) K(i\lambda)  \,, \\
\eta_{\nu_1 + \nu_2 }(\lambda) &\equiv &  
\lim_{\beta \to 0} \mathfrak{b}(i \lambda) / K(i \lambda)  \,.
\end{eqnarray} 
Therefore, we finally obtain an analytic description for $\eta_m$ from the Y--system.

\section{Calculation of the late-time staggered magnetization}

In this section we describe how to compute the expectation value of the staggered magnetization on a macrostate characterized by the functions $\rho_n(\lambda)$, $\rho_n^h(\lambda)$. The starting point are the general formulae~\cite{kitanine1999form,kitanine2000correlation} for the form factor of the local magnetization $\sigma^z_m$ between two (unnormalized) eigenstates of the transfer matrices $T(u)$ with arbitrary inhomogeneities. 
We proceed in two steps. Fist we specialise the formulae to the case of interest, namely staggered inhomogeneities as in the transfer matrix \eqref{eq:rowtorowTM}, and simplify them. Then we take the thermodynamic limit. 

\subsection{Staggered magnetization in finite volume}

Specializing it to the case of interest here, namely the transfer matrix \eqref{eq:rowtorowTM}, the un-normalized form factor of the spin-operator reads as 
\begin{eqnarray}
\langle \{ p' \} | \sigma_m^z| \{ p \} \rangle 
&=&
\prod_{k=1}^{n}  \prod_{j=1}^{m-1} \frac{\sinh(p_k- i \frac{\gamma}{2} - (-1)^j \frac{i x}{2})}{\sinh(p_k+ i \frac{\gamma}{2} - (-1)^j \frac{i x}{2})}\frac{\sinh(p'_k+i \frac{\gamma}{2} - (-1)^j \frac{i x}{2})}{\sinh(p'_k-i \frac{\gamma}{2} - (-1)^j \frac{i x}{2} )}  \prod_{j=1}^n \frac{\sinh( p'_j -  i \frac{\gamma}{2} - (-1)^m \frac{i x}{2})}{\sinh(p_j +  i \frac{\gamma}{2} - (-1)^j \frac{i x}{2}))}
\nonumber \\
& & \times
\frac{\det_n(H(\{p'\},\{p\})-2 P'_m(\{p'\},\{p\}))}{\prod_{j>k} \sinh(p'_k-p'_j)\prod_{\alpha<\beta} \sinh(p_\beta-p_\alpha)}
\label{MailletFormula}
\end{eqnarray}
where
\begin{eqnarray}
(H(\{p'\},\{p\}))_{ab} &=& \frac{\sinh i \gamma}{\sinh(p'_a - p_b)} \left(\prod_{j \neq a}^n \sinh(p'_j - p_b + i \gamma) -    \left(
\frac{f_x^-(p_b)}{f_x^+(p_b)}
 \right)^{L\over 2} \prod_{j \neq a}^n \sinh(p'_j - p_b - i \gamma)  \right)
\label{Hab}
\\ 
(P'_m(\{p'\},\{p\})_{ab} &=& \sinh(i\gamma) \frac{\prod_{k=1}^n \sinh(p_k-p_b + i \gamma)}{\sinh(p'_a- (-1)^m \frac{i x}{2} - i\frac{\gamma}{2})\sinh(p'_a- (-1)^m \frac{i x}{2} + i\frac{\gamma}{2})}
\label{Pab} \,,
\end{eqnarray}
and where we recall the notation $f_x^{\pm}(p)=\sinh (p+ \frac{i x}{2}\pm i \frac{\gamma}{2})\sinh (p- \frac{i x}{2}\pm i \frac{\gamma}{2})$ introduced in the main text.
The norm of the eigenstates $\ket{\{p\}}$, $\ket{\{p'\}}$ is in turn given by the Gaudin formula \cite{kitanine2000correlation}
\begin{equation}
\braket{\{p\} | \{p\}} 
= (-2\pi i \sinh(i\gamma))^n \prod_{a \neq b}  \frac{\sinh(p_a - p_b+ i \gamma)}{\sinh(p_a - p_b)} \det_n G \,,
\end{equation}
where $G$ is the Gaudin matrix 
\begin{align}
[G]_{i, j}= & L \delta_{i,j} \left(\frac{1}{2}[a(p_i+ \tfrac{i x}{2},{\tfrac \gamma 2})+a(p_i- \tfrac{i x}{2},{\tfrac \gamma 2})]- \frac{1}{L}\sum_{k=1}^n a(p_i-p_k,\gamma) \right)+ a(p_i-p_j,\gamma), 
\label{eq:Gaudin}
\end{align}
and  we defined 
\be
a(z,\gamma)=\frac{\sin(2 \gamma)}{2\pi \sinh(z+i\gamma)\sinh(z-i\gamma)} = \frac{\sin(2\gamma)}{\pi (\cosh(2z)-\cos(2\gamma))}.
\ee
In the limit $\{p'\} \to \{p\}$, eq. \eqref{MailletFormula} simplifies to 
\begin{equation} 
\langle \{ p \} | \sigma_m^z| \{ p \} \rangle 
= 
\frac{\lim_{ \{p'\} \to \{p\} }  \det_n\left(H(\{p'\},\{p\})-2 P'_m(\{ p' \},\{ p\})\right)}{\prod_{a  \neq  b} \sinh(p_b-p_a)}
\end{equation} 
The limit of \eqref{Pab} is straightforward, while for \eqref{Hab} a little care has to be taken for the diagonal terms : the factor $\sinh(p'_a - p_a)$ in the denominator vanishes when $p'_a \to p_a$, but the numerator vanishes as well by virtue of the quantization condition \eqref{eq:bethe_eq}. We find
\begin{eqnarray} 
(P'_m)_{ab} &=& \sinh(i\gamma)  \frac{\sinh(i\gamma)\prod_{j \neq b} \sinh(p_j-p_b - i \gamma) }{\sinh(p_a- (-1)^m \frac{i x}{2} - i \frac{\gamma}{2})\sinh(p_a- (-1)^m \frac{i x}{2} + i \frac{\gamma}{2})},
\\
H_{ab} &=&-2\pi i \sinh(i\gamma) \prod_{j \neq b} \sinh(p_j-p_b - i \gamma) G_{ab} \,.
\end{eqnarray} 
We can then follow the same further steps as described in \cite{kitanine1999form} for the homogeneous case. Namely, dividing by the Gaudin formula for the scalar product, we obtained for the normalized expectation value   
\begin{equation} 
\frac{\langle \{ p \} | \sigma_m^z| \{ p \} \rangle }{\langle \{ p \} | \{ p \} \rangle }
= 
\frac{\det \left( G  + 2 P_m  \right)}{\det G} \,,
\label{eq:Mailletlimit}
\end{equation}
where 
\begin{equation} 
(P_m)_{ab} =   \frac{ i \sinh(i\gamma)}{2\pi \sinh\left(p_a - (-1)^m \frac{i x}{2} + i \frac{\gamma}{2}\right)\sinh\left(p_a - (-1)^m \frac{i x}{2} - i \frac{\gamma}{2}\right)} = - a(p_a- (-1)^m {\tfrac {i x}{2}} ,{\tfrac \gamma 2})
\end{equation}
is a rank-one matrix. Defining 
\be
[v_m]_b = - a(p_b- (-1)^m {\tfrac {i x}{2}} ,{\tfrac \gamma 2}), \qquad [w]_b = 1,
\ee
we can write $P_m$ as 
\be
P_m = v_m \cdot w^T\,.
\ee
Using now the multiplicative property of the determinant and a standard result for rank-1 perturbations to the identity matrix we have 
\be
\det[G + 2  v_m \cdot w^T ] = \det[G]\det[I + 2 G^{-1}  v_m \cdot w^T] = \det[G] (1 + 2  w^T G^{-1}  v_m)\,. 
\ee
Therefore \eqref{eq:Mailletlimit} simplifies to 
\be
\frac{\braket{\{p\}|\sigma_{2k+m}^z|\{p\}}}{\braket{\{p\}|\{p\}}}= 1 + 2  w^T G^{-1}  v_m\,.
\label{eq:simplified}
\ee
Interestingly, this equation has a form very similar to a current in finite volume (cf. Eq. (3.13) in  Ref.~\cite{borsi2021current}). 

\subsection{Thermodynamic limit}

Let us now take the thermodynamic limit of \eqref{eq:simplified}. For definiteness we consider the gapless regime $p_a=\lambda_a \in\mathbb R$ where $\gamma/\pi\in\mathbb Q$ and give the result for the gapped case in the end. 

Our first step is to rewrite \eqref{eq:simplified} using the string hypothesis (see Sec. 5 in \cite{borsi2021current})
\be
\frac{\braket{\{\lambda\}|\sigma_{2k+m}^z|\{\lambda\}}}{\braket{\{\lambda\}|\{\lambda\}}} = 1 +2 \sum_{p,q=1}^{N_s} \sum_{\alpha,\beta} w_p(\lambda_{\alpha}^{(p)}) [G_s^{-1}]_{(\alpha p),(\beta q)} v_{m,q}(\lambda_\beta^{(q)})+\text{subleading}\,.
\label{eq:currentformulaFV}
\ee
where we introduced 
\be
f_p(z) = \sum_{a=1}^{n_p} f\!\left(z+i\frac{\gamma}{2}(n_p+1-2a)+i\frac{(1-\upsilon_p)\pi}{4}\right)
\ee
while $n_p$ and $\upsilon_p$ are the parity and length of the string (cf. Sec.~\ref{sec:gaplessTBA} and Sec.~9.2.1 of~\cite{takahashi2005thermodynamics}). In particular, we have 
\be
v_{m,p}(\lambda) \equiv  - a_p^{(m)}(\lambda) =  - a_p(\lambda - (-)^m \tfrac {i x} 2),\qquad w_p(\lambda) = n_p,
\label{eq:definitions} 
\ee
where we set 
\be
a_p(\lambda) \equiv a_p(\lambda,\gamma/2) = \frac{\upsilon_p \sin(n_p \gamma)}{\pi (\cosh(2\lambda)-\upsilon_p \cos(n_p \gamma))}\,.
\ee 
Moreover, we introduced the ``reduced" Gaudin matrix 
\begin{align}
[G_s]_{(\alpha p),(\beta q)} = &L \delta_{\alpha,\beta}\delta_{p,q} \left[\frac{1}{2}[a_p(\lambda_\alpha^{(p)} + \tfrac {i x} 2,{\tfrac \gamma 2})+a_p(\lambda_\alpha^{(p)}- \tfrac {i x} 2,{\tfrac \gamma 2})]- \frac{1}{L}\sum_{k=1}^{N_s}\sum_{\nu} a_{pk}(\lambda_\alpha^{(p)}-\lambda_\nu^{(k)}) \right] \notag\\
&+ a_{pq}(\lambda_\alpha^{(p)}-\lambda_\beta^{(q)}), 
\label{eq:GaudinString}
\end{align}
and 
\be
a_{pq}(z) = \sum_{b=1}^{n_q}\sum_{a=1}^{n_p} a\!\left(z+i\frac{\gamma}{2}(n_p+1-2a)-i\frac{\gamma}{2}(n_q+1-2b)+i\frac{\pi (\upsilon_q-\upsilon_p)}{4},\gamma \right) \,.
\ee
Then we take the thermodynamic limit using (see, e.g., Eq. 5.37 in Ref.~\cite{borsi2021current})
\begin{align}
\lim_{\rm th}\sum_{p,q=1}^{N_s} \sum_{\alpha,\beta} w_p(\lambda_{\alpha}^{(p)}) [G_s^{-1}]_{(\alpha p),(\beta q)} v_{m,q}(\lambda_\beta^{(q)}) &= \sum_{p=1}^{N_s} \int {{\rm d}\lambda}\, w_p(\lambda) \vartheta_p(\lambda) \sigma_p v_{m,p}^{\rm eff}(\lambda) \notag\\
&= \sum_{p=1}^{N_s} \int {{\rm d}\lambda}\, w^{\rm eff}_p(\lambda) \sigma_p \vartheta_p(\lambda) v_{m,p}(\lambda)\,,
\end{align}
where we introduced the string parameter $\sigma_j={\rm sgn}(q_j)$ (see page 139 of \cite{takahashi2005thermodynamics}), the filling function 
\be
\vartheta_p(\lambda) = \frac{\rho_p(\lambda)}{\rho^{t}_p(\lambda)},
\ee
and the function $f_p^{\rm eff}(x)$ solving  
\be
f_p^{\rm eff}(\lambda) = f_p(\lambda) - \sum_{q=1}^{N_s} \int {{\rm d}\mu}\, a_{pq}(\lambda-\mu)\vartheta_q(\mu) \sigma_q f_q^{\rm eff}(\mu). 
\label{eq:feff}
\ee
In summary we find 
\begin{align}
\lim_{\rm th} \frac{\braket{\{\lambda\}|\sigma_{2k+m}^z|\{\lambda\}}}{\braket{\{\lambda\}|\{\lambda\}}} &= 1 -2  \sum_{p=1}^{N_s} \int {{\rm d}\lambda}\, n^{\rm eff}_p \vartheta_p(\lambda) \sigma_p a^{(m)}_{p}(\lambda)\\
 &= 1 -2  \sum_{p=1}^{N_s} \int {{\rm d}\lambda}\, n_p \vartheta_p(\lambda) \sigma_p a^{(m)\,{\rm eff}}_{p}(\lambda)\,. 
 \label{eq:finalresult}
\end{align}
Note that from \eqref{eq:feff} and \eqref{eq:tbagapless} we see 
\be
\frac{1}{2}(a^{(0)\,{\rm eff}}_{p}(\lambda)+ a^{(1)\,{\rm eff}}_{p}(\lambda))= \sigma_p \rho_{t,p}(\lambda),
\ee
which implies 
\begin{align}
\lim_{\rm th} \frac{1}{2}\left(\frac{\braket{\{\lambda\}|\sigma_{2k}^z|\{\lambda\}}}{\braket{\{\lambda\}|\{\lambda\}}}+ \frac{\braket{\{\lambda\}|\sigma_{2k+1}^z|\{\lambda\}}}{\braket{\{\lambda\}|\{\lambda\}}}\right) = 1 - 2  \sum_{p=1}^{N_s} \int {{\rm d}\lambda}\, n_p \rho_{p}(\lambda)\,.
\end{align}
This correctly reproduces the value of the magnetization on the stationary state described by $\{\rho_{p}(\lambda)\}$.

Note that for the special points \eqref{eq:special_points}, the relevant values of $n_p$, $\upsilon_p$, and $q_p$ are reported in Sec.~\ref{sec:gaplessTBA}. For convenience of the reader we also list them here 
\begin{align}
&n_{p}=
\begin{cases}
   p        &  1\leq p\leq \nu_{1} - 1 \\
   1+(p-\nu_{1})\nu_{1} & \nu_{1}\leq p \leq \nu_{1}+\nu_{2}-1 \\
   \nu_{1} & p=\nu_1 + \nu_2\,.
\end{cases}\\
&\upsilon_{p}=
\begin{cases}
   1       &  1\leq p \leq \nu_{1} - 1 \\
   -1     &   p=\nu_{1}  \\
(-1)^{p-\nu_1}&   \nu_{1} +1 \leq p \leq  \nu_1+\nu_2\,.
\end{cases}\\
&q_p = \begin{cases}
   \frac{1+\nu_{1}\nu_{2}}{\nu_{2}}-p        &  1\leq p \leq \nu_{1} - 1 \\
   \frac{1}{\nu_{2}}(p-\nu_{1})-1 & \nu_{1}\leq p \leq \nu_{1}+\nu_{2}-1 \\
   \frac{1}{\nu_{2}} & p= \nu_{1}+\nu_{2}\,.
\end{cases}
\end{align}

Considering now the gapped case $|\Delta| \geq 1$ and performing an analogous derivation one finds again an equation of the form Eq.~\eqref{eq:finalresult} with the replacements
\be
N_s=\infty,\quad n_p = p,\quad   \sigma_p= 1, \quad a_p(\lambda,\gamma) \mapsto a_p(\lambda,\eta) = \frac{1}{\pi} \frac{\sinh(p \eta)}{\cosh(p \eta)-\cos(2\lambda)}, \quad \int_{-\infty}^{\infty}  {\rm d}\lambda \mapsto \int_{-\pi/2}^{\pi/2} {\rm d}\lambda. 
\ee

\section{The non-interacting regime}
\label{SMsec:free}
Whenever 
\begin{itemize}
\item[(i)]  $\Delta\neq 0$ and $\tau=2 \pi n /\Delta$, with $n\in \mathbb Z$. 
\item[(ii)] $\Delta= 0$ and $\tau\in \mathbb R_+$
\end{itemize}
the Floquet operator \eqref{eq:floquet_even_odd} becomes Gaussian. Namely, it can be written as a quadratic form of spinless fermions through a standard Jordan-Wigner transformation 
\begin{align}
U_{o}(\tau) &= \exp{\left[i \frac{\tau}{4} \sum_{n=1}^{L/2} \sigma^{x}_{2n}\sigma^{x}_{2n+1}+ \sigma^{y}_{2n}\sigma^{y}_{2n+1}\right]} = \exp{\left[i \frac{\tau}{2} \sum_{n=1}^{L/2} c^{\dag}_{2n}c^{\phantom{\dag}}_{2n+1}+ c^{\dag}_{2n+1}c^{\phantom{\dag}}_{2n}\right]},\\
U_{e}(\tau) &= \exp{\left[i \frac{\tau}{4} \sum_{n=1}^{L/2} \sigma^{x}_{2n+1}\sigma^{x}_{2n+2}+ \sigma^{y}_{2n+1}\sigma^{y}_{2n+2}\right]} = \exp{\left[i \frac{\tau}{2} \sum_{n=1}^{L/2} c^{\dag}_{2n+1}c^{\phantom{\dag}}_{2n+2}+ c^{\dag}_{2n+2}c^{\phantom{\dag}}_{2n+1}\right]}.
\end{align}
Here we defined the fermionic operators  
\be
c^{\dag}_{n} = \frac{\sigma_n^x+i\sigma_n^y}{{2}}\prod_{j=1}^{n-1} \sigma_j^z,  \qquad c_{n}=\frac{\sigma_n^x-i\sigma_n^y}{{2}}\prod_{j=1}^{n-1} \sigma_j^z \,,
\ee
fulfilling the canonical anti-commutation relations. Considering a Fourier transform of the fermionic operators 
\be
c_k = \frac{1}{\sqrt{L}} \sum_{n=1}^{L} e^{i k n} c_n,
\ee
we find 
\be
U_{o}(\tau) = \exp\left[ \sum_{k>0} \begin{pmatrix} c_k^\dag & c^\dag_{k-\pi} \end{pmatrix} A  
\begin{pmatrix} c_k \\ c_{k-\pi} \end{pmatrix} \right],
\ee
\be
U_{e}(\tau) = \exp\left[ \sum_{k>0} \begin{pmatrix} c_k^\dag & c^\dag_{k-\pi} \end{pmatrix} B  
\begin{pmatrix} c_k \\ c_{k-\pi} \end{pmatrix} \right],
\ee
where the sum is over $k\in \frac{2\pi}{L}\mathbb Z_{L}\cap [0,\pi]$ and we introduced 
\be
A =  i \frac{\tau}{2} \begin{pmatrix} 
\cos(k) & i \sin(k) \\
 -i \sin(k) & -\cos(k) \\
\end{pmatrix}, \qquad  
B =  i \frac{\tau}{2}  \begin{pmatrix} 
\cos(k) & -i \sin(k) \\
 i \sin(k) & -\cos(k) \\
\end{pmatrix}.
\ee
The time-evolution operator can be written as a single exponential of quadratic form of fermions using the BCH formula and the fact that quadratic forms of fermions are closed under commutation algebra. Specifically, we obtain 
\be
U(\tau) = \exp\left[ \sum_k \begin{pmatrix} c_k^\dag & c^\dag_{k-\pi} \end{pmatrix} C  
\begin{pmatrix} c_k \\ c_{k-\pi} \end{pmatrix} \right],
\ee
with  
\be
e^{C} = e^{A} e^{B} = a \1 + i b \sigma^x + i d \sigma^z,
\ee
and 
\be
a = \cos(\tau/2)^2-\cos(2k)\sin(\tau/2)^2, \quad b= \sin^2(\tau/2)\sin(2k),\quad d= \cos(k)\sin(\tau).
\ee
Using the Pauli algebra we find 
\be
C = i \varepsilon_k e^{-\frac{i}{2} \varphi_k \sigma^y} \sigma^z e^{\frac{i}{2} \varphi_k \sigma^y},
\ee
with 
\begin{align}
\varphi_k &= \tan^{-1}\left(\frac{b}{d}\right) = \tan^{-1}\left({\sin(k) \tan(\tau/2)}\right)+\pi\theta_{\rm H}\left(k-\frac{\pi}{2}\right) ,\\
\varepsilon_k &= \tan^{-1} \frac{\sqrt{b^2+d^2}}{a} = \tan ^{-1}\left(\frac{\sqrt{\sin ^4\left(\frac{{\tau}}{2}\right) \sin ^2(2 k)+\sin ^2({\tau}) \cos ^2(k)}}{\cos ^2\left(\frac{{\tau}}{2}\right)-\sin ^2\left(\frac{{\tau}}{2}\right) \cos (2 k)}\right)\,,
\end{align}
and $\theta_{\rm H}(\cdot)$ is the step function. Therefore, defining the Bogoliubov fermions 
\be
\begin{pmatrix} b_k \\ b_{k-\pi} \end{pmatrix} = e^{\frac{i}{2} \varphi_k \sigma^y} \begin{pmatrix} c_k \\ c_{k-\pi} \end{pmatrix}
\ee
we finally obtain
\be 
U(\tau) = \exp\left[ i \sum_k \varepsilon_k(b^\dag_k b_k- b^\dag_{k-\pi} b^{\phantom{\dag}}_{k-\pi}) \right].
\ee

\subsection{Magnetization}

Let us now look at the expectation value of the magnetisation on a Gaussian state after a quantum quench
\begin{align}
\braket{\sigma^z_{j}(t)} &= \frac{2}{L}\sum_k \braket{\begin{pmatrix} c_k^\dag(t) & c^\dag_{k-\pi}(t) \end{pmatrix} \begin{pmatrix} 
1 & (-1)^j \\
(-1)^j & 1 \\
\end{pmatrix} 
\begin{pmatrix} c_k(t) \\ c_{k-\pi}(t) \end{pmatrix}} -1\notag\\
&= \frac{2}{L}\sum_k \braket{\begin{pmatrix} b_k^\dag e^{i \varepsilon_k t} & b^\dag_{k-\pi} e^{- i \varepsilon_k t} \end{pmatrix} 
e^{\frac{i}{2} \varphi_k \sigma^y} 
\begin{pmatrix} 
1 & (-1)^j \\
(-1)^j & 1 \\
\end{pmatrix} 
e^{-\frac{i}{2} \varphi_k \sigma^y} \begin{pmatrix} b_k e^{- i \varepsilon_k t} \\ b_{k-\pi} e^{i \varepsilon_k t} \end{pmatrix}} -1\notag\\
&= (-1)^j \frac{2}{L}\sum_k \braket{\begin{pmatrix} b_k^\dag e^{i \varepsilon_k t} & b^\dag_{k-\pi} e^{- i \varepsilon_k t} \end{pmatrix} 
e^{i \varphi_k \sigma^y} \begin{pmatrix} b_{k-\pi} e^{i \varepsilon_k t} \\ b_k e^{- i \varepsilon_k t} \end{pmatrix}}+\frac{2}{L}\sum_k\braket{b_k^\dag b_k}+\braket{b_{k+\pi}^\dag b_{k+\pi}} -1,
\end{align}
where we used the Pauli algebra. This expression gives a convenient sum representation of the expectation value of the magnetisation at all times. In particular, using that $\varepsilon_k>0$, we find 
\begin{align}
\lim_{t\to\infty}\lim_{L\to\infty}\braket{\sigma^z_{0}(t)}+\braket{\sigma^z_{1}(t)} &= \frac{2}{\pi} \int_0^\pi \!\!\!\! {\rm d}k\,  (\braket{b_k^\dag b^{\phantom{\dag}}_k} +  \braket{b_{k-\pi}^\dag b^{\phantom{\dag}}_{k-\pi}})-2\notag\\
&= \int_{-\pi}^\pi \!\!\!\! {\rm d}k\, 4  \rho(k)-2\,,
\end{align}
and 
\begin{align}
\lim_{t\to\infty}\lim_{L\to\infty}\braket{\sigma^z_{0}(t)}-\braket{\sigma^z_{1}(t)} &= \frac{2}{\pi} \int_0^\pi \!\!\!\! {\rm d}k\, \sin\varphi_k\,  (\braket{b_k^\dag b^{\phantom{\dag}}_k} -  \braket{b_{k-\pi}^\dag b^{\phantom{\dag}}_{k-\pi}})\notag\\
&= \int_{-\pi}^\pi \!\!\!\! {\rm d}k\, \frac{4 \sin k \sinh i x}{\sqrt{1+ \sin(k)^2 \sinh(i x)^2}}\,  \rho(k)\,,
\end{align}
where we expressed $\tau$ in terms of $x$ (cf.~\eqref{eq:def_x}) using that for the curves (i) and (ii) we have 
\be
x=i \operatorname{arcsinh}\left[\tan (\tau/2)\right],
\ee
and introduced the root density 
\be
\rho(k) = \frac{\braket{b_k^\dag b^{\phantom{\dag}}_k}}{2\pi}\,.
\ee
For instance, for a quench from the N\'eel state \eqref{eq:neel} we have 
\be
\braket{b_{k}^\dag b^{\phantom{\dag}}_k} = 2\pi \rho(k) =\frac{1}{2}+\frac{1}{2}\sin(\varphi_k) = \frac{1}{2}+ \frac{\sin k \sinh i x}{2\sqrt{1+ \sin(k)^2 \sinh(i x)^2}}\,,
\ee
and obtain 
\begin{align}
\!\!\!\lim_{t\to\infty}\lim_{L\to\infty}\braket{\sigma^z_{j}(t)} &=  \frac{(-1)^{j}}{\pi} \int_0^\pi \!\!\!\!{\rm d}k \frac{\sin(k)^2 \sinh(i x)^2}{1+ \sin(k)^2 \sinh(i x)^2} = (-1)^{j+1}\left (\frac{1}{|\cosh i x |}-1\right)=(-1)^{j+1}\left (|\cos \tau/2 |-1\right)\,.
\label{eq:free_magn}
\end{align}

\subsection{Current}

Let us now consider the magnetization current~\cite{ljubotina2019ballistic} 
\be
{\hat J_{2n}} = {\mathcal N_x}^{-1} \left( 2 ( \sigma_{2n-1}^+ \sigma_{2n}^- - \sigma_{2n-1}^- \sigma_{2n}^+) + i \sinh(i x) ( \sigma_{2n-1}^z  -  \sigma_{2n}^z)\right)
\label{eq:currentLenart}
\ee
where we introduced 
\be
\mathcal N_x = \frac{i(1 + \cosh 2 i x)}{2 \sinh i x}\,.
\ee
Its expectation value on a Gaussian state reads as 
\begin{align}
\braket{\hat J_{2n}} &= \frac{2}{\mathcal N_x L}\sum_k \braket{\begin{pmatrix} b_k^\dag e^{i \varepsilon_k t} & b^\dag_{k-\pi} e^{- i \varepsilon_k t} \end{pmatrix} 
( - i \sin(k) e^{i \varphi_k \sigma^y} \sigma^z + i (-1)^j \cos(k)\sigma^y) \begin{pmatrix} b_{k} e^{-i \varepsilon_k t} \\ b_{\pi-k} e^{i \varepsilon_k t} \end{pmatrix}}\notag\\
&-\frac{2i\sinh i x}{\mathcal N_x L}\sum_k \braket{\begin{pmatrix} b_k^\dag e^{i \varepsilon_k t} & b^\dag_{k-\pi} e^{- i \varepsilon_k t} \end{pmatrix} 
e^{i \varphi_k \sigma^y} \begin{pmatrix} b_{k-\pi} e^{i \varepsilon_k t} \\ b_k e^{- i \varepsilon_k t} \end{pmatrix}}\,.
\end{align}
Therefore, in the thermodynamic limit followed by the limit of infinte time we find 
\be
\lim_{t\to\infty} \lim_{L\to\infty}\braket{{\hat J_{2n}}} =  \int_{-\pi}^\pi \!\!\!\! {\rm d}k\, \frac{-2 \sin k \sinh i x}{\sqrt{1+ \sin(k)^2 \sinh(i x)^2}}\,  \rho(k) \left[\frac{2i/\sinh i x-2i\sinh i x}{\mathcal N_x}\right]
\ee
Noting 
\be
\varepsilon'_k = \frac{-2 \sin k \sinh i x\, {\rm sgn}(\cos(k))}{\sqrt{1+ \sin(k)^2 \sinh(i x)^2}}\,,
\ee
and 
\be
 \left[\frac{2i/\sinh i x-2i\sinh i x}{\mathcal N_x}\right] = 2, 
\ee
we have 
\be
\lim_{t\to\infty} \lim_{L\to\infty}\braket{{\hat J_{2n}}} = \int_{-\pi}^\pi \!\!\!\! {\rm d}k\,   2 \varepsilon'_k \rho(k)\,,
\ee
which, as expected, is the standard current formula of GHD~\cite{bertini2016transport,castro2016emergent}.

\section{Additional numerical results}
In this section we provide the numerical details of the iTEBD simulations employed to obtain the data in Fig.~\ref{fig:realtime} and some additional iTEBD results for $\Delta = 2.5$. Moreover, we compute the discrete free dynamics for $\tau = 2 \pi/\Delta$ via free fermion techniques and compare it with the analytical prediction of Sec.~\ref{SMsec:free}. Finally, we test our main result Eq.~\eqref{eq:finalresultmain} at some root of unity points by computing numerically the dGGE at finite size via exact diagonalization.

	\begin{figure}[h!]
	\centering
		\includegraphics[scale=0.45]{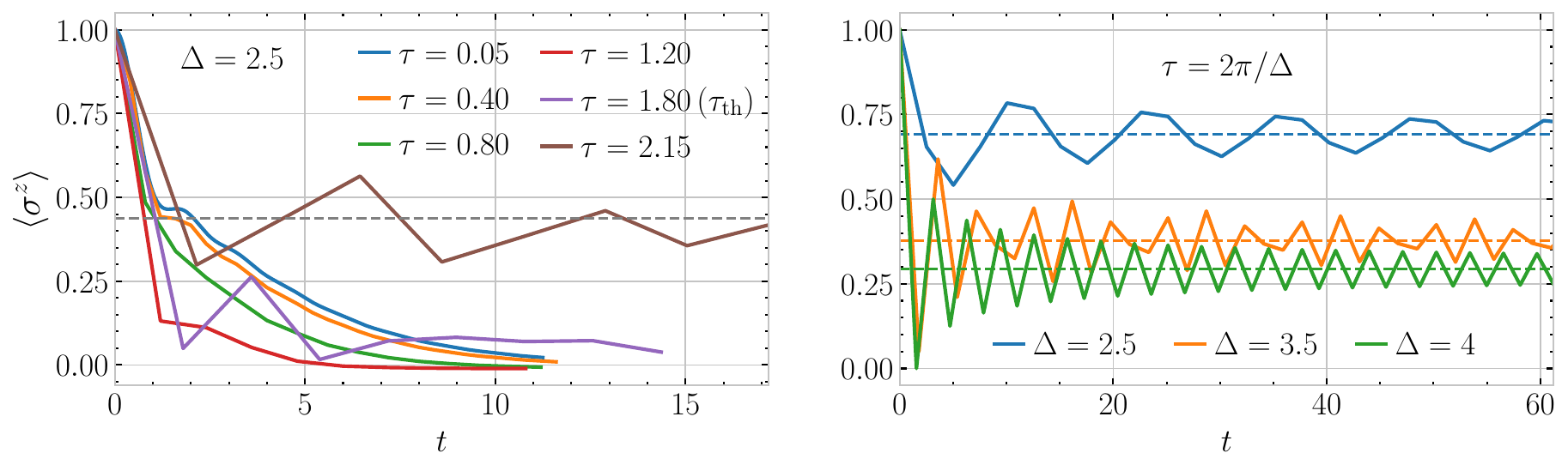}
		\caption{Left: Discrete interacting dynamics of $\sigma^z$ starting from the N\'eel state computed with iTEBD for $\Delta =2.5$ and several values of $\tau$ across the Trotter transition $\tau_{\mathrm{th}} \simeq 1.80$. For $\tau \le \tau_{\mathrm{th}}$ the magnetization approaches $0$, while it is consistent with the analytical prediction Eq.~\eqref{eq:finalresultmain} for $\tau > \tau_{\mathrm{th}}$. Right: Discrete free dynamics corresponding to $\tau = 2 \pi/\Delta$ for several values of $\Delta$ computed with Gaussian techniques. The dashed lines are the analytical value reported in Eq.~\eqref{eq:free_magn}. }
		\label{fig:numerical1}
	\end{figure}

 	\begin{figure}[h!]
	\centering
		\includegraphics[scale=0.45]{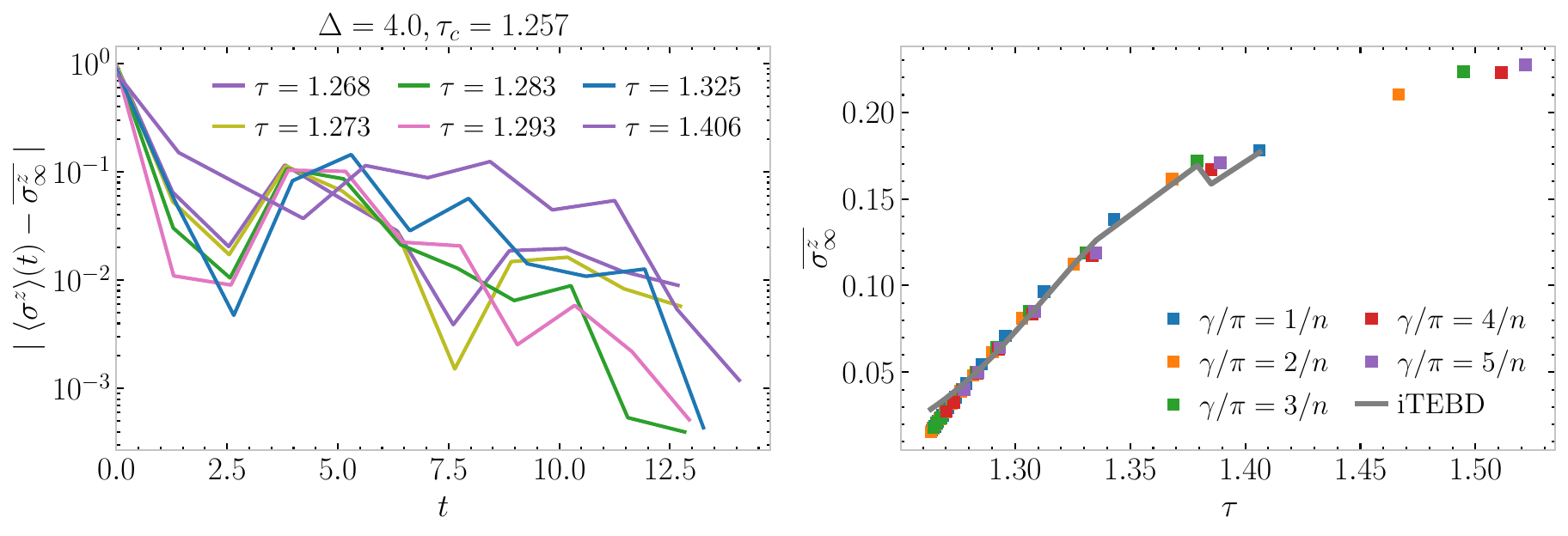}
		\caption{Left: Difference between the numerical value of $\braket{\sigma^z(t)}$ and the analytic asymptotic value as a function of time. Right: symbols show the analytic predictions for $\overline{\sigma^z_\infty}$, while the continuous gray curve is obtained stopping the iTEBD simulation at the maximum available time $t\sim 13$. The gray curve is not shown for larger values of $\tau$, due to manifest large finite-time effects.}
		\label{fig:numerical_phase_transition}
	\end{figure}

\subsection{iTEBD simulations}
The result in Fig.~\ref{fig:realtime} have been obtained by performing iTEBD simulations of the discrete dynamics Eq.~\eqref{eq:floquet_even_odd} applied to the N\'eel state. The bond dimension is increased during the time evolution such that the truncation error is kept below $10^{-10}$. The simulation is stopped when the required bond dimension grows above $7000$. The smaller is $\Delta (> 1)$, the shorter is the time reached with this stopping criterion and the fastest is the approach to the thermal equilibrium value $\overline{\sigma^z} = 0$ for $\tau < \tau_{\mathrm{th}}$. In Fig.~\ref{fig:numerical1} (left) we show that $\langle \sigma^z (t) \rangle$ rapidly approaches zero for $\tau \le \tau_{\mathrm{th}}$, while it oscillates around the non-zero infinite-time prediction obtained from Eq.~\eqref{eq:finalresultmain} (dashed line) for the root of unity point $\gamma/\pi = 1/3$ ($\tau \simeq 2.15$), cf. Eq.~\eqref{eq:def_x}. Further numerical tests of our analytic predictions for $\overline{\sigma^z}$ are reported in Fig.~\ref{fig:numerical_phase_transition}.

\subsection{Gaussian discrete dynamics}
As explained in Sec.~\ref{SMsec:free}, when $\tau = 2 \pi/\Delta$ the Floquet operator is Gaussian. Therefore, the dynamics can be computed with an amount of resources that only scales linearly with time, allowing us to reach times much larger than in the interacting case. In Fig.~\ref{fig:numerical1} (right) we show the Gaussian discrete dynamics of the magnetization for $\Delta = 2.5,3.5,4$, against its infinite-time prediction Eq.~\eqref{eq:free_magn} (dashed lines).

\subsection{dGGE at finite size}

	\begin{figure}
	\centering
		\includegraphics[scale=0.45]{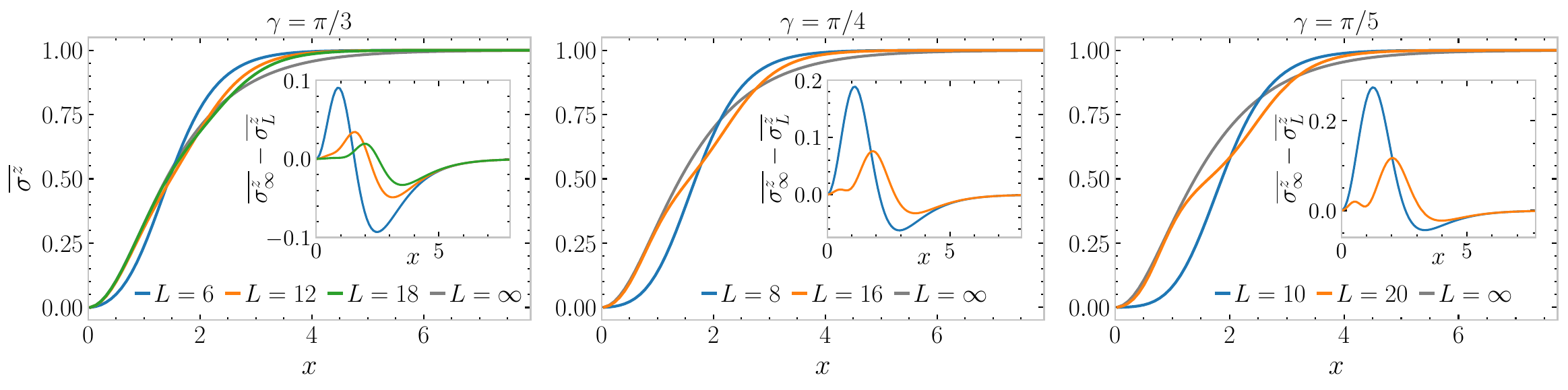}
		\caption{Comparison between the finite-size infinite-time expectation values of $\sigma^z$ extracted from the dGGE computed numerically on a chain with periodic boundary conditions and the analytical result Eq.~\eqref{eq:finalresultmain} ($L=\infty$) at the root of unity points $\gamma/\pi = 1/3, 1/4, 1/5$. The insets show the difference between the infinite-size prediction and the finite-size numerical values.  }
		\label{fig:numerical2}
	\end{figure}

A complementary approach to check the correctness of Eq.~\eqref{eq:finalresultmain} consists in computing numerically the dGGE by exact diagonalization of the Floquet operator Eq.~\eqref{eq:floquet_even_odd}, and using it to extract the finite-size infinite-time expectation values $\overline{\sigma^z_L}$. We perform this calculation by exploiting all the symmetries of the Hamiltonian (except from reflection symmetry), namely $2$-sites translation and $U(1)$ symmetries, which allow to fully diagonalize the Floquet operator up to $L=20$. In Fig.~\ref{fig:numerical2} we report the comparison between Eq.~\eqref{eq:finalresultmain} at the root of unity points $\gamma/\pi = 1/3, 1/4, 1/5$ and the value extracted from the numerical dGGE, as function of $x$, cf. Eq.~\eqref{eq:def_x}. As we observed strong finite-size effects in the dGGE result when $L$ is not multiple of $2 k$, with $\gamma = \pi/k$, we only plot the results for $L=2 k$, which are consistent with an approach to the exact prediction in the thermodynamic limit. Note that the limit $x\to \infty$ corresponds to the dual-unitary point of the dynamics, $\tau=\pi$, which fixes the N\'eel state. Accordingly, the staggered magnetization is constant in time, and its late time limit is equal to $1$. 
\ \\
\ \\
\section{More general initial states}

Our predictions for the existence of a locally-detectable Trotter transition were derived for the case of the N\'eel state, where we could carry out analytic computations. However, we claim that they hold more generally (although analytic predictions might be difficult to obtain for other initial states). In particular, we show that a similar transition is expected for initial states with the same symmetries of the N\'eel state, namely for states breaking the translation ($\mathcal{T}$) and spin-flip ($\mathcal{S}$) symmetries individually, but which are invariant under the combined symmetry $\mathcal{T}\mathcal{S}$. 

Let us consider an initial state $\ket{\Psi_0}$, which is two-site shift invariant, breaks the symmetries $\mathcal{T}$ and $\mathcal{S}$, but satisfies
\begin{equation}\label{eq:initial_symmetry}
	\mathcal{T}\mathcal{S}\ket{\Psi_0}=\ket{\Psi_0}\,.
\end{equation}
Consider now a quench from $\ket{\Psi_0}$, where the system is evolved according to the Trotterized Hamiltonian with $\Delta>1$. We show that for $\tau<\tau_{\rm th}(\Delta)$ the late-time limit of the staggered magnetization is zero. To this end, we will assume that, in this regime, all the conserved charges are spin-flip invariant, except for the total magentization $S_z$. This result is established in the Hamiltonian case~\cite{ilievski2016quasilocal}, and we assume that it holds in our setting as well (consistently with our analytic computations for the N\'eel state). Therefore, the dGGE takes the general form
\begin{equation}
	\rho_{\rm GGE}(\beta_0) = e^{\beta_0 S_z} \widetilde{\rho}_{\rm GGE}\,,
\end{equation}
where $\widetilde{\rho}_{\rm GGE}$ is invariant under $\mathcal{S}$. For the initial state~\eqref{eq:initial_symmetry}, the GGE must be invariant under $\mathcal{T}\mathcal{S}$, yielding $\beta_0=0$. In turn, this implies that the expectation value of the staggered magnetization must be zero. Indeed, defining
\begin{equation}
	A=\sum_{j}(-1)^{j}s_j^z\,,
\end{equation}
we have ${\rm Tr} [A \widetilde{\rho}_{\rm GGE}]=-{\rm Tr} [\mathcal{S}A\mathcal{S}^\dagger  \widetilde{\rho}_{\rm GGE}]=-{\rm Tr} [A \widetilde{\rho}_{\rm GGE}]$. On the other hand, for $\tau>\tau_{\rm th}(\Delta)$, additional conservation laws appear, breaking both $\mathcal{T}$ and $\mathcal{S}$. Therefore, denoting by $Q$ one such charge, $(\mathcal{T} \mathcal{S} )Q(\mathcal{T} \mathcal{S} )^\dagger\neq - Q$, so that the associated Lagrange multiplier is not forced to be zero. Excluding fine-tuned cases, this implies a non-zero asymptotic value for the staggered magnetization for $\tau>\tau_{\rm th}(\Delta)$. In conclusion, quenches from initial states~\eqref{eq:initial_symmetry} are expected to display a Trotter transition similar to that predicted analytically for the N\'eel state.

\end{document}